\documentclass[aps,reprint,prb,floatfix,footinbib,superscriptaddress]{revtex4-2}
\usepackage{hyperref}
\hypersetup{
  breaklinks=true, 
  colorlinks=true, 
  citecolor=blue,
  urlcolor=blue,
  linkcolor=blue,
  pdfusetitle=true, 
}
\usepackage{braket} 
\usepackage{float}
\usepackage{epsfig}
\usepackage{graphicx}
\usepackage{amsmath}
\usepackage{amsfonts}
\usepackage{amssymb}
\usepackage{bm}
\usepackage{bbold}
\usepackage{epsfig}
\usepackage{graphicx}
\usepackage{color}
\usepackage{pictex}
\usepackage{mathtools}
\usepackage{extarrows}
\usepackage{lipsum}
\usepackage{footmisc}
\usepackage{xcolor}
\usepackage{ulem}

\usepackage{comment}

\begin{document}
\title{A comparative study
of perturbative and nonequilibrium
Green's function approaches for Floquet sidebands in periodically driven
quantum systems}

\newcommand{\affiliationMPSD}{
Max Planck Institute for the Structure and Dynamics of Matter,
Center for Free-Electron Laser Science (CFEL),
Luruper Chaussee 149, 22761 Hamburg, Germany
}

\newcommand{\affiliationBremen}{
Institute for Theoretical Physics and Bremen Center for Computational Materials Science,
University of Bremen, 28359 Bremen, Germany
}

\author{Karun Gadge}
\affiliation{Institute for Theoretical Physics, Georg-August-University G\"{o}ttingen, Friedrich-Hund-Platz 1, D-37077 G\"{o}ttingen, Germany}

\author{Marco Merboldt}
\affiliation{I. Physikalisches Institut, Georg-August-Universit\"at G\"ottingen, Friedrich-Hund-Platz 1, D-37077 G\"{o}ttingen, Germany}

\author{Michael Schüler}
\affiliation{PSI Center for Scientific Computing, Theory and Data, Paul Scherrer Institute, Villigen PSI, Switzerland}
\affiliation{Department of Physics, University of Fribourg, Fribourg, Switzerland}

\author{Jan Philipp Bange}
\affiliation{I. Physikalisches Institut, Georg-August-Universit\"at G\"ottingen, Friedrich-Hund-Platz 1, D-37077 G\"{o}ttingen, Germany}

\author{Wiebke Bennecke}
\affiliation{I. Physikalisches Institut, Georg-August-Universit\"at G\"ottingen, Friedrich-Hund-Platz 1, D-37077 G\"{o}ttingen, Germany}

\author{Michael A. Sentef}
\affiliation{\affiliationBremen}
\affiliation{\affiliationMPSD}

\author{Marcel Reutzel}
\affiliation{I. Physikalisches Institut, Georg-August-Universit\"at G\"ottingen, Friedrich-Hund-Platz 1, D-37077 G\"{o}ttingen, Germany}
\affiliation{Department of Physics, Philipps-Universität Marburg, D-35032 Marburg, Germany}
\affiliation{mar.quest | Marburg Center for Quantum Materials and Sustainable Technologies,
Hans-Meerwein-Straße 6, D-35032 Marburg, Germany}

\author{Stefan Mathias}
\affiliation{I. Physikalisches Institut, Georg-August-Universit\"at G\"ottingen, Friedrich-Hund-Platz 1, D-37077 G\"{o}ttingen, Germany}
\affiliation{International Center for Advanced Studies of Energy Conversion (ICASEC), University of Göttingen, Göttingen, Germany}

\author{Salvatore R. Manmana}
\affiliation{Institute for Theoretical Physics, Georg-August-University G\"{o}ttingen, Friedrich-Hund-Platz 1, D-37077 G\"{o}ttingen, Germany}

%
%
\begin{abstract}

We compare two complementary theoretical approaches to compute and interpret Floquet sidebands in periodically driven quantum materials:
a first-order perturbative approach (1st order perturbative Born approximation, PB1) and time-dependent nonequilibrium Green’s functions (tdNEGF).
Using graphene as a model Dirac system, we disentangle in pump-probe setups Floquet-dressed initial states, Volkov-dressed final states (also known as laser-assisted photoelectric effect, LAPE), and their interference. 
We quantify how photoemission matrix elements, polarization, incidence angle, and near-surface screening shape the momentum-resolved sideband intensity observed in tr-ARPES.
PB1 yields an analytical expression for the momentum-dependent sideband intensity and for graphene it captures the correct symmetry trends, such as the magnitude of the intensities when considering the interference between the Floquet and the Volkov states and photoemission matrix elements.
tdNEGF reproduces the full energy–momentum–resolved spectra, including hybridization gaps and spectral-weight redistribution.
We find qualitative agreement between PB1 and tdNEGF once matrix elements are included; quantitative differences arise near hybridization regions and at specific angles where higher-order processes and self-energies are essential.
Thus, for systems with simple band structures and away from these regions, the two approaches can be used in a complementary way.
\end{abstract}
\maketitle


%
%
%
\section{Introduction}\label{Intro}
The concept of Floquet engineering---using time-periodic driving fields to create and manipulate novel states of matter---has attracted wide attention across condensed matter physics~\cite{Basov2017,Oka2019,delaTorre2021,Bao2021,Goldman2014,Eckardt2017,Mentink2015,Valmispild2024,Li2020,Murakami2025,Giovannini2019,Rudner2020}.
A broad range of experiments have reported signatures of Floquet physics, including transport 
measurements~\cite{DayMcIver2024,Grifoni1998,liu2025}
and optical probes~\cite{Shimano2024,Hsieh2022,Bao2025}. 
While these studies demonstrate light-induced modifications of materials' properties, they typically provide only indirect evidence for the existence of Floquet–Bloch states, since they do not resolve the transient electronic band structure in energy and momentum space.
Time- and angle-resolved photoemission spectroscopy (tr-ARPES)~\cite{Sobota2021,Boschini2024} offers the most direct window into Floquet physics by mapping the band structure of driven solids with femtosecond resolution. 
Light-induced Floquet sidebands have been observed in semiconductors, topological insulators, and semimetals~\cite{Wang2013,Mahmood2016,DeGiovannini2016,Gedik2013,Gierz2021,Merboldt2025,Choi2025,Bielinski2025,Fragkos2025,Bao2025}, but interpreting these features requires care.
The measured spectra contain overlapping contributions from Floquet–Bloch states inside the crystal and Volkov-dressed final states (also known as the laser-assisted photoelectric effect, LAPE~\cite{Glover1996,Schins1996,Saathoff2008}), as well as their interference~\cite{Keunecke2020,Park2014,Merboldt2025,Choi2025,Bao2025,Fragkos2025,Yen2025}.
Distinguishing these mechanisms demands a detailed understanding of photoemission matrix elements, polarization, incidence angle, and near-surface screening, supported by robust theoretical modeling.

In this work, we present a systematic benchmark of two complementary theoretical approaches for computing and interpreting Floquet sidebands in periodically driven quantum materials.
The first is a perturbative method (1st order perturbative Born approximation, PB1) introduced by S.T.~Park \cite{Park2014}, which provides analytical amplitude maps and captures symmetry-trends in the sideband intensities (see, e.g.,~\cite{ Choi2025,Keunecke2020}).
The second method employs time-dependent nonequilibrium Green’s functions (tdNEGF) \cite{Mahmood2016,Freericks2015,Sentef2015,Ito2023,Schler2021,Merboldt2025}, yielding full energy- and momentum-resolved spectra that naturally include higher-order processes, dynamical renormalization, and finite lifetime effects.
Using graphene as a prototypical Dirac material \cite{Bostwick2006,CastroNeto2009}, we identify the regimes in which PB1 provides a reliable description of the transient photoemission response, 
and those where fully dynamical simulations (such as tdNEGF) are required for quantitative comparison with experiment. 

The Floquet description of the time-periodic Hamiltonian goes back to Shirley’s formulation of quasienergies and stroboscopic evolution~\cite{Shirley1965}.
On the Volkov side, laser-assisted photoemission (LAPE) and dressed free-electron final states have been developed in the context of strong-field physics of atoms and molecules and were adapted to solids and surfaces~\cite{Reiss1980,MiajaAvila2006}.
In solids, Floquet and Volkov channels coexist and
interfere—a point emphasized theoretically and observed
experimentally~\cite{Park2014,Bao2025,Merboldt2025,Choi2025}.
Understanding this interference is central to interpreting modern pump-probe tr-ARPES experiments, as the relative phase and amplitude of the two channels depends sensitively on the  geometry, the polarization, and near-surface dielectric screening.
The photoemission matrix elements mostly shape the angular distribution of intensity maps; hence, 
they must be handled consistently when comparing 
theoretical approaches with each other and with experiment~\cite{Hwang2011,Moser2017,Krasovskii2021}.
In addition, equally important are the electromagnetic boundary conditions at the sample surface.
We incorporate these using Fresnel reflection and transmission coefficients for the incident pump, accounting for refraction and field mixing between in-plane and out-of-plane components~\cite{Neppl2015,Chen2017,Tao2016}.
However, idealized Fresnel optics may not fully capture the microscopic screening and substrate-dependent response of real samples such as graphene on dielectric supports.
The theoretical frameworks must account for possible field mixing or imperfect
surface response.
To bridge this gap, we introduce phenomenological screening parameters that independently renormalize the effective Volkov (vacuum) and Floquet (in-sample) field amplitudes (also see Refs.~\onlinecite{Neppl2015,Merboldt2025}).
The parameters adapt the effective fields used in the 
calculations, allowing for sensitivity checks against assumptions regarding substrate permittivity, providing a controlled framework for connecting idealized theory with realistic experimental environments.

The remainder of the paper is organized as follows:
In Sec.~\ref{sec:drive} we discuss basic aspects of the periodically driven system and the geometry of the setup considered in this paper.
In Sec.~\ref{sec:FVI-intro}, we introduce the theoretical background of Floquet and Volkov states and describe the corresponding interference channels relevant to pump–probe photoemission for the PB1 in Sec.~\ref{subsec:PB1-intro} and for the tdNEGF in Sec.~\ref{subsec:tdNEGF}. 
Section~\ref{sec:results} presents results from both approaches across various geometry, polarization angles, and field-strength regimes, as well as direct comparison to experimental tr-ARPES data on graphene.
We conclude with the discussion in
Sec.~\ref{sec:discussion_outlook}.
Further details on the matrix elements within the PB1 method are discussed in Appendix~\ref{app:matrix_element_PB1}, and Appendix~\ref{app:GaussianFitting} discusses a smoothening scheme used in some of the Figures. 

\section{Periodic driving and geometry of the setup}
\label{sec:drive}
Throughout the paper, we treat systems of electrons on a lattice, which do not interact with each other, and which are periodically driven by the time-dependent vector potential of incoming monochromatic laser pulses.
The coupling between the light-field and the electrons is incorporated via the Peierls substition. 

For the PB1 approach, we compute the dynamics of the quantum mechanical state under such an external periodic drive by solving the time-dependent Schr\"odinger equation (TDSE) with a time-periodic Hamiltonian,
\begin{equation}
i \hbar \frac{\mathrm{d}}{\mathrm{d} t}|\psi(t)\rangle=\mathcal{H}(t)|\psi(t)\rangle \\
\text { with } \mathcal{H}(t)=\mathcal{H}(t+T) \,.
\end{equation} 
For simplicity, we work with pure states and formally work at temperatures $T=0$. 
For the tdNEGF approach, we directly solve the equations of motion for the single-particle density matrix, which also allows us to treat finite temperatures; this approach will be explained in more detail in Sec.~\ref{subsec:tdNEGF}.
Using Floquet's theorem~\cite{Floquet1883}, one finds that the time evolution operator for the dynamics in a time-interval $t \in [t_0,t_0+T]$ can be written as $\mathcal{U}\left(t_0+T, t_0\right)=e^{-i \mathcal{H}_{\mathrm{F}} \, T / \hbar}$, 
with $\mathcal{H}_F$ hermitian 
and time-independent; this ensures that $\mathcal{U}\left(t_0+T, t_0\right)$ is unitary.
The
eigenvalues are expressed in the form $e^{-i\epsilon _n T/\hbar}$ where the $\epsilon _n$ are multi-valued and defined by the integer multiples of $\hbar \omega$, where $\omega = 2 \pi/T$.
In direct analogy with the \textit{quasi-momentum} that arises in systems possessing spatial periodicity, the  $\epsilon _n$ are referred to as `quasi-energies'.
For a time-periodic Hamiltonian, solutions are known in the literature as `Floquet states' $\ket{\psi(t)}=e^{-i\epsilon _n (t-t_0)/\hbar} \ket{u_n(t)}$, here $\ket{u_n(t)}$ is the time-periodic Floquet mode, i.e. the eigenfunction of the one-period evolution operator with periodicity $T$.
They are the time-domain analogue of Bloch states in spatially periodic systems.
\subsection{Interface effects}\label{subsec:interface_effect}
\begin{figure}[t]
    \centering
\includegraphics[width=\linewidth]{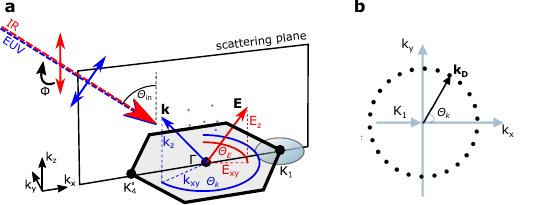} %
    \caption{Setup geometry and relevant  vectors: (a) The scattering plane is along the $\Gamma-K_1$ direction. 
    Both IR drive and EUV probe impinge under $\theta_{\mathrm{in}}$. In this coordinate frame, the in-plane field is parametrized as $(\textbf{E}_{xy},\theta_E)$ and the momenta as $(\textbf{k}_{xy},\theta_k)$; similarly $\textbf{E}_z$ and $\textbf{k}_z$ are out of plane components. The polarization is defined by the angle $\phi$. 
    (b) Sketch of the $(k_x,k_y)$-plane parametrized around the $K_1-$ point using the vector $\mathbf{k_D}$. }
    \label{fig:k-gamma}
\end{figure}

The geometry of the setup is displayed in Fig.~\ref{fig:k-gamma}.
We take the incident electric field to be monochromatic and real valued,
\begin{equation}
\mathbf{E}(t)=\Re\left[\mathbf{E}_{xyz} e^{-i\omega t}\right]
           =\mathbf{E}_{xyz} \cos(\omega t),
           \label{eq:E_t}
\end{equation}
with amplitude
\begin{equation}
\mathbf{E}_{xyz}=
E_0
\begin{pmatrix}
\cos\theta_{\mathrm{in}}\cos\phi\\[2pt]
\sin\phi\\[2pt]
\sin\theta_{\mathrm{in}}\cos\phi
\end{pmatrix},
\label{eq:E0_xyz}
\end{equation}
where $E_0$ is the field magnitude, $\theta_{\mathrm{in}}$ the incidence angle, and
$\phi$ the polarization angle. 
Let the interface lie in the $(x, y)-$plane with unit normal $\hat{\mathbf{n}}=\hat{\mathbf{z}}$,
and the plane of incidence is chosen as the $(x, z)-$plane. 
The unit polarization
vectors are denoted as $\hat{\mathbf{s}}$ for the one perpendicular to the plane of incidence, and $\hat{\mathbf{p}}$ for the one in the plane of incidence, and are given by
\begin{equation}
\hat{\mathbf{s}}=\hat{\mathbf{y}},\qquad
\hat{\mathbf{p}}_j=(\cos\theta_j,\,0,\,\kappa_j\sin\theta_j)\,.
\end{equation}
Here $j=\{\mathrm{in},r,t\}$ denote incident, reflected, and transmitted waves;
$\theta_j$ is the corresponding propagation angle, and
$\kappa_{\mathrm{in}}=+1$, $\kappa_r=-1$, $\kappa_t=+1$.
The decomposition of the incident field is therefore
\begin{equation}
\mathbf{E}_{\rm in}(t)=
E_0\!\left[\cos\phi\,\hat{\mathbf{p}}_{\rm in}+\sin\phi\,\hat{\mathbf{s}}\right]\cos(\omega t).
\label{eq:Ei_sp}
\end{equation}
We use Snell’s law, $n_1\sin\theta_{\mathrm{in}}=n_2\sin\theta_t$, to match the phase
across the interface, where $n_1$ and $n_2$ are the refractive indices of the
incident and transmitted media.
The boundary conditions at the interface define the Fresnel amplitude coefficients.
We denote by $r_s$ and $r_p$ the reflection
coefficients for $s$- and $p$-polarized fields, respectively (ratio of reflected to
incident field amplitudes), and by $t_s$ and $t_p$ the corresponding transmission
coefficients (ratio of transmitted to incident amplitudes). Explicitly,
\begin{equation}
\begin{aligned}
r_s&=\frac{n_1\cos\theta_{\mathrm{in}}-n_2\cos\theta_t}
          {n_1\cos\theta_{\mathrm{in}}+n_2\cos\theta_t},\qquad
t_s=\frac{2n_1\cos\theta_{\mathrm{in}}}
          {n_1\cos\theta_{\mathrm{in}}+n_2\cos\theta_t},
\\[4pt]
r_p&=\frac{n_1\cos\theta_t-n_2\cos\theta_{\mathrm{in}}}
          {n_1\cos\theta_t+n_2\cos\theta_{\mathrm{in}}},\qquad
t_p=\frac{2n_1\cos\theta_{\mathrm{in}}}
          {n_1\cos\theta_t+n_2\cos\theta_{\mathrm{in}}}.
\end{aligned}
\label{eq:Fresnel}
\end{equation}
The incident, reflected,
and transmitted fields are
\begin{equation}
\begin{aligned}
&\mathbf{E}_{\rm in} = 
E_0\cos(\omega t) \left(\cos\phi\,\hat{\mathbf{p}}_{\rm in}+\sin\phi\,\hat{\mathbf{s}}\right),\nonumber \\
&\mathbf{E}_r = 
E_0\cos(\omega t) \left(\cos\phi\,r_p\,\hat{\mathbf{p}}_r+\sin\phi\,r_s\,\hat{\mathbf{s}}\right),\nonumber \\
&\mathbf{E}_t = 
E_0\cos(\omega t) \left(\cos\phi\,t_p\,\hat{\mathbf{p}}_t+\sin\phi\,t_s\,\hat{\mathbf{s}}\right).
\end{aligned}
\label{eq:EiErEt_sp}
\end{equation}
In Cartesian components,
\begin{equation}
\begin{aligned}
&\mathbf{E}_{\rm in} = E_0\cos(\omega t)
\big(\cos\phi\cos\theta_{\mathrm{in}},\;\sin\phi,\;\cos\phi\sin\theta_{\mathrm{in}}\big),\nonumber \\
&\mathbf{E}_r = E_0\cos(\omega t)
\big(\cos\phi\,r_p\cos\theta_{\mathrm{in}},\;
     \sin\phi\,r_s,\;
    -\cos\phi\,r_p\sin\theta_{\mathrm{in}}\big),\nonumber \\
&\mathbf{E}_t = E_0\cos(\omega t)
\big(\cos\phi\,t_p\cos\theta_t,\;
     \sin\phi\,t_s,\;
     \cos\phi\,t_p\sin\theta_t\big).
\end{aligned}
\label{eq:EiErEt_xyz}
\end{equation}

Fig.~\ref{fig:k-gamma} shows a sketch of our setup, 
where we relate the geometry of the experimental setup to the hexagonal 2D Brillouin zone in $k-$space.
Fig.~\ref{fig:k-gamma}(a) illustrates the infrared (IR) pump and the extreme-ultraviolet (EUV) probe pulse setup
for incidence along the $\Gamma - K_1$ direction. 
The pump field impinges onto the surface at incidence angle $\theta_{\mathrm{in}}$.
The in-plane electric field 
in the $xy$-plane is parametrized as $(E_{xy},\theta_E)$, while the momentum-component in this plane is parametrized as $(k_{xy},\theta_k)$.
Accordingly, $E_z$ and $k_z$ denote the out-of-plane $z$-components.
Figure~\ref{fig:k-gamma}(b) shows
the $(k_x,k_y)$-plane in the vicinity of the $K_1-$point, where in this paper we analyze the corresponding angular intensity distribution.

\section{Floquet, Volkov and Interference states}\label{sec:FVI-intro}

In a pump–probe setup, emitted photoelectrons may absorb additional photons. This leads to the laser-assisted photoemission effect (LAPE)~\cite{Keunecke2020,Park2014,Merboldt2025,Choi2025}.
This phenomenon has been studied in the literature~\cite{Glover1996,Schins1996,Saathoff2008,GUMHALTER2025100768,Pazourek2013},
where it is understood in terms of Volkov solutions for free electrons, commonly referred to as Volkov states: a special case of Floquet states applicable to unbound electrons.
In both the Floquet and Volkov scenarios, the periodic drive results in a set of replica energy states, spaced by the photon energy of the driving field \cite{Park2014}. 
These can be interpreted as “echoes” of the equilibrium energy spectrum, shifted by integer multiples of the drive frequency.

We employ the perturbative approach and the tdNEGF approach in order to analyze the contributions to the spectral function stemming from the Floquet-, and Volkov-sector, respectively, and their interference.

\subsection{Park's ansatz with linearized bands (PB1)}\label{subsec:PB1-intro}
In the following we highlight the main points of the approach presented by S.T. Park in Ref.~\onlinecite{Park2014}, as needed for the discussion in the present paper.

We consider the interaction of an electron with a classical light field; the
time-dependent Hamiltonian is given as
\begin{equation}
H(t)=\frac{1}{2m}\big(\mathbf{P}-q\,\mathbf{A}(t)\big)^{2},
\end{equation}
where $\mathbf{P}$ is the momentum operator of the electron, $\mathbf{A}(t)$ is a spatially uniform vector potential, and $q$ is the electron charge.
The Volkov state is
given as a solution for free electrons in an electromagnetic
field
\cite{Reiss1980,Bergou1980,Madsen2005bl}:
\begin{equation}
\Psi_{\mathbf{P}}^{\mathrm{V}}(\mathbf{r},t)
=\frac{1}{(2\pi\hbar)^{3/2}}
e^{\left[i\left(\mathbf{P}\cdot\mathbf{r}/\hbar+\Theta(t)\right)\right]}\, ,
\end{equation}
where
\begin{equation}
\Theta(t)
=-\frac{1}{\hbar}\int_{-\infty}^{t}dt'\,
\varepsilon_{\mathbf{P}}(t'),\qquad
\varepsilon_{\mathbf{P}}(t')
=\frac{\big(\mathbf{P}-q\mathbf{A}(t')\big)^{2}}{2m}\,.
\end{equation}

The electrical field of the monochromatic laser pulses is associated to the vector potential via $\mathbf{E}(t)=-\partial \mathbf{A}(t)/\partial t$, where $\mathbf{E}(t) = \mathbf{E}_{xyz,\,\mathrm{ph}} \cos(\omega_\mathrm{ph} t)$.
By 
expanding $e^{i\Theta(t)}$,
we obtain the familiar $\hbar\omega$-spaced sidebands with Bessel weights that appear in laser-assisted photoemission.
The Volkov state is, hence, given by~\cite{Reiss1980,Bergou1980,Madsen2005bl}:
\begin{equation}
\Psi^{\mathrm{V}}_{\mathbf{P}}(\mathbf{r},t)=
e^{\left[\frac{i}{\hbar}\,\mathbf{P}\cdot\mathbf{r} - i \omega t \right]}
\sum_{j=-\infty}^{+\infty} (-i)^j
J_{j}(\alpha)\,
e^{i\left( j\omega_{\mathrm{ph}} t\right)}\,.
\label{eq:Volkov_solution}
\end{equation}
Here, $\omega_{\mathrm{ph}}$ is frequency of the driving field (photon frequency) and $\omega=\mathbf{P}^2/2m\hbar$.
We then define the \textit{Volkov parameter} by~\cite{Park2014,Choi2025,Keunecke2020}:
\begin{equation}
\alpha = \frac{q\,\boldsymbol{\nu}_{\alpha}\cdot\mathbf{E}_{xyz,\,\mathrm{ph}}}{\hbar\omega^2_\mathrm{ph}}\,.
\label{eq:AlphaBeta_parameters}
\end{equation}
Here, $q$ is the charge and
$\boldsymbol{\nu}_{\alpha}$ is
the velocity of the photoelectron~\cite{Park2014,Choi2025,Keunecke2020}, $\mathbf{E}_{xyz,\,\mathrm{ph}}$ is the driving field,
and $\hbar\omega_{ph}$ is the photon energy.
As follows from Eqs.~\eqref{eq:Volkov_solution} and~\eqref{eq:AlphaBeta_parameters},
the intensity amplitude of the first sideband due to the Volkov contribution alone is
\begin{equation}
I_1^{\rm{Volkov}} = J_1(\alpha)^2 \, .
\label{eq:Volkov_parameters}
\end{equation}
In addition, one can approximate the Bessel-functions as $J_1(x) \approx x$ for the leading order contribution so that in leading order the intensity of the first sideband is given by $I_1^{\rm Volkov} = \alpha^2$.

We parameterize our equations at the $K_1-$point (see Fig.~\ref{fig:k-gamma}) and use the field vector, following Eqs.~\eqref{eq:E_t} and~\eqref{eq:E0_xyz} (see also~\cite{Choi2025,Merboldt2025,Keunecke2020}). 
In this setup, the polarization angle $\phi=0^{\circ}$ ($90^{\circ}$) corresponds to $p$- ($s$-) polarized light. 
The incidence angle $\theta_{\mathrm{in}}$ describes the transformation to the laboratory frame.
The Volkov parameter involves the electron velocity, for this we use 
$\boldsymbol{\nu}_{\alpha}=\mathbf{P}/m_e$ with
$\mathbf{P}=\hbar\mathbf{k}$, where $m_e$ is the electron mass.
This yields
\begin{equation}
\boldsymbol{\nu}_{\alpha}
=\frac{\hbar}{m_e}
\begin{pmatrix}
k_x + k_D\cos\theta_k \\
k_D\sin\theta_k \\
k_z
\end{pmatrix},
\end{equation}
where $\mathbf{k_D} = (k_D\cos\theta_k, \,k_D\sin\theta_k)$ is
the distance vector to the $K_1$-point, and $k_z$ is the out-of-plane component. 
We now calculate $\alpha$ from Eq.~\eqref{eq:AlphaBeta_parameters}. 
The photoelectron carries a full three-dimensional momentum $\mathbf{P}$, so that the three-dimensional polarization of the driving field must be taken into account.
We therefore use the effective external field at the surface, see Subsec.~\ref{subsec:interface_effect}, as a superposition of the incident and the reflected components 
\begin{align}
\mathbf{E}_{\mathrm{eff,out}}
&=\mathbf{E}_{\mathrm{in}}+\mathbf{E}_{r} = \mathbf{E}_{xyz,\,\mathrm{ph}} \cos(\omega_{\mathrm{ph}}t) \nonumber\\[2pt]
&=E_{0}\Big[
\cos\phi\,\cos\theta_{\mathrm{in}}(1+r_{p})\,\hat{\mathbf{x}}
+\sin\phi\,(1+r_{s})\,\hat{\mathbf{y}}\nonumber\\[2pt]
&+\cos\phi\,\sin\theta_{\mathrm{in}}(1-r_{p})\,\hat{\mathbf{z}}
\Big] \cos(\omega_{\mathrm{ph}} t).
\label{eq:Eout}
\end{align}
For the Volkov amplitude, we thus find
\begin{align}
\alpha
&=\frac{e}{m_{e}\,\omega^2_\mathrm{ph}}\,E_{0}\Big[
\cos\phi\,\cos\theta_{\mathrm{in}}(1+r_{p})\big(k_{x}+k_{D}\cos\theta_{k}\big)
\nonumber\\
&
+\;\sin\phi\,(1+r_{s})\,k_{D}\sin\theta_{k}
\nonumber\\
&
+\;\cos\phi\,\sin\theta_{\mathrm{in}}(1-r_{p})\,k_{z}
\Big].
\label{eq:alpha_final}
\end{align}
The out-of-plane component 
is given by $k_z=\sqrt{\frac{2m_e}{\hbar^2} (E_{\rm pr} - E_b - W) - (k_x^2 - k_y^2)}$, where $E_{\rm pr}= \hbar \omega_\mathrm{pr}$ is the probe photon energy, $E_b$ is the binding energy of the state of the electron experienced in the material, and $W$ is the work function in the context of tr-ARPES experiments.

Next, we compute the Floquet-only contribution. 
We now focus on graphene, which is usually modeled as a system of noninteracting electrons on a 2D honeycomb lattice.
In the following, we first consider the linearized Hamiltonian around the $K-$points and work in coordinates $(k_x,k_y)$. 
As illustrated in Fig.~\ref{fig:k-gamma},
there are two possible conventions for measuring the values of
$k_{x,y}$: i) with respect to the $\Gamma$-point, or ii) with respect to one of the $K-$points.
In the following, we focus on the $K_1-$point, but we choose the origin of the $(k_x,k_y)-$coordinates to be in the $\Gamma-$point.

The unperturbed, linearized Hamiltonian is then given by 
\begin{equation}
\mathbf{H}_0 =
\begin{pmatrix}
H^+_0 & 0 \\
0 & H^-_0
\end{pmatrix}.
\label{eq:undrivenH}
\end{equation}

\begin{equation}
\begin{aligned}
H_0^{\pm}
&= \pm \hbar v_F \left( k_x \sigma_x + k_y \sigma_y \right) \\
&= \pm \hbar v_F \tilde{k}
\begin{pmatrix}
0 & e^{-i \theta_k} \\
e^{+i \theta_k} & 0
\end{pmatrix}.
\end{aligned}
\end{equation}

The superscript $\pm$ 
denotes the valley isospin index, $\sigma_{x,y}$ are the corresponding Pauli matrices.
For the $K_1-$point, we can write $\tilde{k} = |\textbf{k} - \textbf{K}_1|=\sqrt{k_x^2 + k_y^2}$ and $\theta_k  = \tan^{-1} (k_y/k_x)$, where
$v_F$ is the Fermi velocity.
The spinors $\Psi_\Lambda^{+}=e^{i \mathbf{k} \cdot \mathbf{r}} \hat{\psi}_\Lambda^{+}(\mathbf{k})$, $\Psi_\Lambda^{-}=e^{i \mathbf{k} \cdot \mathbf{r}} \hat{\psi}_\Lambda^{-}(\mathbf{k})$ are the four eigenstates of the Hamiltonian, 
\begin{equation}
\begin{aligned}
& \Psi_\Lambda^{+}=e^{i \mathbf{k} \cdot \mathbf{r}} \hat{\psi}_\Lambda^{+}(\mathbf{k})=e^{i \mathbf{k} \cdot \mathbf{r}} \frac{1}{\sqrt{2}}\left(\begin{array}{c}
1 \\
+\Lambda e^{i \theta_{k}} \\
0 \\
0
\end{array}\right), \\
& \Psi_\Lambda^{-}=e^{i \mathbf{k} \cdot \mathbf{r}} \hat{\psi}_\Lambda^{-}(\mathbf{k})=e^{i \mathbf{k} \cdot \mathbf{r}} \frac{1}{\sqrt{2}}\left(\begin{array}{c}
0 \\
0 \\
1 \\
-\Lambda e^{i \theta_{k}}
\end{array}\right)\,,
\end{aligned}
\end{equation}
with $\Lambda = \pm $ denoting the band index, while  $ \hat{\psi}_\Lambda^{\pm}$ is the spinor vector.
In its eigenbasis, we can write the Hamiltonian in the diagonal form
\begin{equation}
    H_0^{\xi}= \xi \hbar v_F \tilde{k}
\left(\begin{array}{cc}
+1 & 0 \\
0 & -1
\end{array}\right)\,,
\end{equation}
where $\xi=\pm$ is the valley isospin.

Now, we will make our system time periodic by irradiating it with monochromatic light, which will interact with the electrons in the system. 
We use the definition of the electric field given in Eq.~\eqref{eq:E_t}, where the driving field is an 
IR pulse.
In minimal coupling, the momentum takes the form
$\mathbf{k}\;\longrightarrow\;\mathbf{k}-q\,\mathbf{A}(t),$
where $q$ is the electron charge and the electric field is related to the vector potential through $\mathbf{E}(t)=-\frac{\partial \mathbf{A}(t)}{\partial t}$.
We treat the time-dependent part of the Hamiltonian as a perturbation and write 
\begin{equation}
H(t)=H_{0}^{\pm}+H^{\prime}(t)\,.
\label{eq:perturbedH}
\end{equation}
The in-plane vector potential is defined as $\mathbf{A}(t)=\big(A_{x}(t),A_{y}(t)\big),$
constructed from the effective in-plane electric field
\begin{equation}
\mathbf{E}_{\mathrm{eff,xy}}
=
E_{0}\cos(\omega_{\mathrm{IR}} t) (\cos\phi\,t_{p}\cos\theta_{t}\,\hat{\mathbf{x}}
+
\sin\phi\,t_{s}\,\hat{\mathbf{y}}).
\label{eq:Et_xy_revised}
\end{equation}
This field is obtained from the vector potential
\begin{equation}
\mathbf{A}(t)=\begin{pmatrix}
A_x(t) \\
A_y(t)
\end{pmatrix}=
\frac{E_{0}}{\omega_{\mathrm{IR}}}
\begin{pmatrix}
\cos\phi\,t_{p}\cos\theta_{t} \\
\sin\phi\,t_{s}
\end{pmatrix}
\sin(\omega_{\mathrm{IR}} t)\,.
\label{eq:Axy_revised}
\end{equation}
The quantities $t_p$ and $t_s$ are the Fresnel transmission coefficients for $p$- and $s$-polarized light, respectively, at the
vacuum--sample interface; they depend on incidence angle, frequency, and the sample refractive index (concerning the refractive index of a monolayer see Ref.~\onlinecite{Neppl2015}).
The angle $\theta_t$ is the angle at which the transmitted light leaves the sample, 
and which is introduced formally via Snell's law. 
We use the effective refractive index as in ~\cite{Merboldt2025} for $n_2(\omega)$, so that 
Snell's law reads $n_{1}\,\sin\theta_{\mathrm{in}}=n_{2}(\omega)\,\sin\theta_{t}$, with $n_1=1$ for vacuum and $n_2(\omega)$ the sample refractive index.

For the transmitted field interacting with the electrons in the graphene layer, only the in-plane components are taken into account.
The perturbation term in Eq.~\eqref{eq:perturbedH} then takes the form
\begin{equation}
H^{\prime}(t)
=-q v_{F}
\begin{pmatrix}
0 & A_{x}(t)-iA_{y}(t)\\
A_{x}(t)+iA_{y}(t) & 0
\end{pmatrix}\,.
\label{eq:Hprime_revised}
\end{equation}
When representing the perturbation matrix $H^{\prime}(t)$ in the eigenbasis of the unperturbed Hamiltonian $H^{\pm}_0$, 
we obtain
\begin{equation}\label{eq:perturbation_matrix}
H^{\prime}(t)= \left(\begin{array}{cc}
A^{\|} & i A^{\perp} \\
-i A^{\perp} & -A^{\|}
\end{array}\right) \, .
\end{equation}
$A^{\|}$ is the component of the vector potential parallel to $\mathbf{k}$ in the $(x,y)$-plane, $A^{\perp}$ the one orthogonal to it.
We find 
\begin{align}
A^{\parallel}(t)
&=
\underbrace{
\beta\,E_{0}\!\left[t_{p}\cos\theta_{t}\,\cos\phi\,\cos\theta_{k}+t_{s}\sin\phi\,\sin\theta_{k}\right]}_{=:\,\beta^{\parallel}}\; \nonumber \\
& \times \sin(\omega_{\mathrm{IR}} t)\,,
\label{eq:A_parallel}
\end{align}
and 
\begin{align}
A^{\perp}(t)
&=
\underbrace{
\beta\,E_{0}\!\left[t_{p}\cos\theta_{t}\,\cos\phi\,\sin\theta_{k}-t_{s}\sin\phi\,\cos\theta_{k}\right]}_{=:\,\beta^{\perp}}\; \nonumber \\
& \times\sin(\omega_{\mathrm{IR}} t)\,,
\label{eq:A_perp}
\end{align}
where we define $\beta =-q v_f/\hbar\omega^2_{\mathrm{IR}}$. 
Hence, we can write 
\begin{equation}
H^{\prime}(t)
= \hbar\omega_{\rm IR}\big[ \beta^{\|} \sin{(\omega_{\rm IR}t)\sigma_{z}\big] - \hbar\omega_{\rm IR} \big[\beta^{\perp} \sin{(\omega_{\rm IR}t)}}\sigma_{y}\big] \,.
\end{equation}
We obtain
\begin{align}
H(t)
=& \hbar\left[\Lambda\nu_F \tilde{k}+  \omega_{\rm IR} \beta^{\|} \sin(\omega_{\rm IR}t)\right]\sigma_{z} \nonumber \\
&- \hbar\omega_{\rm IR} \left[\beta^{\perp} \sin{(\omega_{\rm IR}t)}\right]\sigma_{y}\, .
\end{align}
We follow Ref.~\onlinecite{Park2014}, where the diagonal part is used to treat the Floquet sidebands, while the non-diagonal part describes Rabi oscillations.
In the following, we focus on the contribution from the diagonal part.
We obtain for the temporal evolution of the pseudo-eigenvectors 
\begin{equation}
\Psi^{\pm}_{\Lambda}(t)
=
\exp\left[
-\frac{i}{\hbar}
\int_{0}^{t} \! dt'\,
H_{\Lambda\Lambda}(t')
\right]\Psi^{\pm}_{\Lambda}(0),
\end{equation}
where $H_{\Lambda\Lambda}$ is the diagonal part of $H(t)$. We obtain

\begin{equation}\label{eq:floquet_solution}
\begin{aligned}
\Psi_\Lambda^{\pm}(t)
& =\Psi_\Lambda^{\pm} e^{-i  \omega_0 t +i\beta^{\|}} \sum_{m} i^{m} J_m\left( \beta^{\|}\right) e^{-i m \omega_{\mathrm{IR}} t} \, ,
\end{aligned}
\end{equation}
where $\Psi_\Lambda^{\pm}$ is the unperturbed eigenstate wave function and $\omega_0 =\hbar \nu_F \tilde{k}$ is the unperturbed energy.
We define the Floquet coefficients
$b_m \equiv  J_m\left( \beta^{\|}\right),$
where $\beta^{\|}$ is the Floquet parameter defined in Eq.~\eqref{eq:A_parallel}.
Similarly to Eq.~\eqref{eq:Volkov_parameters}, the intensity of the first sideband due to Floquet-driving of the electrons inside the material is 
\begin{equation}
    I_1^{\rm Floquet} = J_1(\beta^{\parallel})^2 \approx \left|\beta^\parallel\right|^2 \,.
\label{eq:Floquet_parameters}
\end{equation}
Next, we account for Floquet–Volkov interference by treating the transition from a Floquet state to a Volkov state as the final state.
Following Ref.~\onlinecite{Park2014}, the transition amplitude is calculated using Born approximation within first order time-dependent perturbation theory.
The intensity of the first sideband arising from the interference between the Volkov and Floquet contributions can be written, similarly to Eqs.~\eqref{eq:Volkov_parameters} and ~\eqref{eq:Floquet_parameters}, as 
\begin{equation}
    I_1^{\rm Interference} = J_1(\beta^\| - \alpha)^2 \approx  \left|\beta^\| - \alpha \right|^2 \equiv |\gamma|^2 \,,
    \label{eq:interference}
\end{equation}
where we have defined the interference parameter $\gamma=\beta^\| - \alpha$.
Hence, one main result in leading order perturbation theory is that the intensity of the sideband is determined by interference terms between the Volkov and the Floquet contribution, and not only by the bare parameters $\alpha$ and $\beta^\|$.
This will be crucial for the analysis of the results throughout the paper. 
However, in order to be able to directly compare to experimental results of tr-ARPES measurements, one needs to also take into account matrix elements of the photoemission process $M$~\cite{Choi2025,Hwang2011}.
Using Eq.~\eqref{eq:interference}, the intensity amplitude of the first sideband
in PB1 is then given by~\cite{Choi2025,Park2014,Hwang2011}
\begin{equation}\label{eq:photoemission_mit}
    I_1^{\rm tr-ARPES} \propto \vert M \vert^2 \times \vert \gamma \vert^2 \, .
\end{equation}
For a monolayer of graphene, the transition matrix element is given by $M = \braket{\psi_{\mathbf{k},\text{free}}|H'|\psi_{\Lambda,\mathbf{k}}}$, where $\ket{\psi_{\mathbf{k},\text{free}}}$ is the final state of the photoelectron (vacuum state) and $\ket{\psi_{\Lambda,\mathbf{k}}}$ is one of the graphene eigenstates with band index $\Lambda=\pm 1$.
The choice of transition matrix elements, following Ref.~\onlinecite{Hwang2011} (see Appendix~\ref{app:matrix_element_PB1} for a derivation), depends on the  polarization field and incidence geometry of the tr-ARPES setup. 
In PB1 for monolayer graphene, the simplest choice is to assume the final state is a plane wave with equal weights on
the two carbon sublattices, leading to
$|M|^2\sim\sin^2(\theta_k/2)$ or $|M|^2\sim\cos^2(\theta_k/2)$,
depending on whether the photoemitted electron was at energies below or above the Dirac point (see Appendix~\ref{app:matrix_element_PB1}).
As we will see in the next section, it is essential to include these matrix elements for reproducing the dark corridor consistently in both experiment and theory, thereby enabling meaningful comparison.
\subsection{tdNEGF}\label{subsec:tdNEGF}
To simulate the tr-ARPES signal, we adopt a time‐dependent nonequilibrium Green’s function (tdNEGF) formalism.
For simplicity we treat the electrons in graphene as independent particles, ignoring any self-energy effects. 
We can thus employ the generalized Kadanoff-Baym ansatz (GKBA) to circumvent the explicit calculation of the two-time Green's functions.
We first propagate the single-particle density matrix in momentum space under the
time-dependent Hamiltonian,
\begin{equation}
\frac{d}{dt}\rho(\mathbf{k},t)
=-\frac{i}{\hbar}\big[H(\mathbf{k},t),\rho(\mathbf{k},t)\big].
\end{equation}
Using the time-dependent density matrix $\rho(\mathbf{k},t)$, we compute the time diagonal of the lesser Green's function via
the relation $G^{<}(\mathbf{k},t,t')=i\,\rho(\mathbf{k},t)$, while the off-diagonal time dependence is, within the GKBA, obtained from the equation of motion
\cite{Schler2020,Schler2021}:
\begin{equation}
\left[i\hbar\,\partial_{t}-H(\mathbf{k},t)\right]
G^{<}(\mathbf{k},t,t')=0 \,.
\end{equation}
We then compute the photoemission intensity via~\cite{PruschkePRL2009,Freericks2015,Sentef2015,Mahmood2016,Ito2023,Schler2021,Merboldt2025}
\begin{align}\label{eq:intensity_NEGF}
    I(k,E,&\tau) =
    \text{Im} \sum_{aa^{\prime}} M^*_{a}(\mathbf{k},E)M_{a^{\prime}}(\mathbf{k},E)\int_0^{\infty} dt \int_0^t dt^{\prime} \nonumber \\
    &s(t,\tau)s(t',\tau)\, 
    e^{i (E-\hbar\omega_\mathrm{pr})(t-t')}\,
    e^{-i\Theta(\mathbf{k},t,t')} G^<_{aa^{\prime}}(\mathbf{k},t,t')\, .
\end{align}
Here, $s(t,\tau)$ is the probe envelope at delay $\tau$, $M_{a}(\mathbf{k},E)$
denotes the photoemission matrix elements  with respect to band $a$, $E$ is the final state energy,
$\hbar\omega_{\mathrm{pr}}$ is the
probe-photon energy, and $\Theta(\mathbf{k},t,t')$ is the Volkov phase.
Note that the effect of the matrix-elements
can be switched off in Eq.~\eqref{eq:intensity_NEGF} by setting their values to one and then replace $\sum_{a a'}$ by the trace  ${\rm Tr}[G^<]$, as for instance done in Ref.~\onlinecite{Sentef2015}.
Note that in order to account for finite-temperature broadening of the band structure, a finite-temperature state at $T\approx 80$K is initialized when constructing the occupations and the single-particle density matrix. 
This only affects the occupations close to the Fermi energy. 
The temperature broadening $\sim k_B T$ of the Fermi edge is, however, much smaller than the energy resolution imposed by the duration of the probe pulse. 
There are no other broadening effects.
Furthermore, the photoemission matrix elements within the tdNEGF approach are calculated with a similar method as in Ref.~\onlinecite{Schler2020}.
This framework allows a controlled separation of Floquet, Volkov (LAPE), and
interference contributions:
(i) Floquet only: propagate with the pump-periodic Hamiltonian
$H(\mathbf{k},t)$, but set $\Theta=0$;
(ii) Volkov only: replace $H(\mathbf{k},t)\rightarrow H_{0}(\mathbf{k})$
(no Floquet modulation);
(iii) Interference: use the full expressions in
Eq.~\eqref{eq:intensity_NEGF}.
We compare the computed first-order sideband intensity in the $(k_x,k_y)$ plane.

\section{results}\label{sec:results}
\begin{figure*}
    \centering
  \includegraphics[width=0.8\linewidth]{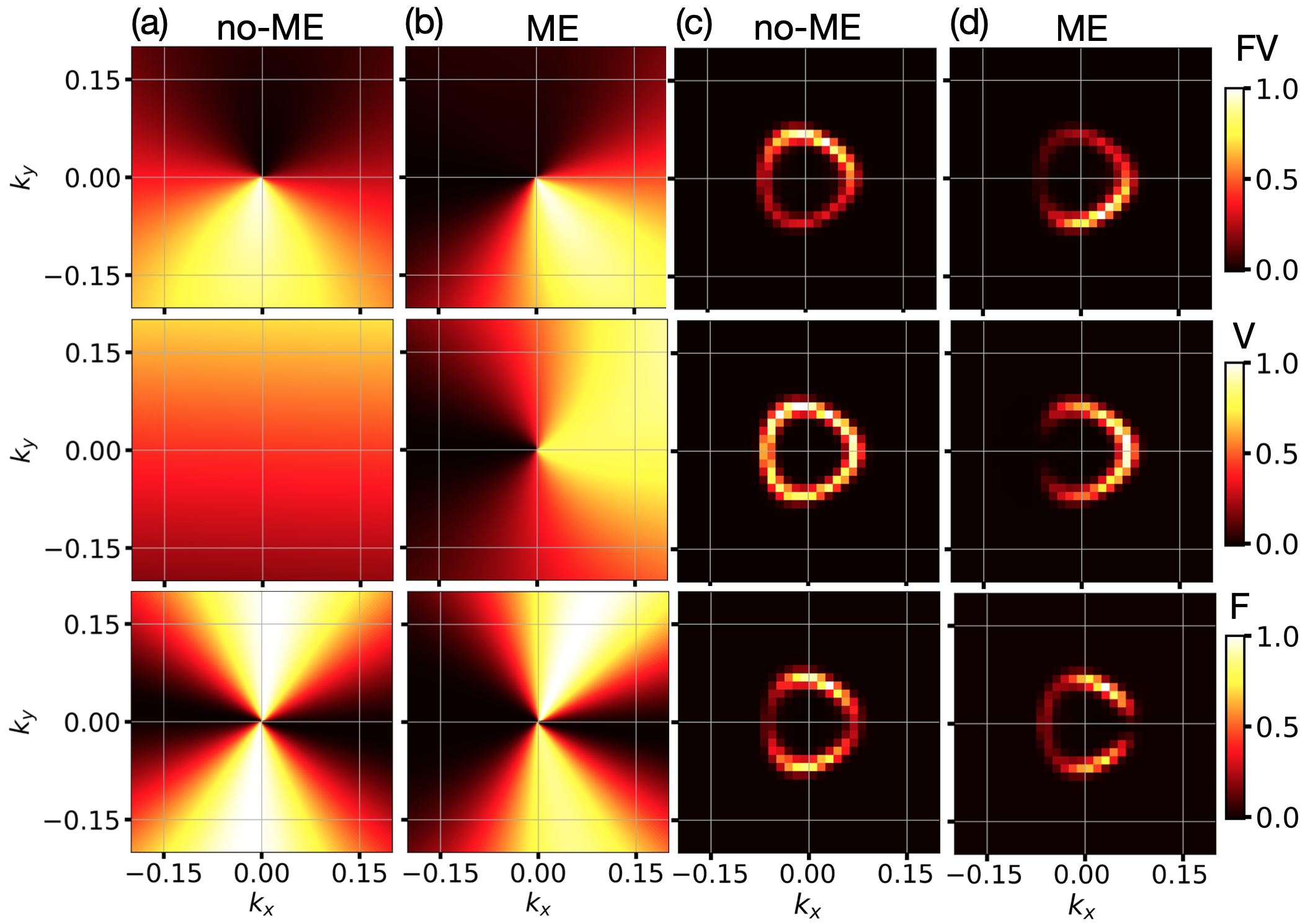}
    \caption{First Floquet sideband intensity in the $(k_x,k_y)$ plane as obtained from the perturbative approach (PB1) and from the tdNEGF, respectively. 
    Columns (a,b) show PB1 results; columns (c,d) show tdNEGF results extracted from constant-energy cuts ($E–E_F=0.63eV$) that isolate the first sideband.
    Top, middle, and bottom rows correspond to Floquet–Volkov (FV), Volkov (V), and Floquet (F) contributions, respectively. 
    Columns (a,c) are without photoemission matrix elements (no-ME); columns (b,d) include matrix elements (ME). 
    Parameters used $\theta_{\mathrm{in}}=68^{\circ}$, $\phi=85^{\circ}$, and field strength $4~\mathrm{MV/cm}$. 
    All intensities are normalized to unity.}\label{fig:Park_NEGF_overview_WME}
\end{figure*}
\subsection{Intensity distribution in PB1 and tdNEGF}\label{subsec:results_intensity_PB1_tdNEGF}

Unless stated otherwise, we use an incidence angle $\theta_{\mathrm{in}}=68^{\circ}$, a polarization angle $\phi=85^{\circ}$, a driving pulse center frequency $\hbar \omega_{\mathrm{IR}} = 0.65~\mathrm{eV}$ and a field strength of $4~\mathrm{MV/cm}$, which corresponds to the experiments as in Ref.~\cite{Merboldt2025}.
Fig.~\ref{fig:Park_NEGF_overview_WME} shows the first sideband intensity, for the PB1 using Eq.~\eqref{eq:photoemission_mit} and for the tdNEGF  using Eq.~\eqref{eq:intensity_NEGF}.
For each approach, we display results with (ME) and without (no-ME) photoemission matrix elements. 
The top, middle, and bottom rows correspond to Floquet–Volkov (FV), Volkov (V), and Floquet (F) contributions respectively, in the $(k_x,k_y)$ plane. 
All intensities are normalized to unity.
We follow the convention from Ref.~\onlinecite{Hwang2011} (see also ~\cite{Moser2017}) for monolayer graphene and use $|M|^2\sim\sin^2(\theta_k/2)$ for intensities above the Dirac point in PB1.
We emphasize an important aspect of the perturbative framework:
PB1 yields the angular distribution of the intensity in the $(k_x,k_y)$ plane for a selected sideband.
While this provides momentum-resolved information, it does not by itself reconstruct the full nonequilibrium band structure. 
Rather, it represents the momentum-dependent transition probability associated with that sideband, and may exhibit anisotropies reflecting the symmetry and driven dynamics of the system. 

Column (a) of Fig.~\ref{fig:Park_NEGF_overview_WME} shows the intensities of the first Floquet sideband when neglecting the matrix elements as a function of $(k_x,k_y)$ in the vicinity of the $K_1$-point.
The top panel shows the results when considering the interference between the Volkov and the Floquet channel, the other two panels are for the Volkov-only and the Floquet-only contribution, respectively. 
A dark-corridor–like suppression~\cite{Shirley1995,Gierz2011,Kuemmeth2009,MuchaKruczyski2008} is visible in all maps. 
We also observe a clear up–down asymmetry in the FV and V maps in momentum space, whereas the F maps appear up–down symmetric; we will quantify this further below.
Column (b) presents the  case when including the matrix elements, using the before-mentioned approximation $|M|^2\sim\sin^2(\theta_k/2)$.
The dark corridor shifts relative to column (a), and the overall intensity distribution becomes more asymmetric.
This demonstrates the direct impact of photoemission matrix elements on the momentum-resolved intensity distribution.

We now turn to the tdNEGF results in Fig.~\ref{fig:Park_NEGF_overview_WME}. 
Columns (c) and (d) display the momentum maps for the first sideband in the $(k_x,k_y)$ plane for the no-ME and ME cases respectively, using the same $\theta_{\mathrm{in}}$ and $\phi$ as above.
Unlike columns (a) and (b), now the first Floquet sidebands are extracted from constant-energy cuts from the  spectra of the driven graphene, leading to the ring-like patterns in momentum space~\cite{Merboldt2025,Schler2020}.
In column (c), the FV maps exhibit an up–down asymmetry whose sense is opposite to that of PB1 in column (a) (no-ME). 
The characteristic dark corridor is barely visible or absent.
The V and F maps appear essentially up–down symmetric within visual resolution; we assess the asymmetries via angular lineouts further below.
Including ME in tdNEGF, column (d), reinstates a dark-corridor–like suppression (also visible in V) and yields FV intensities with an asymmetry opposite to the no-ME case.
Comparison of columns (b) and (d) now reveals that both approaches, when including the matrix elements, provide qualitatively very similar results, including the dark regions and the direction of the asymmetries.
This is interesting, since the FV-results without considering the matrix elements lead to opposite predictions of both approaches.
In addition, while in PB1 the matrix elements enter simply as a multiplicative factor in the analytical expressions for $I_1$, the situation in tdNEGF is more involved.
In graphene, in particular, the bands are slightly renormalized along the $k_x$-direction, which modifies the relative phases of the contributions from the A and B sublattices.
For this reason, we do not expect the same intensity trends when comparing the F, V and FV channels with and without ME to each other, but rather each situation has to be computed independently. 
However, upon including the ME, both PB1 and tdNEGF yield the same qualitative intensity distribution in the F, V, and FV channel, respectively.
This indicates that generically for the  investigation of Floquet sidebands also for other materials it is crucial to include the matrix elements. 
Luckily, for graphene a minimal modeling of the matrix elements as discussed in Ref.~\onlinecite{Hwang2011} appears to suffice, but for other materials it might be important to provide a material-specific estimation of the photo-emission matrix elements.  

Next, we will compare PB1 and tdNEGF more directly by analyzing angular lineouts to quantify asymmetries of the signals and differences between the two approaches.

\subsubsection{Angular  intensity distribution}\label{subsec:azimuthal_lineouts}
\begin{figure}
    \centering
  \includegraphics[width=0.95\linewidth]{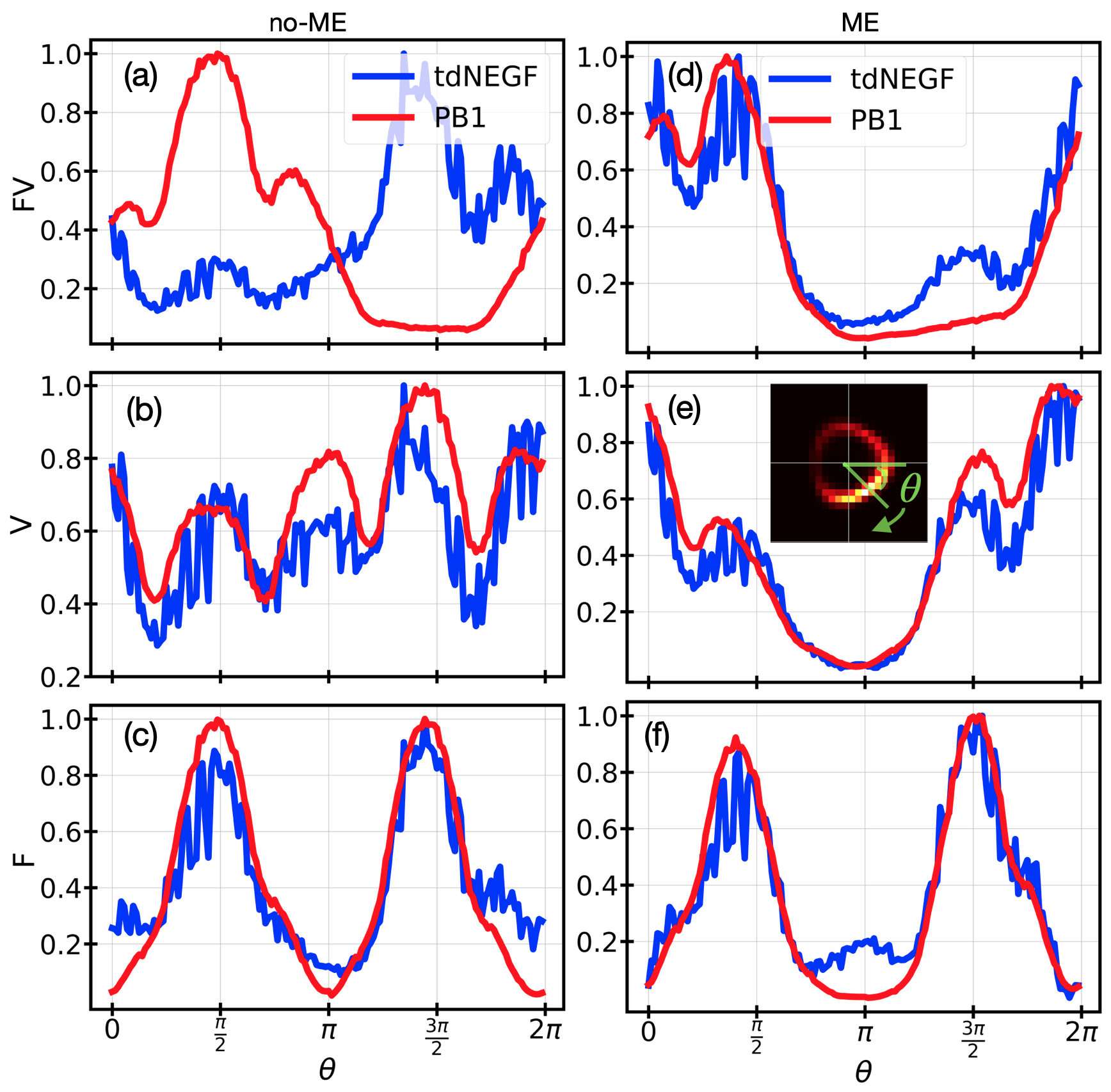}
    \caption{Azimuthal ($\theta$) dependence of the first sideband intensity for Floquet–Volkov (FV), Volkov (V), and Floquet (F) contributions, corresponding to the momentum maps in Fig.~\ref{fig:Park_NEGF_overview_WME}.
    The blue lines represent tdNEGF results, and the red lines the PB1 results.
    Each curve is normalized to its own maximum. 
    Left panel: results without photoemission matrix elements (no-ME).
    Right panel: results including matrix elements (ME). 
    The inset in (e) illustrates the way the azimuthal data is obtained. 
    }
    \label{fig:cross-sec-Park_NEGF_overview_WME}
\end{figure}

The inset in Fig.~\ref{fig:cross-sec-Park_NEGF_overview_WME}(e) illustrates how the intensity as a function of azimuthal angle $\theta$ is obtained by summing over intensities within narrow angular sectors (bins in $\theta$-direction). 
For both methods, we employ a sliding angular binning procedure in momentum space.
An angular bin of width $\theta = 20^\circ$ is defined, and its starting angle is incremented in steps of $2.5^\circ$.
At each increment, we sum over all $(k_x,k_y)$ points shown in the plots, whose angle lies within the current bin.
If not stated otherwise, this procedure is used for all azimuthal angular intensity distributions presented in this work.

Figure~\ref{fig:cross-sec-Park_NEGF_overview_WME} shows the so-obtained angular dependence of the first-sideband intensity for the FV, V, and F contributions, respectively, corresponding to the momentum maps in Fig.~\ref{fig:Park_NEGF_overview_WME}, for both no-ME and ME cases. 
For a consistent comparison, each curve is normalized to its own maximum, since, on general grounds, we cannot expect that both methods obtain the same amplitudes of the signals. 
Note that, if not mentioned otherwise, for the sake of clarity and direct comparison, we  display the raw data in the lineouts obtained from the bins without further smoothening or other post processing, which leads to the visible noise in the plots.
However, we expect this noise to be reduced when using a finer $k$-grid.

In Fig.~\ref{fig:cross-sec-Park_NEGF_overview_WME}(a) (FV, no-ME), both approaches exhibit a clear asymmetry around $\theta=\pi$, but with opposite sense: 
tdNEGF yields higher intensity for $\theta>\pi$, whereas PB1 is larger for $\theta<\pi$.
Figures~\ref{fig:cross-sec-Park_NEGF_overview_WME}(b) and (c) show the results for V and F for the no-ME case.
While in detail the results from both methods differ, in their overall behavior, we find the same directional asymmetry in the intensity distribution. 
Including the matrix elements (right panel) modifies these trends. 
In Fig.~\ref{fig:cross-sec-Park_NEGF_overview_WME}(d) (FV, ME), the asymmetry reverses relative to the tdNEGF no-ME case in Fig.~\ref{fig:cross-sec-Park_NEGF_overview_WME}(a).
Overall, with the ME, results in Figs.~\ref{fig:cross-sec-Park_NEGF_overview_WME}(d) –\ref{fig:cross-sec-Park_NEGF_overview_WME}(f) show good qualitative agreement between PB1 and tdNEGF. 
A notable exception appears in Fig.~\ref{fig:cross-sec-Park_NEGF_overview_WME}(f) (F, ME): tdNEGF predicts a finite intensity at $\theta=\pi$, while PB1 yields a vanishing signal there.
This difference is consequential because the experimentally accessible observable for F and V is a quantum–path–integrated signal~\cite{Merboldt2025}.
As seen in Fig.~\ref{fig:cross-sec-Park_NEGF_overview_WME}(e) (V, ME), both methods give vanishing intensity near $\theta=\pi$.
Consequently, in the FV channel, the interference enforces intensity $(\theta=\pi)$ to $0$ within PB1, whereas tdNEGF retains a finite value originating entirely from the F contribution.
This distinction could provide an experimental handle on Floquet effects and sideband formation.
It indicates that a simple ME choice such as $|M|^2\sim\sin^2(\theta_k/2)$ is generally insufficient for realistic situations.

\begin{figure}[t]
    \centering
    \includegraphics[width=\linewidth]{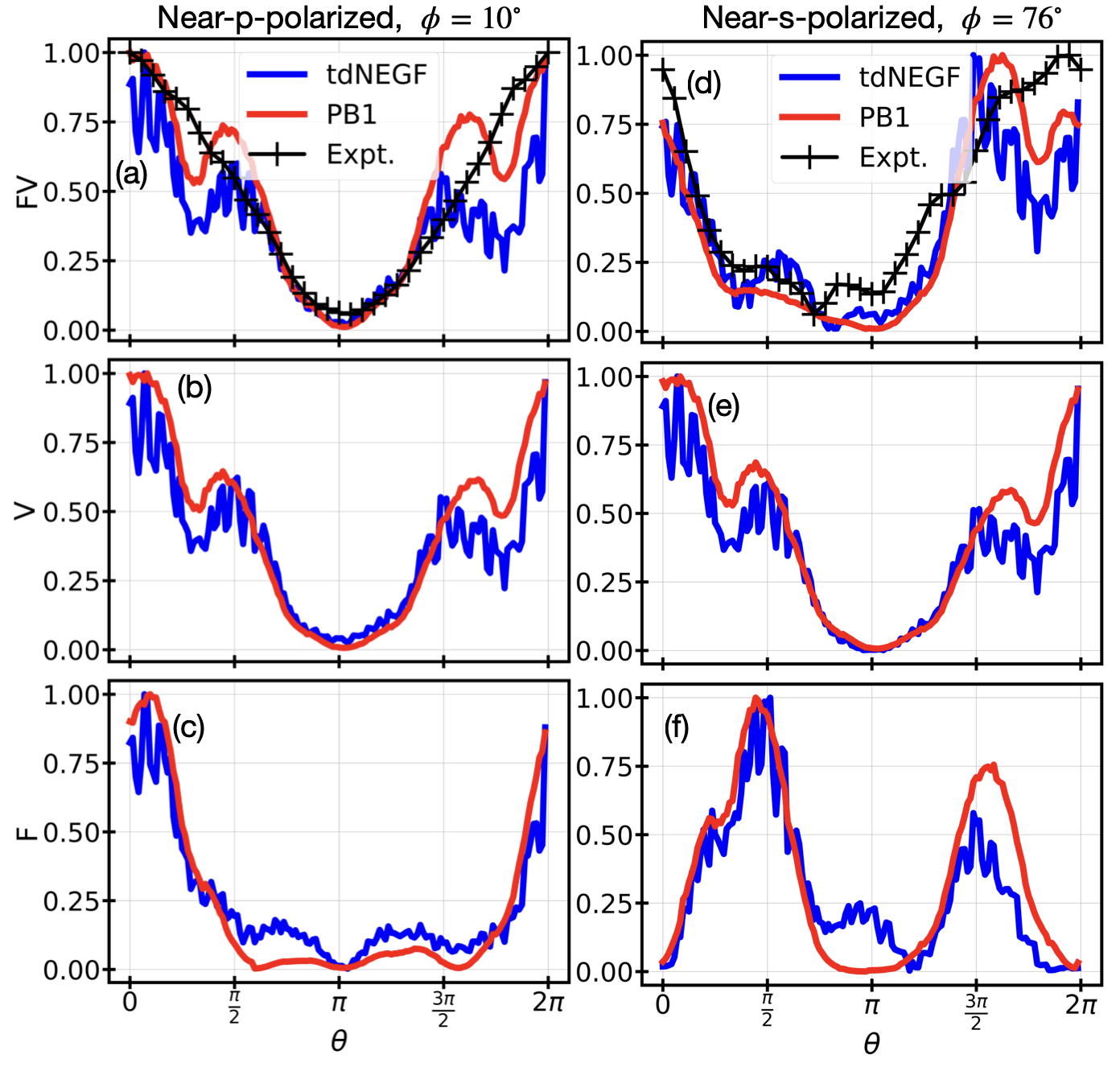}
    \caption{Comparison between theoretical photoemission intensities obtained by tdNEGF and PB1, respectively,
    and experimental tr-ARPES data for monolayer graphene as published in Ref.~\onlinecite{Merboldt2025}.
    The plots show the angular dependence of the normalized photoemission intensity for the first sideband, shown for two different values of the polarization of the incident light: near p-polarized light ($\phi=10^{\circ}$, left column) and near s-polarized light ($\phi=76^{\circ}$, right column).
    Black crosses represent the experimental data, the blue lines represent tdNEGF results, and the red lines the PB1 results.
    Theoretical results include contributions from Floquet-Volkov (FV) in (a) and (d), Volkov only (b) and (d), and Floquet  data in (c) and (f).
    Parameters are $\theta_{\mathrm{in}}=68^{\circ}$, $(f_F,\,f_V)=(0.5,\,0.5)$, and ME are included.
    }
    \label{fig:direct-comparison}
\end{figure}
\subsection{Direct comparison with experiment}\label{subsec:comparison_with_experiment}

In order to better model the experimental situation, we adapt the driving field to incorporate the strength at the surface and within the sample for the monolayer graphene setup studied in this paper.
We follow Ref.~\onlinecite{Merboldt2025} and use as effective field acting on the photoelectron 
\begin{equation}\label{eq:eff_out}
    \mathbf{E}_{\text{eff,out}} = f_V\,\,(\mathbf E_{\rm in} + \mathbf E_{\rm r})\,,
\end{equation}
while the effective driving field for electrons inside the graphene monolayer is interpolated as
\begin{equation}\label{eq:eff_xy}
    \mathbf{E}_{\text{eff,xy}} = f_F( \mathbf E_{\rm in} + \mathbf E_{\rm r}) + (1-f_F)\mathbf E_{\rm t}\,, 
\end{equation}
where $f_V$ denotes the screening parameter for the Volkov contribution and $f_F$ the screening at the monolayer.
Ideally, one would expect a perfect surface, i.e. $(f_V,\,f_F)=(1.0,\,0.0)$; atomically thin graphene can, however, not be treated as a hard vacuum-matter interface.
We use the scaling factors $(f_V,f_F)$ to approximately account for the uncertainty of the effective fields, anchoring to experimental observations.
Good agreement with dielectric screening on the atomic length scale is obtained for $f_V=0.5$~\cite{Neppl2015}.
This gives us 
\begin{align}
\alpha
&= f_{V}\,\frac{e\,E_{0}}{m_{e}\,\omega^2_\mathrm{ph}}
\Big[
\cos\phi\,\cos\theta_{\mathrm{in}}(1+r_{p})\big(k_{x}+k_{D}\cos\theta_{k}\big)
\nonumber\\
&\qquad\qquad\quad
+\;\sin\phi\,(1+r_{s})\,k_{D}\sin\theta_{k}
\nonumber\\
&\qquad\qquad\quad
+\;\cos\phi\,\sin\theta_{\mathrm{in}}(1-r_{p})\,k_{z}
\Big] \, ,
\label{eq:screened_alpha}
\end{align}
and
\begin{align}
\beta^{\parallel}
&= f_{F}\,\beta E_{0}\!\left[
\cos\phi\,\cos\theta_{\mathrm{in}}(1+r_{p})\cos\theta_{k} \right. \nonumber\\[4pt]
&\quad \quad \quad  \left. +\sin\phi\,(1+r_{s})\sin\theta_{k}
\right]
\nonumber\\[4pt]
&\quad + (1-f_{F})\, \beta E_{0}\!\left[
t_{p}\cos\phi\,\cos\theta_{t}\cos\theta_{k}
+t_{s}\sin\phi\,\sin\theta_{k}
\right].
\label{eq:screened_beta}
\end{align}

To assess the usefulness of the theoretical approaches, we compare in the following the photoemission intensity results obtained from the two approaches with tr-ARPES data on driven graphene from Ref.~\onlinecite{Merboldt2025}.
As discussed in detail in Refs.~\onlinecite{Merboldt2025,Choi2025}, these comparisons underscore that both Floquet and Volkov contributions—and their mutual path interference—are essential for reproducing the measured tr-ARPES signal.

Figure~\ref{fig:direct-comparison} shows the angular distribution of the intensity of the first sideband for two different values of the polarization of the driving-field: nearly-$p$ polarized ($\phi=10^{\circ}$) and nearly-$s$ polarized ($\phi=76^{\circ}$). 
For all cases, we choose parameters $\theta_{\mathrm{in}}=68^{\circ}$, $f_F=f_V=0.5$, and in the theoretical results the ME are included.
The bottom and middle panels show the Floquet (F) and Volkov (V) components from PB1 and tdNEGF, respectively. 
In the top panels of Fig.~\ref{fig:direct-comparison}, we directly compare the results of both theoretical approaches to the ones obtained in the experiment.
First, we compare the left column for the near p-polarized $\phi = 10^{\circ}$ case in Fig.~\ref{fig:direct-comparison}.
In Fig.~\ref{fig:direct-comparison}(c),
we observe a slight mismatch between the two theoretical approaches in the angular dependence of the intensity distribution in the Floquet channel. 
However, in Fig~\ref{fig:direct-comparison}(b) in the Volkov channel, the results are very similar to each other, noticeably around $\theta \sim \pi$.
The intensity distribution deviates slightly when going further away from $\pi$, but the overall qualitative behavior remains the same.
In Fig~\ref{fig:direct-comparison}(a), we show the interference results are directly compared to the experimental data from Ref.~\onlinecite{Merboldt2025}. 
Both theoretical approaches reproduce well the experimental data, in particular around $\theta = \pi$.

Next, in the right column of  Fig.~\ref{fig:direct-comparison} we show the results for the near s-polarized case with $\phi = 76^{\circ}$.
In Fig.~\ref{fig:direct-comparison}(f)
the intensity distribution in the Floquet channel as obtained by both theoretical approaches shows a similar asymmetry with respect to $\theta = \pi$.
However, PB1 shows no intensity around $\theta \approx \pi$, while the tdNEGF results clearly show a finite intensity at this point.
This indicates mixing of the bands, which can be captured by tdNEGF formalism, but not by the simple perturbative PB1 approach.
However, Fig.~\ref{fig:direct-comparison}(e) shows that in the Volkov channel both methods again provide very similar results, which in addition are similar to the results in the case of $\phi = 10^{\circ}$ polarization. 
Again, we get the best agreement between both approaches for $\theta \sim \pi$.
In Fig.~\ref{fig:direct-comparison}(d), we again compare the theoretical results to the experimental results of Ref.~\onlinecite{Merboldt2025}.
While the overall agreement is very good, in particular at $\theta \sim \pi$ we notice that the intensity measured in experiment is notably higher than the one obtained by both theoretical approaches.
However, the tdNEGF shows a trend towards the stronger signal, which stems from the finite intensities in the Floquet channel. This raises the question, whether in the s-polarized case a finite intensity in experimental data at $\theta \sim \pi$ is already indicative for Floquet sidebands. 
However, the PB1 results, which have no contribution from the Floquet sector, have also a finite, but smaller intensity, so that such a finding does not suffice to unambiguously confirm the presence of a Floquet sideband. 
Instead, a detailed understanding of the angular distribution of the intensities is needed, as discussed in Refs.~\onlinecite{Merboldt2025,Choi2025}. 

In this section we saw that qualitatively, both approaches reproduce the experimental findings on the angular modulations of the intensity distribution and capture the general trends of the interference between Floquet- and Volkov- final states. 
Quantitative differences between each other, and to the experimental results, however, remain, and are polarization-dependent.
The tdNEGF approach more accurately captures the overall line shape, particularly the asymmetry and the intensity variation across $\theta$ in the vicinity of $\theta=\pi$ in the nearly s-polarized case.
The differences between the two theoretical approaches can further increase when considering  higher-order processes, self-energy effects, and spectral broadening that are naturally captured in the tdNEGF formalism, but which are absent in simple perturbative treatments like PB1. 
Next, we check how both methods perform when changing different parameters at hand.

\begin{figure*}[t]
    \centering
    \includegraphics[width=0.9\linewidth]{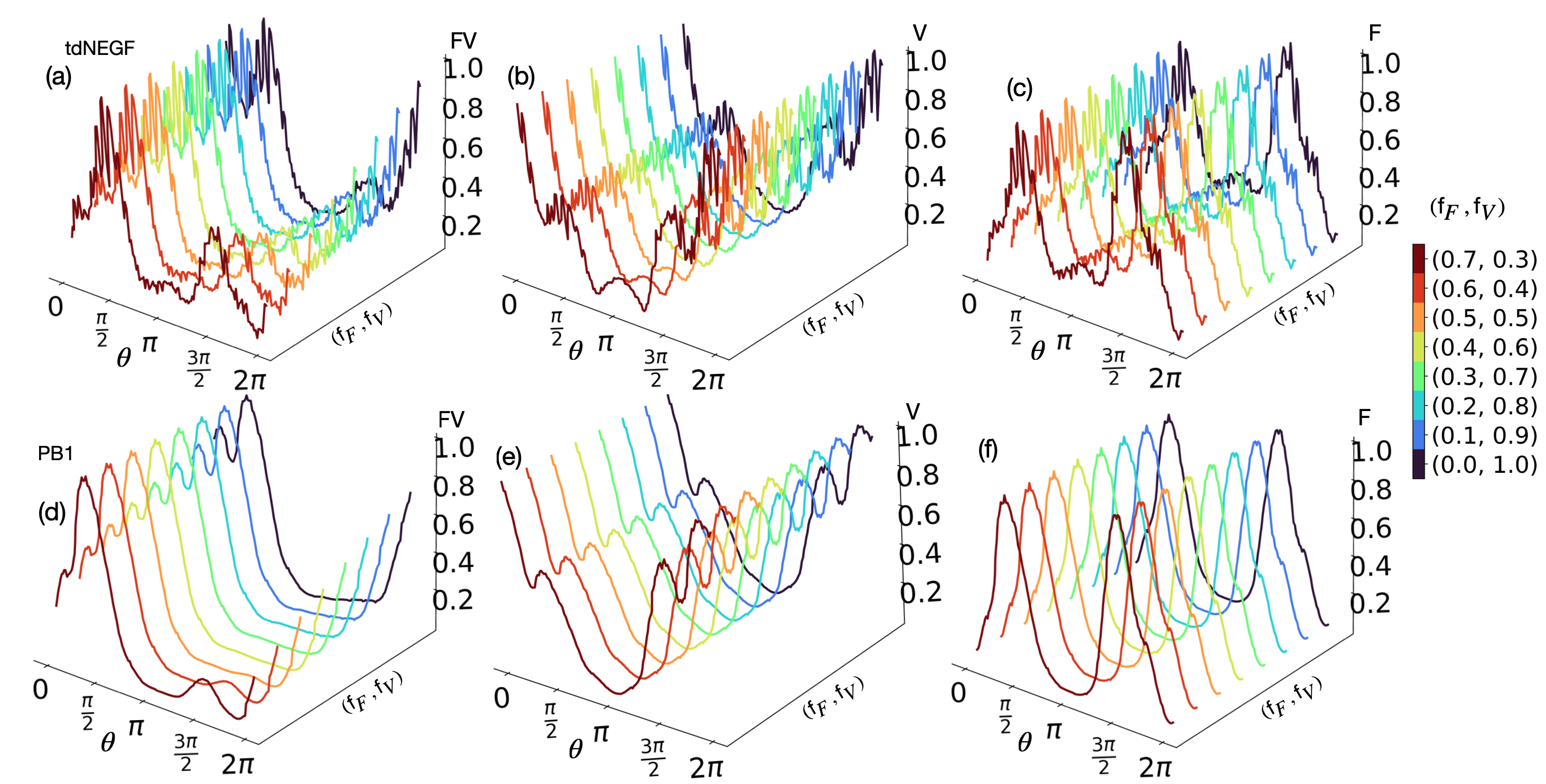}
    \caption{Sensitivity of the first sideband angular intensity to screening.
    Top: tdNEGF; bottom: PB1.
    Matrix elements are included throughout. 
    The incidence and polarization are fixed to $\theta_{\mathrm{in}}=68^{\circ}$ and $\phi=85^{\circ}$.
    Screening pairs $(f_F,f_V)$, defined in Sec.~\ref{subsec:comparison_with_experiment} [Eqs.~\eqref{eq:eff_out}, \eqref{eq:eff_xy}], are varied from the “perfect-surface’’ limit $(0.0,1.0)$ (black) to $(0.7,0.3)$ (dark red) and encoded by the color bar; all curves are normalized to unity.}
    \label{fig:screnn_approximation}
\end{figure*}

\subsection{Sensitivity to parameters}\label{subsec:sensitivity_to_parameters}

\subsubsection{Screening approximation}

We compute the first-sideband intensity with both approaches for different choices of screening parameters $f_V$ and $f_F$, which were introduced in Eqs.~\eqref{eq:eff_out} and~\eqref{eq:eff_xy}.
As further discussed in Sec.~\ref{subsec:comparison_with_experiment}, in this modeling the driving field is adapted to account for surface and vacuum responses, and in the idealized case of a perfect surface one has $(f_V, \,f_F)=(1.0,\,0.0)$.
In the following, we fix the incidence angle $\theta_{\mathrm{in}}=68^{\circ}$, the polarization angle $\phi=85^{\circ}$, and include ME in both theoretical approaches.

Figure~\ref{fig:screnn_approximation} shows the direct impact of different choices of the screening parameters.
The top row shows the tdNEGF results, while the bottom row shows the PB1 results.
The numerical values of the values of the screening parameters $(f_F,f_V)$ are encoded by the color bar. 
The plots show results for a ``perfect surface” $(f_F,f_V)=(0.0,1.0)$ (black lines) and the effect of varying the values up to $(0.7,0.3)$ (dark red lines).
All curves are normalized to unity.
Overall, the intensity profiles are rather robust with respect to changing the screening parameters, with the most visible changes occurring in the FV results.
In the tdNEGF results of  Fig.~\ref{fig:screnn_approximation}(a), moving away from the perfect-surface limit—i.e., allowing field components to mix—enhances the intensity for $\theta>\pi$.
Even at the perfect-surface setting, there remains finite weight near $\theta=3\pi/2$; the subsequent enhancement is therefore naturally attributed to a stronger Volkov contribution.

The PB1 results, shown in Fig.~\ref{fig:screnn_approximation}(d) for FV with the same color scheme, exhibit a related trend: as mixing increases, a secondary peak  near $\theta=3\pi/2$ appears, whereas  near the perfect-surface limit (or weak screening), this peak is suppressed.
Noticeably, we observe suppression of the intensity at $2\pi$ as going away from perfect-surface limit.
For stronger screening, e.g., $(f_F,f_V)=(0.7,0.3)$, both methods yield qualitatively similar angular distributions—strong intensity around $\theta=\pi/2$ and weak intensity around $\theta=3\pi/2$—while differing quantitatively in the ratio of the peak amplitudes.

\subsubsection{Polarization-angle dependence}

\begin{figure}[t]
    \centering
    \includegraphics[width=\linewidth]{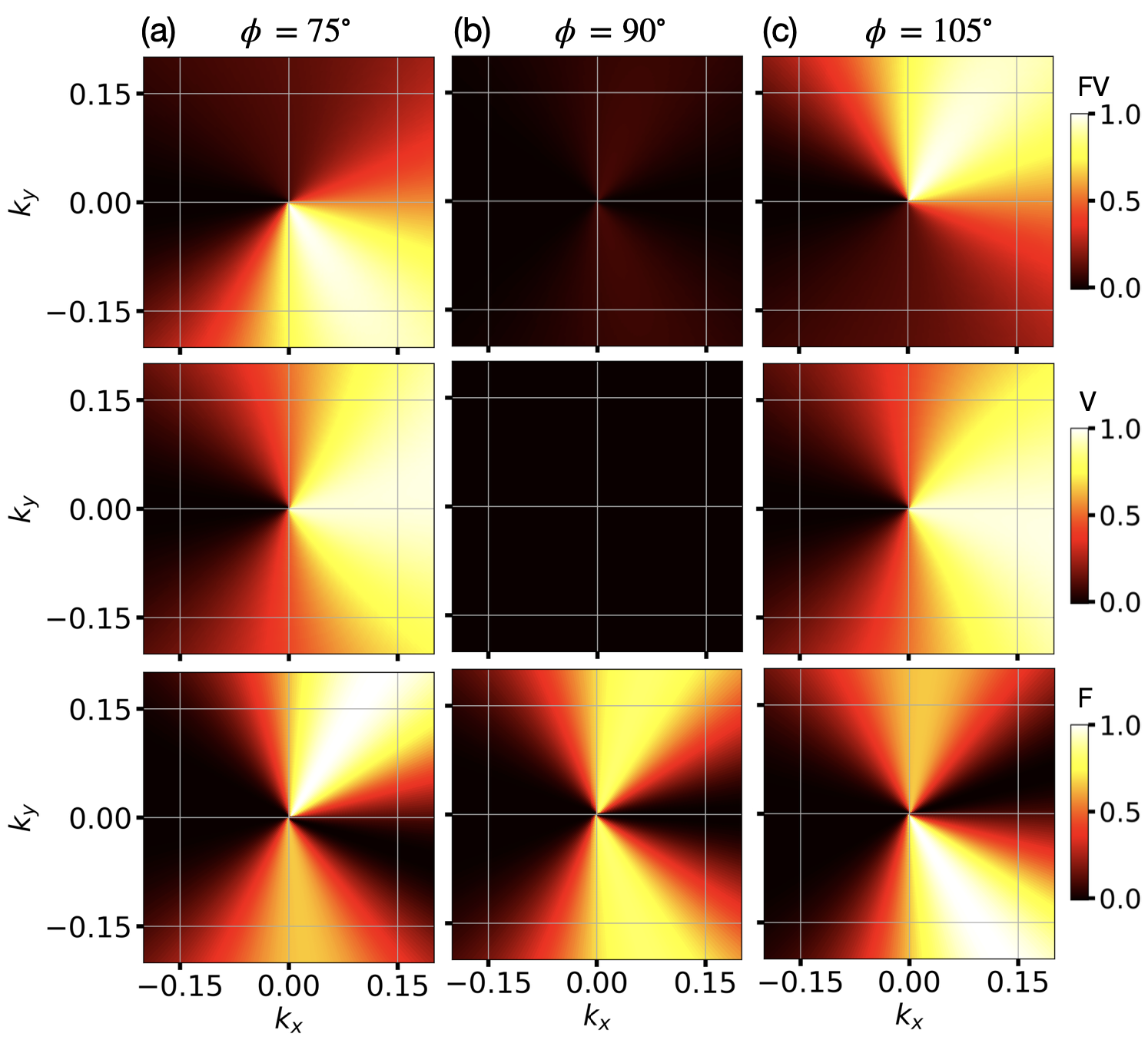}
    \caption{Polarization-angle dependence of the first sideband intensity within PB1. 
    Momentum-resolved maps in the $k_x$–$k_y$ plane are shown for $\phi=75^{\circ}$, $90^{\circ}$, and $105^{\circ}$ (columns), with $\theta_{\mathrm{in}}=68^{\circ}$, $f_F=f_V=0.5$, and photoemission matrix elements included (ME).
    Rows correspond to Floquet–Volkov (FV, top), Volkov (V, middle), and Floquet (F, bottom) contributions.}\label{fig:Polarization_angle_dependence}
\end{figure}

We now examine how the first-sideband intensity varies with the polarization angle $\phi$.
Our focus is the evolution of the up–down asymmetry for polarization angles smaller and larger than $\phi=90^{\circ}$ as in Ref.~\cite{Merboldt2025}, but here we will also compare intensities to each other for different polarization angles.
Figure~\ref{fig:Polarization_angle_dependence} shows results obtained with PB1 for the momentum-resolved maps in the $(k_x, k_y)$ plane for the same parameters as in Ref.~\onlinecite{Merboldt2025} with fixed $\theta_{\mathrm{in}}=68^{\circ}$,  screening parameters $f_F=f_V=0.5$, including  ME, and for $\phi=75^{\circ}$, $90^{\circ}$, and $105^{\circ}$.
However, we now fix the normalization of the amplitudes row-wise in the Figure, so that one can see how the intensity in the F, V, and FV channels, respectively, evolves upon changing $\phi$.
In the FV case, $\phi=75^{\circ}$ produces a clear up–down asymmetry with enhanced intensity in the lower-right quadrant of the $(k_x,k_y)$ plane. 
At $\phi=90^{\circ}$ the distribution becomes nearly symmetric and the intensities are strongly suppressed everywhere, while at $\phi=105^{\circ}$ the sense of the asymmetry flips relative to $\phi=75^{\circ}$, while the magnitudes of the intensities are comparable. 
The V-only case shows a symmetric profile overall, with a strong suppression (nearly vanishing intensity) at $\phi=90^{\circ}$. 
The F-only case mirrors the FV trend: asymmetric for $\phi=75^{\circ}$ and $105^{\circ}$ with opposite sense, and approximately symmetric at $\phi=90^{\circ}$.

\begin{figure}[pht]
    \centering
    \includegraphics[width=\linewidth]{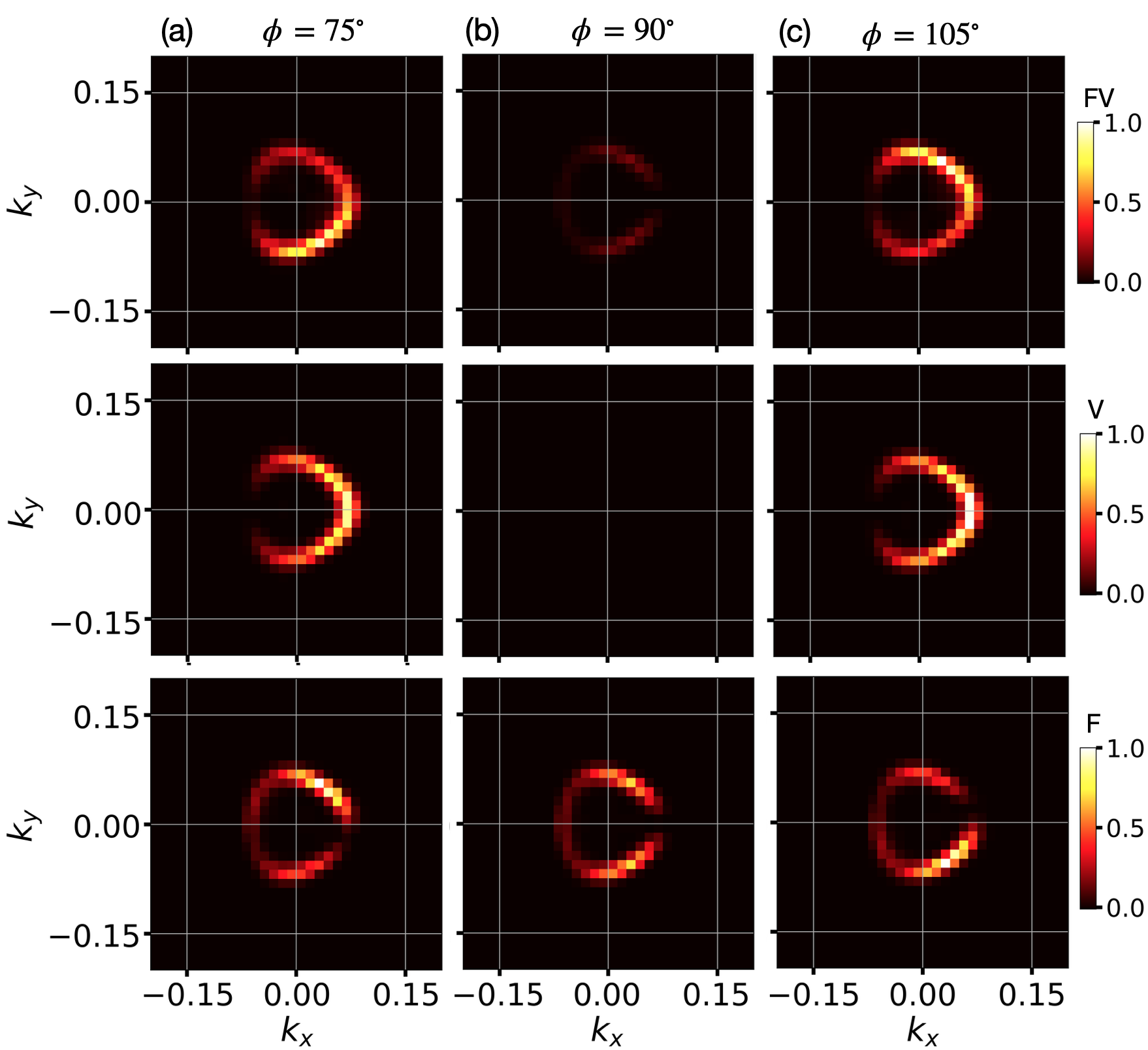}
    \caption{
    Polarization-angle dependence of the first sideband intensity obtained with tdNEGF. 
    Rows correspond to Floquet–Volkov (FV, top), Volkov (V, middle), and Floquet (F, bottom) contributions.
    Momentum-resolved maps in the $k_x$–$k_y$ plane are shown for $\phi=75^{\circ}$, $90^{\circ}$, and $105^{\circ}$ (columns), with $\theta_{\mathrm{in}}=68^{\circ}$, $f_F=f_V=0.5$, and photoemission matrix elements included (ME).}\label{fig:Polarization_angle_dependence_NEGF}
\end{figure}

\begin{figure}[b]
    \centering
    \includegraphics[width=0.8\linewidth]{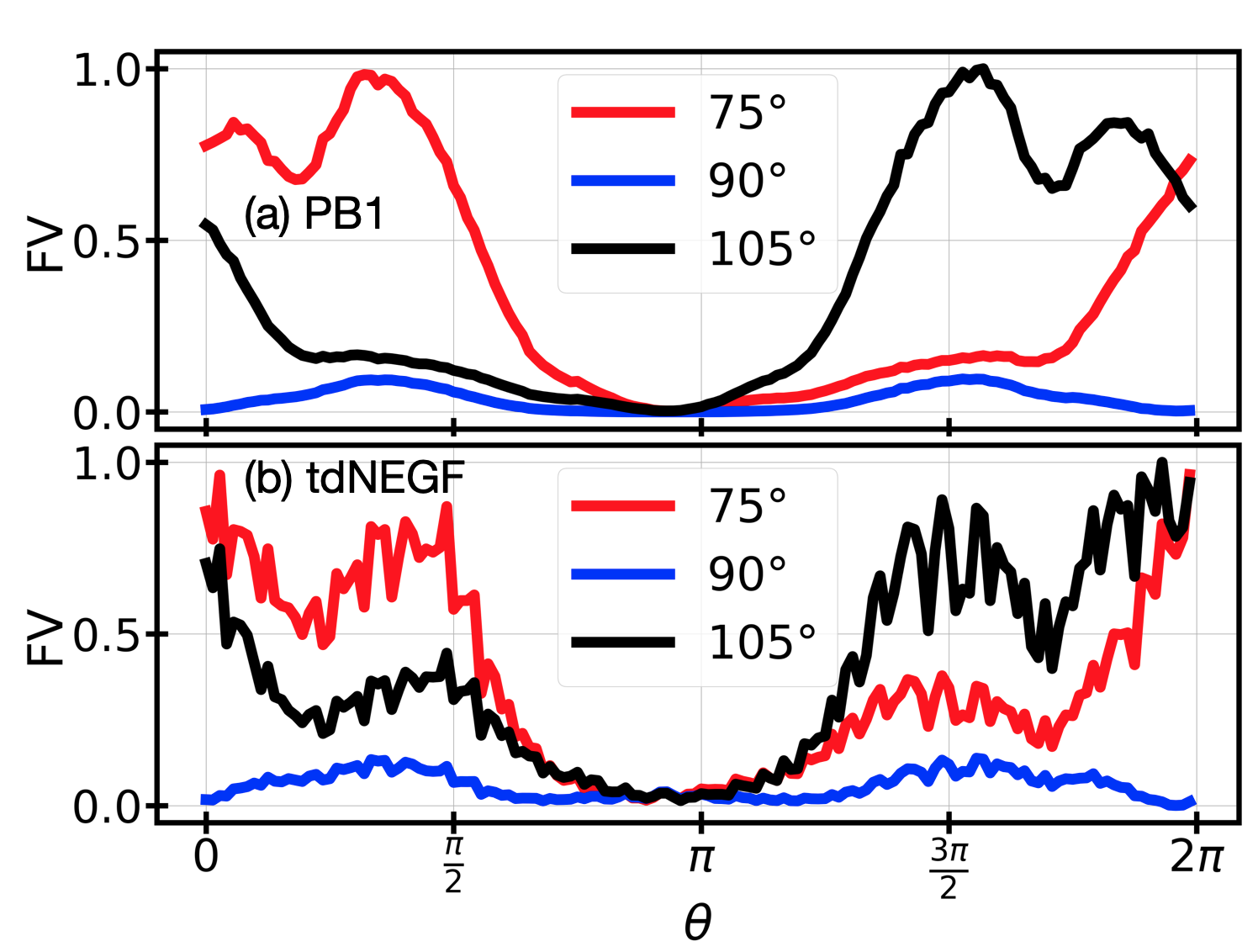}
    \caption{Azimuthal ($\theta$) lineouts for the FV contribution corresponding to the top row of Fig.~\ref{fig:Polarization_angle_dependence} in the top panel and results from Fig.~\ref{fig:Polarization_angle_dependence_NEGF} in the bottom panel.
    Curves for $\phi=75^{\circ}$, $90^{\circ}$, and $105^{\circ}$ are normalized as in Fig.~\ref{fig:Polarization_angle_dependence} and Fig.~\ref{fig:Polarization_angle_dependence_NEGF};  parameters are $\theta_{\mathrm{in}}=68^{\circ}$, $f_F=f_V=0.5$, and ME included.}\label{fig:cross_sec_Polarization_angle_dependence}
\end{figure}

In Fig.~\ref{fig:Polarization_angle_dependence_NEGF} we show the same as in Fig.~\ref{fig:Polarization_angle_dependence}, but as obtained by tdNEGF.
We observe similar behavior: in the FV case, at $\phi = 75^{\circ}$ the distribution exhibits an up–down asymmetry; at $\phi = 90^{\circ}$ it becomes nearly symmetric, but the intensities are strongly suppressed; and at $\phi = 105^{\circ}$ the sense of the asymmetry reverses relative to that at $\phi = 75^{\circ}$, while the magnitudes of the intensities are comparable. 
These observations are consistent with Ref.~\cite{Merboldt2025}, where both experiment and tdNEGF show an asymmetry flip upon crossing $\phi=90^{\circ}$ and a strong suppression of the Volkov channel near $\phi=90^{\circ}$.
Thus, both methods reproduce the qualitative polarization dependence observed in tr-ARPES. 
This is further quantified in Fig.~\ref{fig:cross_sec_Polarization_angle_dependence}, which
displays the intensity distributions in the FV case as a function of azimuthal angle $\theta$.
The $\phi=75^{\circ}$ and $105^{\circ}$ data exhibit opposite asymmetries about $\theta=\pi$, whereas the $\phi=90^{\circ}$ curve is nearly symmetric, but has substantially smaller intensity.

\subsubsection{Incidence-angle dependence}
\begin{figure}[t]
    \centering
    \includegraphics[width=\linewidth]{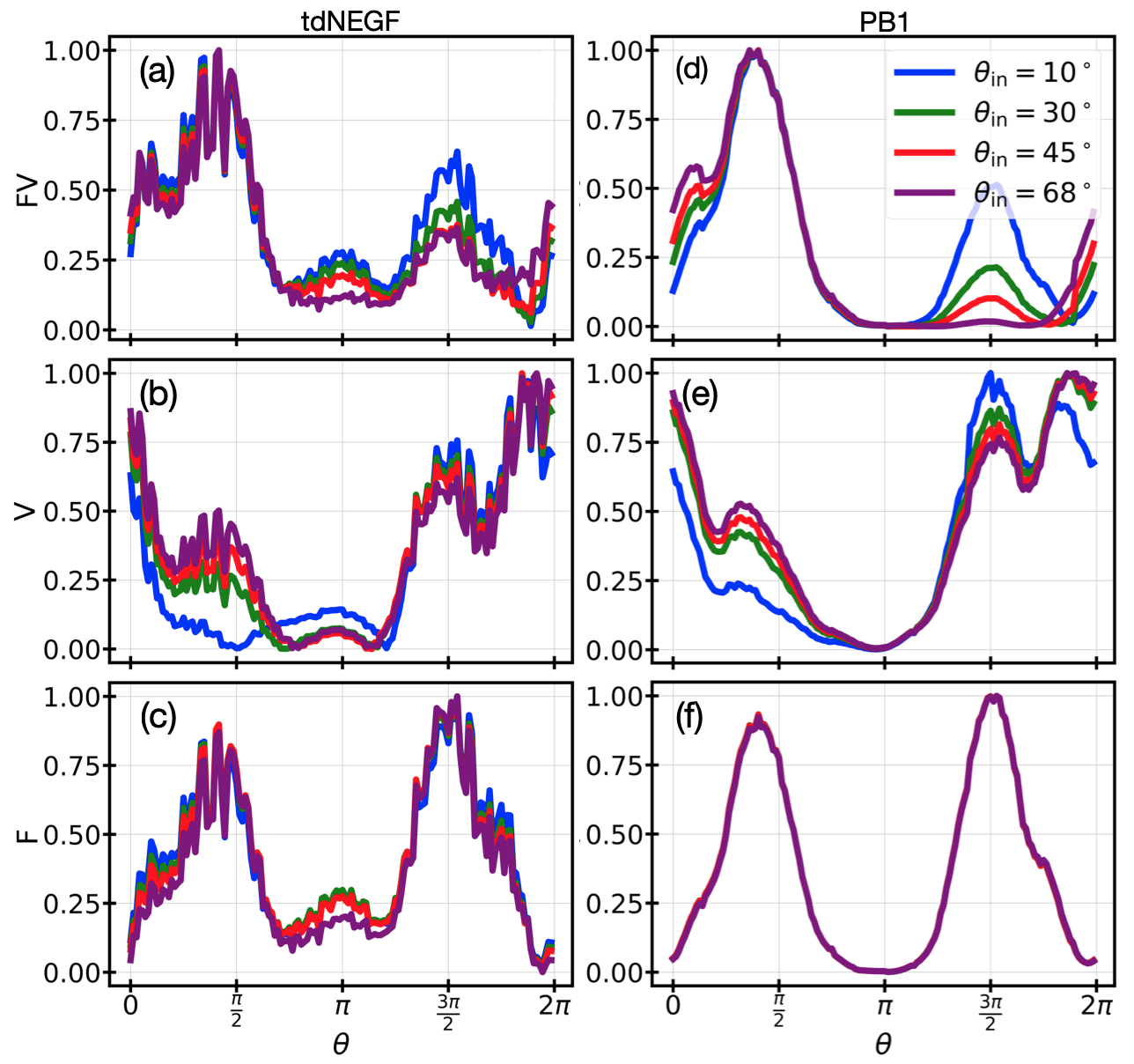}
    \caption{Incidence-angle dependence of the azimuthal intensity dependence of the first sideband. 
    Left column: tdNEGF; right column: PB1. Rows correspond to FV (a,d), V (b,e), and F (c,f) contributions. 
    Curves show $\theta_{\mathrm{in}}=10^{\circ}$ (blue), $30^{\circ}$ (green), $45^{\circ}$ (red), and $68^{\circ}$ (purple) at fixed polarization $\phi=85^{\circ}$ with $f_F=f_V=0.5$ and matrix elements (ME) included;
    each curve is normalized to its maximum.}\label{fig:Incidence_angle_dependence}
\end{figure}

We now fix the polarization angle $\phi = 85^\circ$ and study the behavior at different incidence angles $\theta_{\mathrm{in}}=10^{\circ},\,30^{\circ},\,45^{\circ},\,68^{\circ}$.
For the sake of better readability, we normalize all curves to their respective maxima. Figure~\ref{fig:Incidence_angle_dependence} shows the results for screening $f_F=f_V=0.5$, and including ME.
Panels (a–c) display tdNEGF results for FV, V, and F, respectively; panels (d–f) show the corresponding PB1 results.
Near normal incidence ($\theta_{\mathrm{in}}=10^{\circ},\,30^{\circ}$), both tdNEGF and PB1 exhibit noticeable sensitivity on $\theta_{\mathrm{in}}$ in the FV and V channels, whereas the F channel remains essentially unchanged.
For larger incidence angles ($\theta_{\mathrm{in}}=45^{\circ},\,68^{\circ}$), these trends become largely insensitive to $\theta_{\mathrm{in}}$ within plotting resolution.
However, the PB1 results in the FV channel show an additional peak at $\theta\approx3\pi/2$ that is pronounced at small $\theta_{\mathrm{in}}$ and suppressed at larger $\theta_{\mathrm{in}}$, while tdNEGF [panel (a)] displays a finite peak at $\theta\approx3\pi/2$ for all incidence angles shown here.
In contrast, PB1 shows for the larger incidence angles an increase of intensity at $\theta \approx 2 \pi$, which is absent in the tdNEGF results.
Hence, again there is overall agreement of both methods, but the perturbative PB1 approach seems to incorrectly capture some of the features at large incidence angles.

\subsection{Energy gaps}\label{subsec:energy_gaps}
\begin{figure*}
    \centering
    \includegraphics[width=\linewidth]{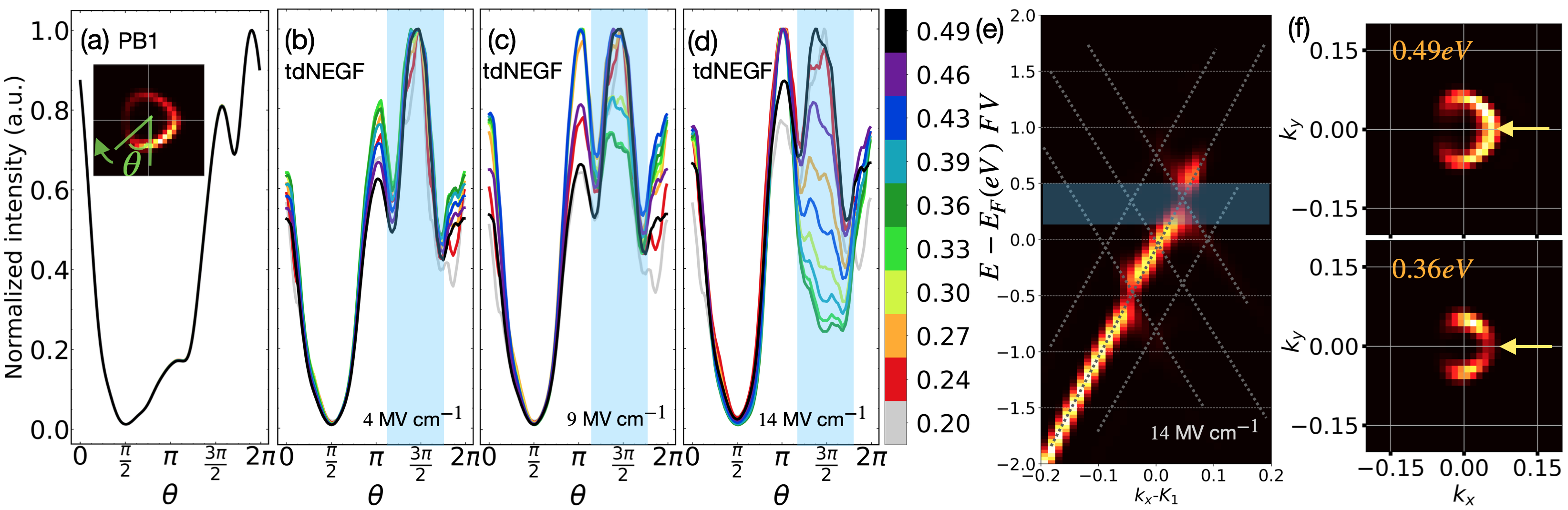}
    \caption{FV-channel intensity at fixed incidence and polarization angles $\theta_{\mathrm{in}}=68^{\circ}$, $\phi=85^{\circ}$, $f_F=f_V=0.5$, and with photoemission matrix elements (ME). The data shown is post-processed using the Gaussian-weighted fitting procedure discussed in Appendix~\ref{app:GaussianFitting} for bin size of $\theta =10^{\circ}$. 
    (a) PB1 intensities are from Eq.~\eqref{eq:photoemission_mit}: the angular profiles are independent of the binding energy ($E_b$) , so lineouts at different ($E_b$) overlap; the color bar in (d) lists the ($E_b$) values used. (b–d) tdNEGF intensities from Eq.~\eqref{eq:intensity_NEGF} at increasing field strengths (4, 9, and 14 MV$/$cm).
    The color bar in (d) maps the constant-energy cuts to the colors of the corresponding angular lineouts.
    (e) Driven graphene spectrum indicating in light blue the hybridization region sampled by the cuts. 
    (f) Intensity distributions at fixed energies taken from (e) at $0.49$eV and $0.36$eV; arrows indicate the hybridization region. 
    All the intensities are normalized to one.}
    \label{fig:energy_cuts}
\end{figure*}

In Fig.~\ref{fig:energy_cuts} we discuss the energy dependence of the intensity distributions. 
This is done by presenting intensity distributions as a function of azimuthal angle $\theta$, but constant energy, at different values of the energy.
We focus on the FV channel and again fix the parameters to the ones of Ref.~\onlinecite{Merboldt2025}, i.e., $\theta_{\mathrm{in}}=68^{\circ}$, $f_F=f_V=0.5$, and we use $\phi=85^{\circ}$ with  ME.

For a better visualization of the angular variation of the intensity at the expected gap location, we smoothen the raw data for the angular distributions: here, we use bins of size $\theta =10^{\circ}$ in width and then increment the starting angle in steps of $2.5^{\circ}$.
This is similar to the procedure in Subsec.~\ref{subsec:azimuthal_lineouts}. 
However, for the sake of a clearer visualization, we now subsequently process the data using a Gaussian-weighted fitting procedure explained in Appendix~\ref{app:GaussianFitting}. 
This allows us to extract smooth angular lineouts.
In the following, we use these fitted distributions to qualitatively assess the variation of the intensity at the gap and to facilitate a direct comparison.
Details of the binning procedure, the fitting methodology, and representative examples of the raw and fitted data discussed in this section are provided in Appendix~\ref{app:GaussianFitting} and Fig.~\ref{fig:gaussianfit}.

Within PB1, the sideband intensity computed from Eq.~\eqref{eq:photoemission_mit} depends solely on the selected sideband (here the first one) and not on the value of the energy see Fig.~\ref{fig:energy_cuts}(a).
This lack of dispersion or hybridization reflects the fact that PB1 retains the transition probability into a given sideband but does not reconstruct the full nonequilibrium band structure.

By contrast, Figs.~\ref{fig:energy_cuts}(b)–\ref{fig:energy_cuts}(d) show tdNEGF results from Eq.~\eqref{eq:intensity_NEGF} for increasing field strengths (4, 9, and $14~\mathrm{MV/cm}$, respectively), which incorporate the band structure.
In Fig.~\ref{fig:energy_cuts}(e) we show the graphene spectrum for the parameters as in Fig.~\ref{fig:energy_cuts}(d), and in Fig.~\ref{fig:energy_cuts}(f) we show the momentum maps for energy cuts at $0.49$eV and $0.36$eV taken from Fig.~\ref{fig:energy_cuts}(e).
Here, we mark in light blue the region of interest in the driven graphene spectrum; the energy $E$ is measured in eV, and $E_F$ is the Fermi energy.
We chose this region, since here we can observe Floquet-induced band mixing and gap openings (hybridization) \cite{Syzranov2008,Oka2009}. 
The different lines in Figs.~\ref{fig:energy_cuts}(b)-(d) correspond to energies $0.2$eV$\leq E \leq 0.49$eV, which all lie within this region of interest, as indicated by the colorbar in Fig.~\ref{fig:energy_cuts}(d).

At Field strength of $4~\mathrm{MV/cm}$ [Fig.~\ref{fig:energy_cuts}(b)]
we observe only weak variations particularly upon changing $E$ around $\theta=3\pi/2$, which is the region
where the opening of the gap is expected [hybridization region, indicated by the vertical light-blue region in Figs.~\ref{fig:energy_cuts}(b)-(d), this corresponds to the expected gap location indicated in the momentum maps in Fig.~\ref{fig:energy_cuts}(f)].

Next, as the field strength increases to 9 and $14~\mathrm{MV/cm}$, respectively, [Figs.~\ref{fig:energy_cuts}(c) and (d)] the energy dependence becomes pronounced: within the hybridization region 
the intensity is suppressed, reducing to $\sim 30\%$ of the original peak value at $14~\mathrm{MV/cm}$.
This is due to the gap opening. Note that, even though the resolution does not suffice to directly observe the gap, this suppression of the intensity is an indicator for its presence and can serve as a guide for experiments.
\begin{figure}
    \centering
    \includegraphics[width=\linewidth]{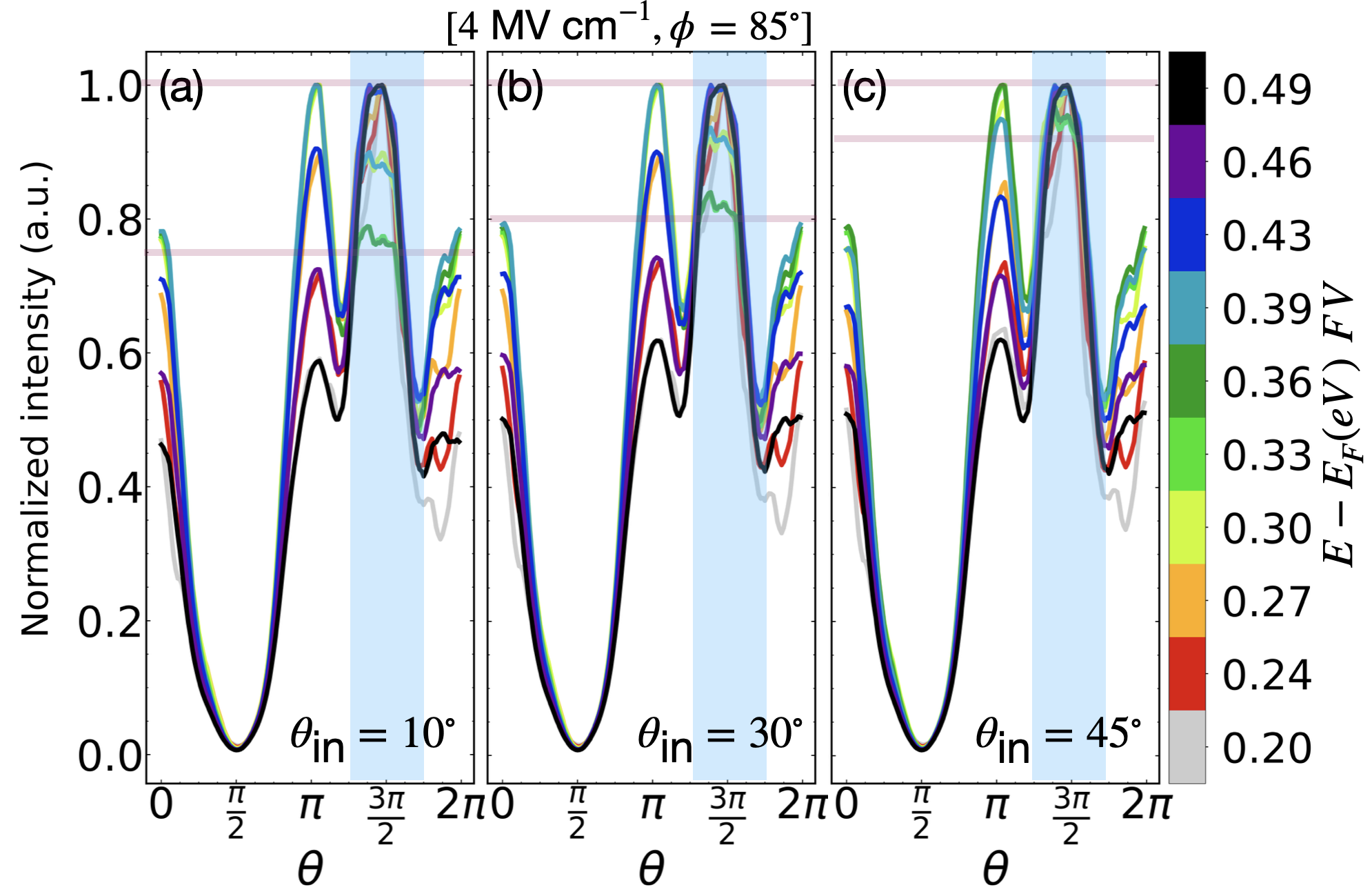}
    \caption{
    FV-channel intensity at fixed parameters matching Fig.~\ref{fig:energy_cuts}(b) (field strength 4MV$/$cm), ME included, $f_F=f_V=0.5$ while varying the incidence angle $\theta_{\mathrm{in}}$.
    The data shown is post-processed using the Gaussian-weighted fitting procedure discussed in Appendix~\ref{app:GaussianFitting} for bin size of $\theta =10^{\circ}$. 
    (a) Near normal incidence, $\theta_{\mathrm{in}}=10^{\circ}$, the angular profile shows a suppression around $\theta=3\pi/2$, consistent with the expected hybridization (gap) region. 
    (b,c) As $\theta_{\mathrm{in}}$ increases to $30^{\circ}$ and $45^{\circ}$, the intensity within the expected gap region rises. Horizontal lines are a guide to the eye for showing the intensity variation.}
    \label{fig:energy_cuts_incidence}
\end{figure}
In Fig.~\ref{fig:energy_cuts_incidence} we present the dependence of the results of Fig.~\ref{fig:energy_cuts}(b) when varying the incidence angle.
We choose this case to further analyse the gap when changing the incidence angle, since Fig.~\ref{fig:energy_cuts}(b) has very weak indications of a gap and we want to see if this can be further enhanced.
As shown in Fig.~\ref{fig:energy_cuts_incidence}(a), for $\theta_{\mathrm{in}}=10^{\circ}$ (near normal incidence) the intensity mainly varies in the region around $\theta=3\pi/2$, consistent with the expected gap location, and is suppressed relative to the peak value.
Note that, in contrast to the results of Fig.~\ref{fig:energy_cuts}(b), which are at $\theta_{\rm in}=68^\circ$, a significant suppression of the intensity is obtained.
Moving away from normal incidence [Figs.~\ref{fig:energy_cuts_incidence}(b) and (c) for $\theta_{\mathrm{in}}=30^{\circ}$ and $\theta_{\mathrm{in}}=45^{\circ}$, respectively] the suppression of the intensity in the expected gap region is much smaller, 
and at $\theta_{\mathrm{in}}=68^{\circ}$ [Fig.~\ref{fig:energy_cuts}(b)], only weak variations remain.
This trend indicates that moving away from normal incidence enhances the Volkov effect, thereby increasing the intensity within the expected hybridization (gap) region, which leads to the vanishing of the gaps.

\begin{figure*}
    \centering
    \includegraphics[width=0.8\textwidth,height=\textheight,keepaspectratio]{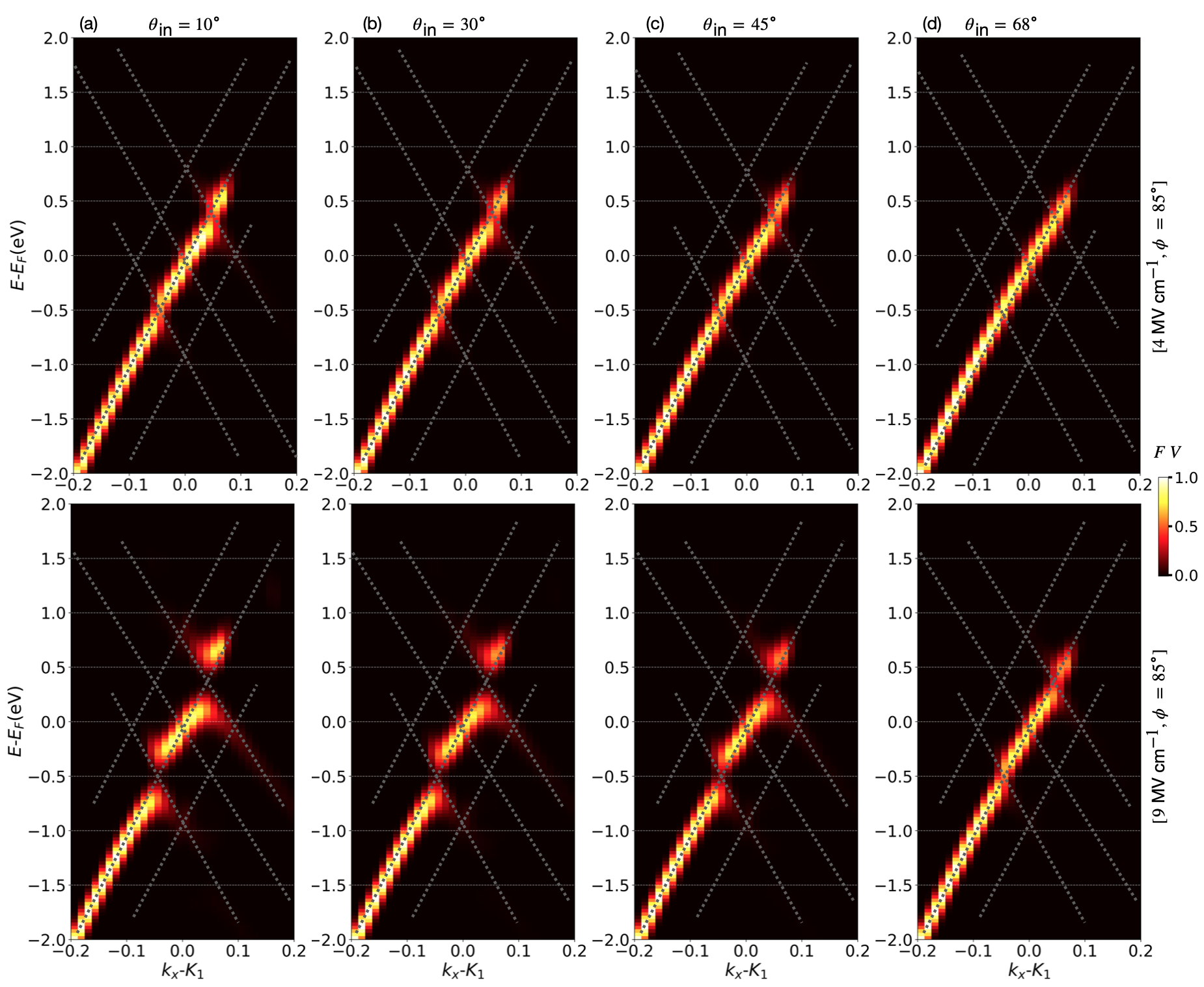}
    \caption{
    Floquet--Volkov (FV) interference in graphene: dependence of the driven band structure on the incidence angle.
    We fix $\phi=85^{\circ}$ and $(f_F,f_V)=(0.5,0.5)$. 
    The top row shows results for $4~\mathrm{MV/cm}$ and the bottom row for $9~\mathrm{MV/cm}$. Panels (a)--(d) correspond to $\theta_{\mathrm{in}}=10^{\circ}, 30^{\circ}, 45^{\circ}, 68^{\circ}$, respectively.
    The field-strength comparison ($4$ vs $9~\mathrm{MV/cm}$) further shows larger apparent gaps at higher fields, complementing Figs.~\ref{fig:energy_cuts} and \ref{fig:energy_cuts_incidence}.}\label{fig:Bandstructure_and_gap_evolution} 
\end{figure*}
This is further discussed in Fig.~\ref{fig:Bandstructure_and_gap_evolution} where we show the tdNEGF results for the band structure in the FV case when varying the field strength and the incidence angle.
We fix the polarization angle at $\phi=85^{\circ}$ and the weights to $(f_F,f_V)=(0.5,0.5)$ as before. 
The top row of Fig.~\ref{fig:Bandstructure_and_gap_evolution} displays results for a field strength of $4~\mathrm{MV/cm}$, and the bottom row for $9~\mathrm{MV/cm}$.
Panels (a)--(d) correspond to incidence angles $\theta_{\mathrm{in}}=10^{\circ}, 30^{\circ}, 45^{\circ},$ and $68^{\circ}$, respectively.
In both field regimes, angles near normal incidence ($\theta_{\mathrm{in}}=10^{\circ}$ and $30^{\circ}$) exhibit a pronounced suppression of the intensity within the expected hybridization (gap) region;
at $9~\mathrm{MV/cm}$ and $\theta_{\mathrm{in}}=10^{\circ}$ the gap-region intensity nearly vanishes.
In contrast, for oblique incidence ($\theta_{\mathrm{in}}=45^{\circ}$ and $68^{\circ}$) we find finite intensity in the gap region, comparable to the peak intensity elsewhere. 
As expected, the apparent gap size increases with field strength when comparing the $4~\mathrm{MV/cm}$ and $9~\mathrm{MV/cm}$ results, and for fixed $\theta_{\mathrm{in}}$ when comparing top and bottom panels.

Note that the residual intensity at the gap does not vanish, which we attribute to finite broadening
in energy resolution, which is due to the energy uncertainty resulting from the short pulse duration, numerical broadening, as well as the finite system size used in the simulations. 
Nevertheless, we do find in our tdNEGF results a significant suppression of the intensities in the hybridization region for small incidence angles or at high field strength and expect that this trend should also be visible in experiments.

\section{Discussion and Outlook}\label{sec:discussion_outlook}
In this paper, we compared the results of the PB1 approach to tdNEGF simulations for the first Floquet sideband intensities by directly computing them for monolayer graphene.
In both methods one can systematically analyze the effect of Floquet-dressed initial states, Volkov states (LAPE), and their interference.
The interference between Volkov- and Floquet-states leads for both methods to results, that, when including matrix elements, reproduce the main features of the experimental findings, in particular the asymmetries of the intensity distributions as a function of $(k_x,k_y)$, which indicate the realization of a Floquet sideband as discussed in Refs.~\onlinecite{Merboldt2025,Choi2025}.
However, while the PB1 approach is able to reproduce qualitatively well the intensity distribution as a function of azimuthal angle, it is not capable of making predictions at different energies, or to describe details of the structure of the Floquet sidebands. 
In particular in hybridization regions, near avoided crossings, and at high field strengths significant quantitative differences to the tdNEGF results arise.
This is due to the more precise treatment of the band structure and of the photoemission intensities within the tdNEGF approach.  

Our results show that, at least for systems with simple band structures such as monolayer graphene, the PB1 and the tdNEGF approaches can be used in a complementary manner.
In a first step, one can use the computationally less expensive PB1 method to explore
trends, e.g. to identify interference regimes and optimize experimental parameters. 
In the next step of the modeling, one needs to incorporate realistic Fresnel coefficients, photoemission matrix elements, and screening parameters of the material at hand.
The so-obtained results can be used to guide the more precise, but also more computationally expensive simulations using tdNEGF.
For typical situations, when interactions are not too strong \cite{Schler2020} (prominently electron-electron scattering, or coupling to phonons), we expect both methods to be able to reproduce the main features of experimental findings.
However, for strongly correlated systems, where due to the interactions the single-particle spectral functions possess a continuum of excitations instead of a band structure, it can be important to apply other numerical methods, such as matrix product states~\cite{Schollwck2011,Paeckel2019} for the investigation of Floquet sidebands~\cite{Gadge2025}.
 
In summary, the tdNEGF should be used instead of the PB1 method whenever strong-field effects, band hybridization, or multiphoton transitions are central.
It also becomes indispensable for quantitative comparisons involving linewidths or dissipation. 
Further system parameters and self-energies can be included~\cite{Schler2020}, which will allow the study of Floquet states in a variety of quantum materials~\cite{delaTorre2021}.

\section{Acknowledgement}
K. G. and S.R.M. acknowledge financial support by Deutsche Forschungsgemeinschaft (DFG, German Research Foundation) Grants No. 436382789, and No. 493420525, via large equipment grants (GOEGrid).
MAS was funded by the European Union (ERC, CAVMAT, project no. 101124492) and by the Deutsche Forschungsgemeinschaft (DFG, German Research Foundation)- 531215165 (Research Unit ‘OPTIMAL’)). Views and opinions expressed are however those of the author(s) only and do not necessarily reflect those of the European Union or the European Research Council. Neither the European Union nor the European Research Council can be held responsible for them.

\section{DATA AVAILABILITY}
The data that support the findings of this article are openly available on Zenodo~\cite{Gadge_data_2026}.
\appendix

\section{Matrix element dependence in PB1}\label{app:matrix_element_PB1}
\begin{figure}[t]
    \centering
    \includegraphics[width=\linewidth]{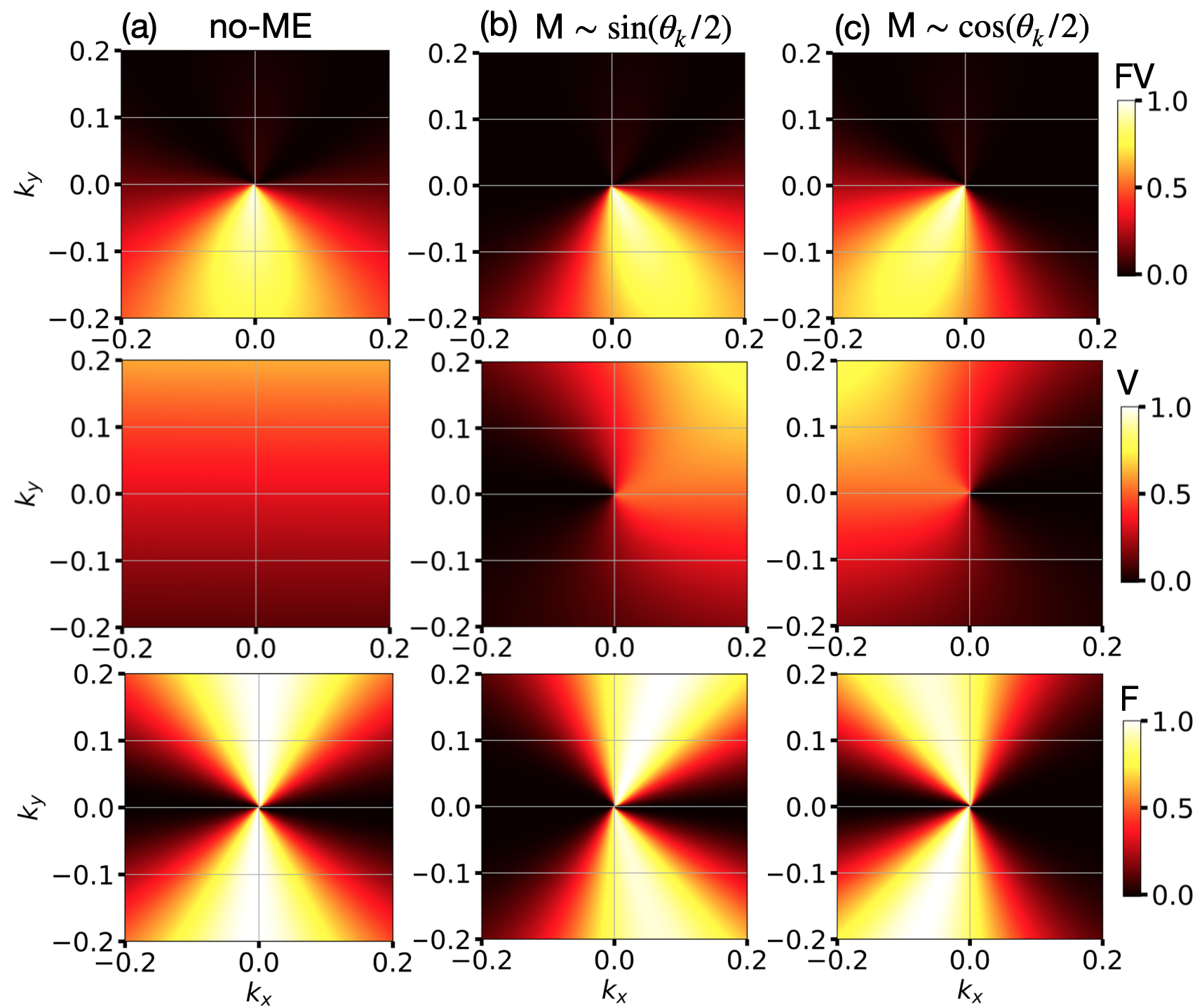}
    \caption{Momentum-resolved intensity distributions plotted in the $(k_x,k_y)$-plane with PB1 for  polarization angle $\phi=85^{\circ}$, $\theta_{\mathrm{in}}=68^{\circ}$, $(f_F,f_V)=(0.5,0.5)$ and different choices of matrix elements.
    In column (a), results are shown with no matrix element (no-ME), in column (b) results are shown with matrix element $\sin{\theta_k/2}$ [ME-sin, see Eq.~\eqref{eq:app_matel_sin}], and in column (c), results are shown with matrix element $\cos{\theta_k/2}$ [ME-cos, see Eq.~\eqref{eq:app_matel_cos}].
    The top row displays the Floquet–Volkov interference (FV), the middle row the Volkov contribution (V), and the bottom row the Floquet contribution (F). }
    \label{fig:Park_matel_dependent}
\end{figure}
\begin{figure}[t]
    \centering
    \includegraphics[width=0.7\linewidth]{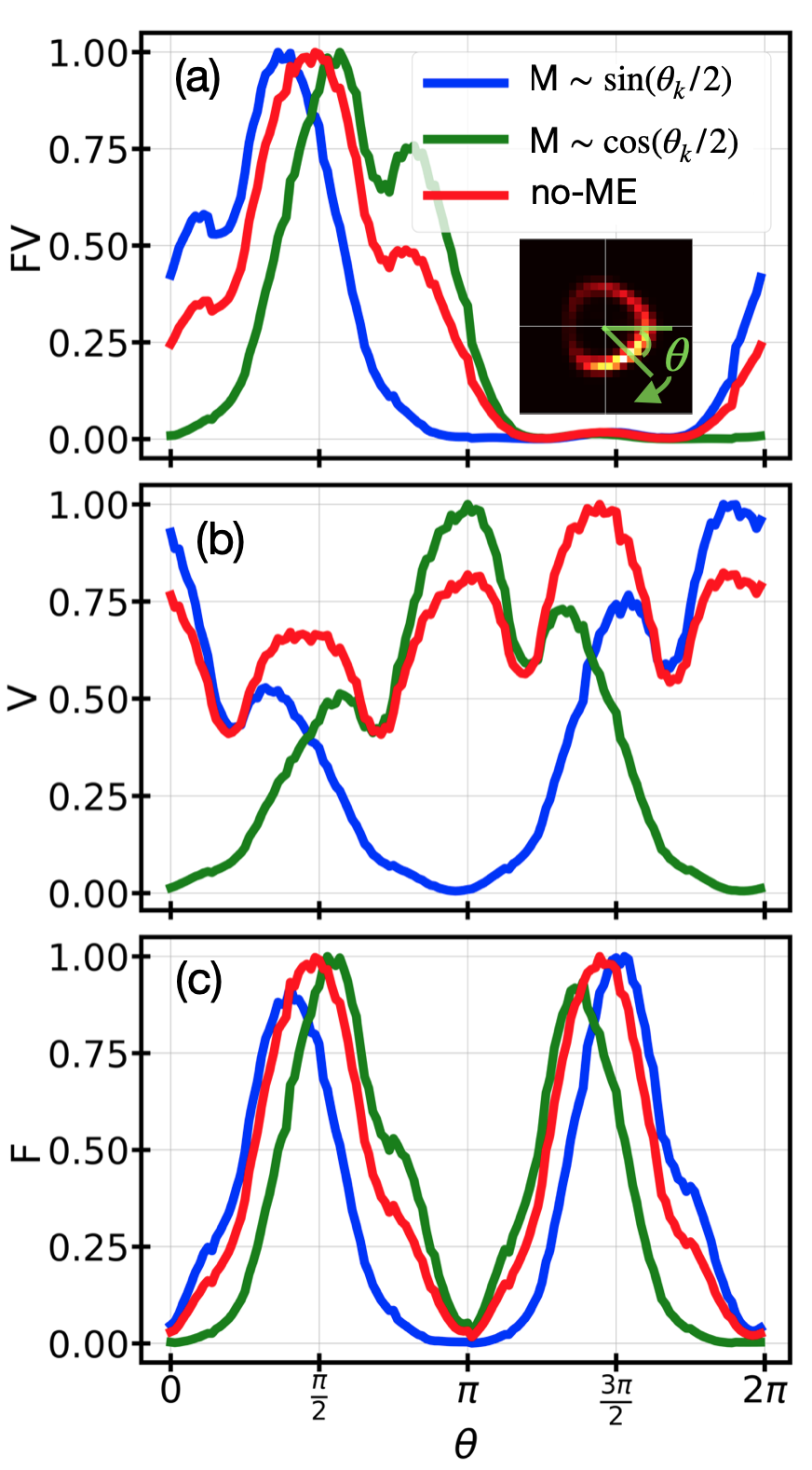}
    \caption{Dependence of the azimuthal $\theta$ intensity distribution on matrix elements in PB1.
    Lineouts are taken from Fig.~\ref{fig:Park_matel_dependent}.
    The inset in (b) illustrates the way the azimuthal data is obtained.
    }
    \label{fig:cross-sec-Park_matel_dependent}
\end{figure}

We summarize in the following how to estimate the matrix elements for monolayer graphene within the PB1 framework, see Ref.~\onlinecite{Hwang2011}.
We emphasize the dependence on the in-plane
momentum angle $\tilde{\theta}_{\mathbf{k}}$, the eigenstates of the
low-energy Dirac Hamiltonian, and the effect of light polarization on the
tr-ARPES intensity (see Ref.~\onlinecite{Hwang2011} for a detailed discussion).

From the Peierls substitution, the light–matter interaction takes the form
\begin{equation}
H^{\mathrm{int}}
= \mathcal{C}\,(A_{x}\sigma_{x}+A_{y}\sigma_{y}),
\qquad
\mathcal{C}=\text{(constant prefactor)} ,
\end{equation}
where $A_{x,y}$ are the Cartesian components of the vector potential.
The  final state of photoelectron is taken to be a plane wave with equal weights on the two carbon sublattices,
\begin{equation}
\ket{f_{k}}
= \frac{1}{\sqrt{2}}
\begin{pmatrix}
1\\
1
\end{pmatrix}.
\end{equation}
For $x$-polarized light we set $\mathbf{A}=(A_{x},0)$, yielding
\begin{equation}
M^{x\text{-pol}}_{s \mathbf{k}}
= \bra{f_{\mathbf{k}}}\, H^{\mathrm{int}} \,\ket{\psi_{s \mathbf{k}}}
= \mathcal{C}A_{x}\,
\bra{f_{\mathbf{k}}}\, \sigma_{x} \,\ket{\psi_{s \mathbf{k}}} \, .
\end{equation}
Here, $s=+$ denotes a state above the Dirac point and $s=-$ a state below the Dirac point; $\ket{\psi_{s \mathbf{k}}}$ is an eigenstate of the undriven Hamiltonian Eq.~\eqref{eq:undrivenH}. 
Following Ref.~\onlinecite{Hwang2011} and setting the hopping amplitude to $1$, the result is
\begin{align}
|M^{x\text{-pol}}_{s \mathbf{k}}|^{2}
&=
\left|
e^{-i\theta_{k}/2}
-(s)\,
e^{+i\theta_{k}/2}
\right|^{2} \\ \nonumber
& \quad =2\big[1-(s)\cos \theta_{k}\big]\, .
\end{align}
\paragraph{Case 1: $s = +$} Here, we choose an electron, which is photoemitted from a state above the Dirac point, and one obtains
\begin{align}
|M^{x\text{-pol}}_{+\mathbf{k}}|^{2}
&\propto (1-\cos \theta_{k})\nonumber\\
&=\sin^{2}\!\left(\frac{\theta_{k}}{2}\right)\,.
\label{eq:app_matel_sin}
\end{align}

\paragraph{Case 2: $s = -$} Here, we choose an electron, which is photoemitted from a state below the Dirac point, and one obtains
\begin{align}
|M^{x\text{-pol}}_{-\mathbf{k}}|^{2}
&\propto (1+\cos\theta_{k})\nonumber\\
&=\cos^{2}\!\left(\frac{\theta_{k}}{2}\right)\,.
\label{eq:app_matel_cos}
\end{align}
Fig.~\ref{fig:Park_matel_dependent} shows the direct effect of the choice of the matrix element for the PB1 results.
We observe that the choice of the matrix element decides the orientation of the dark corridor. 
This is further illustrated by the azimuthal intensity dependence plots in Fig.~\ref{fig:cross-sec-Park_matel_dependent}.

\section{Gaussian-weighted data fitting}\label{app:GaussianFitting}
\begin{figure}[t]
    \centering
    \includegraphics[width=\linewidth]{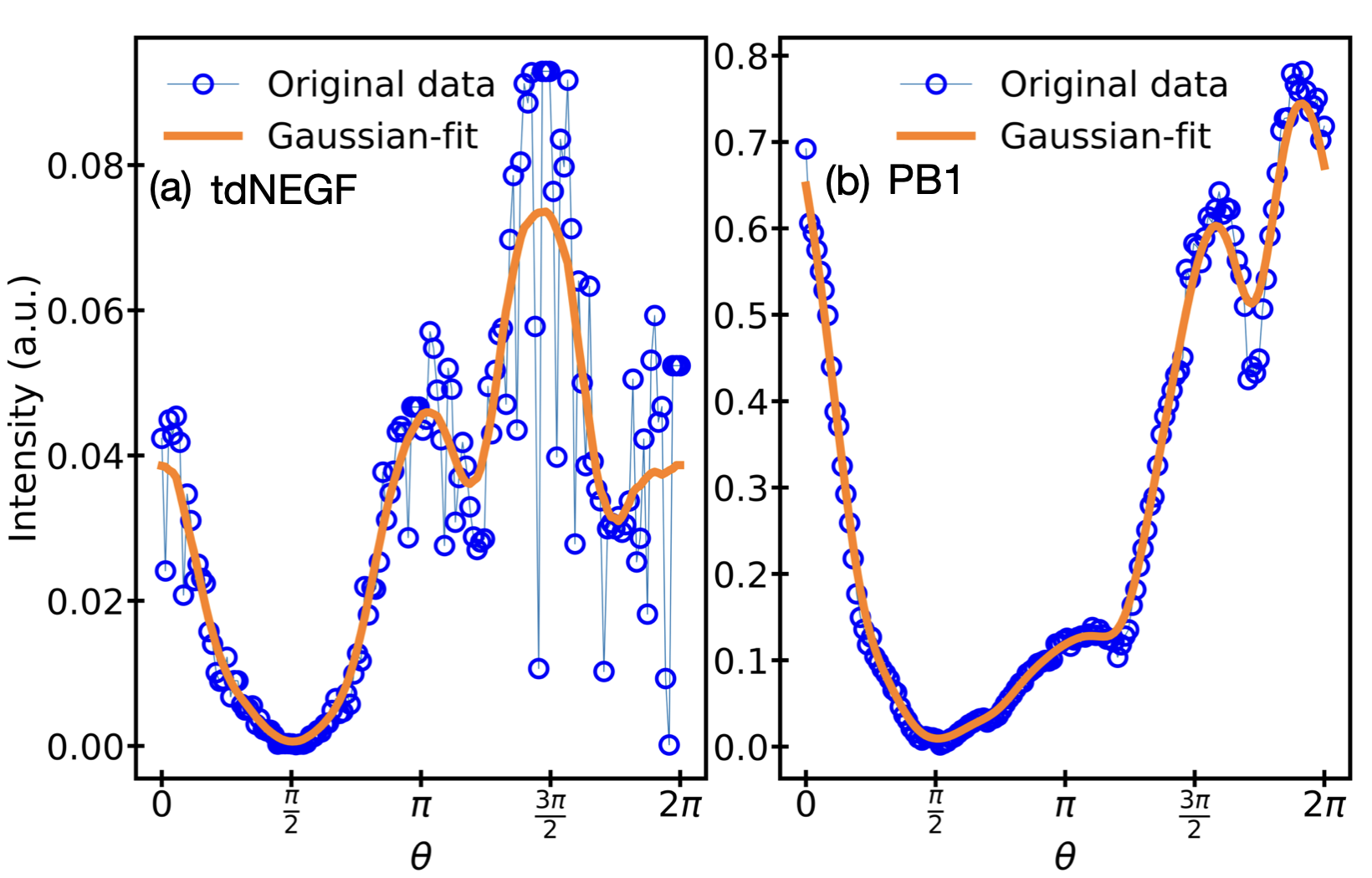}
    \caption{ Azimuthal ($\theta$) lineouts for the data shown in Fig.~\ref{fig:energy_cuts} for tdNEGF ($4~\mathrm{MV/cm}$) in (a) and for PB1 in (b). Blue circles denote the raw data obtained after binning with a $\theta$ bin size of $10^{\circ}$ and summing the intensities within each bin. The orange solid lines show the corresponding Gaussian-weighted fits.
    }
    \label{fig:gaussianfit}
\end{figure}

Curve smoothing is performed using a Gaussian-weighted moving average, which replaces each data point with a weighted average of its neighboring points~\cite{Lindeberg1994,Gonzalez2017}.
Given a discrete signal $y_n$ of length $N$, the smoothed signal $\tilde{y}_n$ is defined as
\begin{equation}
\tilde{y}_n = \sum_{k=-R}^{R} w_k \, y_{(n-k)\,\mathrm{mod}\,N},
\end{equation}
where $R$ is the kernel radius and the weights $w_k$ are given by a normalized discrete Gaussian,
\begin{equation}
w_k = \frac{\exp\!\left(-\frac{k^2}{2\sigma^2}\right)}{\sum_{j=-R}^{R} \exp\!\left(-\frac{j^2}{2\sigma^2}\right)}.
\end{equation}
Here, $\sigma$ controls the width of the Gaussian and therefore the degree of smoothing.
The modulo operation enforces periodic (wrap-around) boundary conditions.
For the discrete data set, we use $R = 8$ (i.e., $2 \cdot \sigma$); in this way, we perform a Gaussian-weighted average at each point, similar to a running average.
In Fig.~\ref{fig:gaussianfit}, we show the raw data and the corresponding Gaussian-weighted fits for the angular cuts presented in the main text Fig.~\ref{fig:energy_cuts}, for the $9~\mathrm{MV/cm}$ tdNEGF case and for the PB1 case.
We use a $\theta$ bin size of $10^{\circ}$ and sum the intensities within each bin. The resulting angular distributions are then fitted using the Gaussian-weighted fitting procedure described above.

\newpage
\bibliography{main2.bib}

\begin{thebibliography}{69}%
\makeatletter
\providecommand \@ifxundefined [1]{%
 \@ifx{#1\undefined}
}%
\providecommand \@ifnum [1]{%
 \ifnum #1\expandafter \@firstoftwo
 \else \expandafter \@secondoftwo
 \fi
}%
\providecommand \@ifx [1]{%
 \ifx #1\expandafter \@firstoftwo
 \else \expandafter \@secondoftwo
 \fi
}%
\providecommand \natexlab [1]{#1}%
\providecommand \enquote  [1]{``#1''}%
\providecommand \bibnamefont  [1]{#1}%
\providecommand \bibfnamefont [1]{#1}%
\providecommand \citenamefont [1]{#1}%
\providecommand \href@noop [0]{\@secondoftwo}%
\providecommand \href [0]{\begingroup \@sanitize@url \@href}%
\providecommand \@href[1]{\@@startlink{#1}\@@href}%
\providecommand \@@href[1]{\endgroup#1\@@endlink}%
\providecommand \@sanitize@url [0]{\catcode `\\12\catcode `\$12\catcode
  `\&12\catcode `\#12\catcode `\^12\catcode `\_12\catcode `\%12\relax}%
\providecommand \@@startlink[1]{}%
\providecommand \@@endlink[0]{}%
\providecommand \url  [0]{\begingroup\@sanitize@url \@url }%
\providecommand \@url [1]{\endgroup\@href {#1}{\urlprefix }}%
\providecommand \urlprefix  [0]{URL }%
\providecommand \Eprint [0]{\href }%
\providecommand \doibase [0]{https://doi.org/}%
\providecommand \selectlanguage [0]{\@gobble}%
\providecommand \bibinfo  [0]{\@secondoftwo}%
\providecommand \bibfield  [0]{\@secondoftwo}%
\providecommand \translation [1]{[#1]}%
\providecommand \BibitemOpen [0]{}%
\providecommand \bibitemStop [0]{}%
\providecommand \bibitemNoStop [0]{.\EOS\space}%
\providecommand \EOS [0]{\spacefactor3000\relax}%
\providecommand \BibitemShut  [1]{\csname bibitem#1\endcsname}%
\let\auto@bib@innerbib\@empty
\bibitem [{\citenamefont {Basov}\ \emph {et~al.}(2017)\citenamefont {Basov},
  \citenamefont {Averitt},\ and\ \citenamefont {Hsieh}}]{Basov2017}%
  \BibitemOpen
  \bibfield  {author} {\bibinfo {author} {\bibfnamefont {D.~N.}\ \bibnamefont
  {Basov}}, \bibinfo {author} {\bibfnamefont {R.~D.}\ \bibnamefont {Averitt}},\
  and\ \bibinfo {author} {\bibfnamefont {D.}~\bibnamefont {Hsieh}},\ }\href
  {https://doi.org/10.1038/nmat5017} {\bibfield  {journal} {\bibinfo  {journal}
  {Nature Materials}\ }\textbf {\bibinfo {volume} {16}},\ \bibinfo {pages}
  {1077–1088} (\bibinfo {year} {2017})}\BibitemShut {NoStop}%
\bibitem [{\citenamefont {Oka}\ and\ \citenamefont {Kitamura}(2019)}]{Oka2019}%
  \BibitemOpen
  \bibfield  {author} {\bibinfo {author} {\bibfnamefont {T.}~\bibnamefont
  {Oka}}\ and\ \bibinfo {author} {\bibfnamefont {S.}~\bibnamefont {Kitamura}},\
  }\href {https://doi.org/10.1146/annurev-conmatphys-031218-013423} {\bibfield
  {journal} {\bibinfo  {journal} {Annual Review of Condensed Matter Physics}\
  }\textbf {\bibinfo {volume} {10}},\ \bibinfo {pages} {387–408} (\bibinfo
  {year} {2019})}\BibitemShut {NoStop}%
\bibitem [{\citenamefont {de~la Torre}\ \emph {et~al.}(2021)\citenamefont
  {de~la Torre}, \citenamefont {Kennes}, \citenamefont {Claassen},
  \citenamefont {Gerber}, \citenamefont {McIver},\ and\ \citenamefont
  {Sentef}}]{delaTorre2021}%
  \BibitemOpen
  \bibfield  {author} {\bibinfo {author} {\bibfnamefont {A.}~\bibnamefont
  {de~la Torre}}, \bibinfo {author} {\bibfnamefont {D.~M.}\ \bibnamefont
  {Kennes}}, \bibinfo {author} {\bibfnamefont {M.}~\bibnamefont {Claassen}},
  \bibinfo {author} {\bibfnamefont {S.}~\bibnamefont {Gerber}}, \bibinfo
  {author} {\bibfnamefont {J.~W.}\ \bibnamefont {McIver}},\ and\ \bibinfo
  {author} {\bibfnamefont {M.~A.}\ \bibnamefont {Sentef}},\ }\bibfield
  {journal} {\bibinfo  {journal} {Reviews of Modern Physics}\ }\textbf
  {\bibinfo {volume} {93}},\ \href
  {https://doi.org/10.1103/revmodphys.93.041002} {10.1103/revmodphys.93.041002}
  (\bibinfo {year} {2021})\BibitemShut {NoStop}%
\bibitem [{\citenamefont {Bao}\ \emph {et~al.}(2021)\citenamefont {Bao},
  \citenamefont {Tang}, \citenamefont {Sun},\ and\ \citenamefont
  {Zhou}}]{Bao2021}%
  \BibitemOpen
  \bibfield  {author} {\bibinfo {author} {\bibfnamefont {C.}~\bibnamefont
  {Bao}}, \bibinfo {author} {\bibfnamefont {P.}~\bibnamefont {Tang}}, \bibinfo
  {author} {\bibfnamefont {D.}~\bibnamefont {Sun}},\ and\ \bibinfo {author}
  {\bibfnamefont {S.}~\bibnamefont {Zhou}},\ }\href
  {https://doi.org/10.1038/s42254-021-00388-1} {\bibfield  {journal} {\bibinfo
  {journal} {Nature Reviews Physics}\ }\textbf {\bibinfo {volume} {4}},\
  \bibinfo {pages} {33–48} (\bibinfo {year} {2021})}\BibitemShut {NoStop}%
\bibitem [{\citenamefont {Goldman}\ and\ \citenamefont
  {Dalibard}(2014)}]{Goldman2014}%
  \BibitemOpen
  \bibfield  {author} {\bibinfo {author} {\bibfnamefont {N.}~\bibnamefont
  {Goldman}}\ and\ \bibinfo {author} {\bibfnamefont {J.}~\bibnamefont
  {Dalibard}},\ }\bibfield  {journal} {\bibinfo  {journal} {Physical Review X}\
  }\textbf {\bibinfo {volume} {4}},\ \href
  {https://doi.org/10.1103/physrevx.4.031027} {10.1103/physrevx.4.031027}
  (\bibinfo {year} {2014})\BibitemShut {NoStop}%
\bibitem [{\citenamefont {Eckardt}(2017)}]{Eckardt2017}%
  \BibitemOpen
  \bibfield  {author} {\bibinfo {author} {\bibfnamefont {A.}~\bibnamefont
  {Eckardt}},\ }\bibfield  {journal} {\bibinfo  {journal} {Reviews of Modern
  Physics}\ }\textbf {\bibinfo {volume} {89}},\ \href
  {https://doi.org/10.1103/revmodphys.89.011004} {10.1103/revmodphys.89.011004}
  (\bibinfo {year} {2017})\BibitemShut {NoStop}%
\bibitem [{\citenamefont {Mentink}\ \emph {et~al.}(2015)\citenamefont
  {Mentink}, \citenamefont {Balzer},\ and\ \citenamefont
  {Eckstein}}]{Mentink2015}%
  \BibitemOpen
  \bibfield  {author} {\bibinfo {author} {\bibfnamefont {J.~H.}\ \bibnamefont
  {Mentink}}, \bibinfo {author} {\bibfnamefont {K.}~\bibnamefont {Balzer}},\
  and\ \bibinfo {author} {\bibfnamefont {M.}~\bibnamefont {Eckstein}},\
  }\bibfield  {journal} {\bibinfo  {journal} {Nature Communications}\ }\textbf
  {\bibinfo {volume} {6}},\ \href {https://doi.org/10.1038/ncomms7708}
  {10.1038/ncomms7708} (\bibinfo {year} {2015})\BibitemShut {NoStop}%
\bibitem [{\citenamefont {Valmispild}\ \emph {et~al.}(2024)\citenamefont
  {Valmispild}, \citenamefont {Gorelov}, \citenamefont {Eckstein},
  \citenamefont {Lichtenstein}, \citenamefont {Aoki}, \citenamefont
  {Katsnelson}, \citenamefont {Ivanov},\ and\ \citenamefont
  {Smirnova}}]{Valmispild2024}%
  \BibitemOpen
  \bibfield  {author} {\bibinfo {author} {\bibfnamefont {V.~N.}\ \bibnamefont
  {Valmispild}}, \bibinfo {author} {\bibfnamefont {E.}~\bibnamefont {Gorelov}},
  \bibinfo {author} {\bibfnamefont {M.}~\bibnamefont {Eckstein}}, \bibinfo
  {author} {\bibfnamefont {A.~I.}\ \bibnamefont {Lichtenstein}}, \bibinfo
  {author} {\bibfnamefont {H.}~\bibnamefont {Aoki}}, \bibinfo {author}
  {\bibfnamefont {M.~I.}\ \bibnamefont {Katsnelson}}, \bibinfo {author}
  {\bibfnamefont {M.~Y.}\ \bibnamefont {Ivanov}},\ and\ \bibinfo {author}
  {\bibfnamefont {O.}~\bibnamefont {Smirnova}},\ }\href
  {https://doi.org/10.1038/s41566-023-01371-1} {\bibfield  {journal} {\bibinfo
  {journal} {Nature Photonics}\ }\textbf {\bibinfo {volume} {18}},\ \bibinfo
  {pages} {432–439} (\bibinfo {year} {2024})}\BibitemShut {NoStop}%
\bibitem [{\citenamefont {Li}\ and\ \citenamefont {Eckstein}(2020)}]{Li2020}%
  \BibitemOpen
  \bibfield  {author} {\bibinfo {author} {\bibfnamefont {J.}~\bibnamefont
  {Li}}\ and\ \bibinfo {author} {\bibfnamefont {M.}~\bibnamefont {Eckstein}},\
  }\bibfield  {journal} {\bibinfo  {journal} {Physical Review Letters}\
  }\textbf {\bibinfo {volume} {125}},\ \href
  {https://doi.org/10.1103/physrevlett.125.217402}
  {10.1103/physrevlett.125.217402} (\bibinfo {year} {2020})\BibitemShut
  {NoStop}%
\bibitem [{\citenamefont {Murakami}\ \emph {et~al.}(2025)\citenamefont
  {Murakami}, \citenamefont {Golež}, \citenamefont {Eckstein},\ and\
  \citenamefont {Werner}}]{Murakami2025}%
  \BibitemOpen
  \bibfield  {author} {\bibinfo {author} {\bibfnamefont {Y.}~\bibnamefont
  {Murakami}}, \bibinfo {author} {\bibfnamefont {D.}~\bibnamefont {Golež}},
  \bibinfo {author} {\bibfnamefont {M.}~\bibnamefont {Eckstein}},\ and\
  \bibinfo {author} {\bibfnamefont {P.}~\bibnamefont {Werner}},\ }\bibfield
  {journal} {\bibinfo  {journal} {Reviews of Modern Physics}\ }\textbf
  {\bibinfo {volume} {97}},\ \href {https://doi.org/10.1103/tkjh-lr83}
  {10.1103/tkjh-lr83} (\bibinfo {year} {2025})\BibitemShut {NoStop}%
\bibitem [{\citenamefont {Giovannini}\ and\ \citenamefont
  {H\"{u}bener}(2019)}]{Giovannini2019}%
  \BibitemOpen
  \bibfield  {author} {\bibinfo {author} {\bibfnamefont {U.~D.}\ \bibnamefont
  {Giovannini}}\ and\ \bibinfo {author} {\bibfnamefont {H.}~\bibnamefont
  {H\"{u}bener}},\ }\href {https://doi.org/10.1088/2515-7639/ab387b} {\bibfield
   {journal} {\bibinfo  {journal} {Journal of Physics: Materials}\ }\textbf
  {\bibinfo {volume} {3}},\ \bibinfo {pages} {012001} (\bibinfo {year}
  {2019})}\BibitemShut {NoStop}%
\bibitem [{\citenamefont {Rudner}\ and\ \citenamefont
  {Lindner}(2020)}]{Rudner2020}%
  \BibitemOpen
  \bibfield  {author} {\bibinfo {author} {\bibfnamefont {M.~S.}\ \bibnamefont
  {Rudner}}\ and\ \bibinfo {author} {\bibfnamefont {N.~H.}\ \bibnamefont
  {Lindner}},\ }\href {https://doi.org/10.1038/s42254-020-0170-z} {\bibfield
  {journal} {\bibinfo  {journal} {Nature Reviews Physics}\ }\textbf {\bibinfo
  {volume} {2}},\ \bibinfo {pages} {229–244} (\bibinfo {year}
  {2020})}\BibitemShut {NoStop}%
\bibitem [{\citenamefont {Day}\ \emph {et~al.}(2024)\citenamefont {Day},
  \citenamefont {Kusyak}, \citenamefont {Sturm}, \citenamefont {Aranzadi},
  \citenamefont {Bretscher}, \citenamefont {Fechner}, \citenamefont
  {Matsuyama}, \citenamefont {Michael}, \citenamefont {Schulte}, \citenamefont
  {Li}, \citenamefont {Hagelstein}, \citenamefont {Herrmann}, \citenamefont
  {Kipp}, \citenamefont {Potts}, \citenamefont {DeStefano}, \citenamefont {Hu},
  \citenamefont {Huang}, \citenamefont {Taniguchi}, \citenamefont {Watanabe},
  \citenamefont {Meier}, \citenamefont {Shin}, \citenamefont {Rubio},
  \citenamefont {Chu}, \citenamefont {Kennes}, \citenamefont {Sentef},\ and\
  \citenamefont {McIver}}]{DayMcIver2024}%
  \BibitemOpen
  \bibfield  {author} {\bibinfo {author} {\bibfnamefont {M.~W.}\ \bibnamefont
  {Day}}, \bibinfo {author} {\bibfnamefont {K.}~\bibnamefont {Kusyak}},
  \bibinfo {author} {\bibfnamefont {F.}~\bibnamefont {Sturm}}, \bibinfo
  {author} {\bibfnamefont {J.~I.}\ \bibnamefont {Aranzadi}}, \bibinfo {author}
  {\bibfnamefont {H.~M.}\ \bibnamefont {Bretscher}}, \bibinfo {author}
  {\bibfnamefont {M.}~\bibnamefont {Fechner}}, \bibinfo {author} {\bibfnamefont
  {T.}~\bibnamefont {Matsuyama}}, \bibinfo {author} {\bibfnamefont {M.~H.}\
  \bibnamefont {Michael}}, \bibinfo {author} {\bibfnamefont {B.~F.}\
  \bibnamefont {Schulte}}, \bibinfo {author} {\bibfnamefont {X.}~\bibnamefont
  {Li}}, \bibinfo {author} {\bibfnamefont {J.}~\bibnamefont {Hagelstein}},
  \bibinfo {author} {\bibfnamefont {D.}~\bibnamefont {Herrmann}}, \bibinfo
  {author} {\bibfnamefont {G.}~\bibnamefont {Kipp}}, \bibinfo {author}
  {\bibfnamefont {A.~M.}\ \bibnamefont {Potts}}, \bibinfo {author}
  {\bibfnamefont {J.~M.}\ \bibnamefont {DeStefano}}, \bibinfo {author}
  {\bibfnamefont {C.}~\bibnamefont {Hu}}, \bibinfo {author} {\bibfnamefont
  {Y.}~\bibnamefont {Huang}}, \bibinfo {author} {\bibfnamefont
  {T.}~\bibnamefont {Taniguchi}}, \bibinfo {author} {\bibfnamefont
  {K.}~\bibnamefont {Watanabe}}, \bibinfo {author} {\bibfnamefont
  {G.}~\bibnamefont {Meier}}, \bibinfo {author} {\bibfnamefont
  {D.}~\bibnamefont {Shin}}, \bibinfo {author} {\bibfnamefont {A.}~\bibnamefont
  {Rubio}}, \bibinfo {author} {\bibfnamefont {J.-H.}\ \bibnamefont {Chu}},
  \bibinfo {author} {\bibfnamefont {D.~M.}\ \bibnamefont {Kennes}}, \bibinfo
  {author} {\bibfnamefont {M.~A.}\ \bibnamefont {Sentef}},\ and\ \bibinfo
  {author} {\bibfnamefont {J.~W.}\ \bibnamefont {McIver}},\ }\href
  {https://doi.org/10.48550/ARXIV.2409.04531} {\bibinfo {title}
  {Nonperturbative nonlinear transport in a floquet-weyl semimetal}} (\bibinfo
  {year} {2024})\BibitemShut {NoStop}%
\bibitem [{\citenamefont {Grifoni}\ and\ \citenamefont
  {H\"{a}nggi}(1998)}]{Grifoni1998}%
  \BibitemOpen
  \bibfield  {author} {\bibinfo {author} {\bibfnamefont {M.}~\bibnamefont
  {Grifoni}}\ and\ \bibinfo {author} {\bibfnamefont {P.}~\bibnamefont
  {H\"{a}nggi}},\ }\href {https://doi.org/10.1016/s0370-1573(98)00022-2}
  {\bibfield  {journal} {\bibinfo  {journal} {Physics Reports}\ }\textbf
  {\bibinfo {volume} {304}},\ \bibinfo {pages} {229–354} (\bibinfo {year}
  {1998})}\BibitemShut {NoStop}%
\bibitem [{\citenamefont {Liu}\ \emph {et~al.}(2025)\citenamefont {Liu},
  \citenamefont {Yang}, \citenamefont {Gaertner}, \citenamefont {Huckabee},
  \citenamefont {Suslov}, \citenamefont {Refael}, \citenamefont {Nathan},
  \citenamefont {Lewandowski}, \citenamefont {Torres}, \citenamefont {Esin},
  \citenamefont {Barbara},\ and\ \citenamefont {Kalugin}}]{liu2025}%
  \BibitemOpen
  \bibfield  {author} {\bibinfo {author} {\bibfnamefont {Y.}~\bibnamefont
  {Liu}}, \bibinfo {author} {\bibfnamefont {C.}~\bibnamefont {Yang}}, \bibinfo
  {author} {\bibfnamefont {G.}~\bibnamefont {Gaertner}}, \bibinfo {author}
  {\bibfnamefont {J.}~\bibnamefont {Huckabee}}, \bibinfo {author}
  {\bibfnamefont {A.~V.}\ \bibnamefont {Suslov}}, \bibinfo {author}
  {\bibfnamefont {G.}~\bibnamefont {Refael}}, \bibinfo {author} {\bibfnamefont
  {F.}~\bibnamefont {Nathan}}, \bibinfo {author} {\bibfnamefont
  {C.}~\bibnamefont {Lewandowski}}, \bibinfo {author} {\bibfnamefont {L.~E.
  F.~F.}\ \bibnamefont {Torres}}, \bibinfo {author} {\bibfnamefont
  {I.}~\bibnamefont {Esin}}, \bibinfo {author} {\bibfnamefont {P.}~\bibnamefont
  {Barbara}},\ and\ \bibinfo {author} {\bibfnamefont {N.~G.}\ \bibnamefont
  {Kalugin}},\ }\href@noop {} {\bibfield  {journal} {\bibinfo  {journal}
  {Nature Communications}\ }\textbf {\bibinfo {volume} {16}},\ \bibinfo {pages}
  {2057} (\bibinfo {year} {2025})}\BibitemShut {NoStop}%
\bibitem [{\citenamefont {Hirai}\ \emph {et~al.}(2024)\citenamefont {Hirai},
  \citenamefont {Okumura}, \citenamefont {Yoshikawa}, \citenamefont {Oka},\
  and\ \citenamefont {Shimano}}]{Shimano2024}%
  \BibitemOpen
  \bibfield  {author} {\bibinfo {author} {\bibfnamefont {Y.}~\bibnamefont
  {Hirai}}, \bibinfo {author} {\bibfnamefont {S.}~\bibnamefont {Okumura}},
  \bibinfo {author} {\bibfnamefont {N.}~\bibnamefont {Yoshikawa}}, \bibinfo
  {author} {\bibfnamefont {T.}~\bibnamefont {Oka}},\ and\ \bibinfo {author}
  {\bibfnamefont {R.}~\bibnamefont {Shimano}},\ }\bibfield  {journal} {\bibinfo
   {journal} {Physical Review Research}\ }\textbf {\bibinfo {volume} {6}},\
  \href {https://doi.org/10.1103/physrevresearch.6.l012027}
  {10.1103/physrevresearch.6.l012027} (\bibinfo {year} {2024})\BibitemShut
  {NoStop}%
\bibitem [{\citenamefont {Shan}\ \emph {et~al.}(2022)\citenamefont {Shan},
  \citenamefont {Ye}, \citenamefont {Chu}, \citenamefont {Lee}, \citenamefont
  {Park}, \citenamefont {Balents},\ and\ \citenamefont {Hsieh}}]{Hsieh2022}%
  \BibitemOpen
  \bibfield  {author} {\bibinfo {author} {\bibfnamefont {J.-Y.}\ \bibnamefont
  {Shan}}, \bibinfo {author} {\bibfnamefont {M.}~\bibnamefont {Ye}}, \bibinfo
  {author} {\bibfnamefont {H.}~\bibnamefont {Chu}}, \bibinfo {author}
  {\bibfnamefont {S.}~\bibnamefont {Lee}}, \bibinfo {author} {\bibfnamefont
  {J.-G.}\ \bibnamefont {Park}}, \bibinfo {author} {\bibfnamefont
  {L.}~\bibnamefont {Balents}},\ and\ \bibinfo {author} {\bibfnamefont
  {D.}~\bibnamefont {Hsieh}},\ }\href
  {https://doi.org/10.1038/s41586-021-04368-4} {\bibfield  {journal} {\bibinfo
  {journal} {Nature}\ }\textbf {\bibinfo {volume} {602}},\ \bibinfo {pages}
  {E19–E19} (\bibinfo {year} {2022})}\BibitemShut {NoStop}%
\bibitem [{\citenamefont {Bao}\ \emph {et~al.}(2025)\citenamefont {Bao},
  \citenamefont {Zhong}, \citenamefont {Fan}, \citenamefont {Cai},
  \citenamefont {Wang}, \citenamefont {Zhou}, \citenamefont {Lin},
  \citenamefont {Zhang}, \citenamefont {Yu}, \citenamefont {Tang},
  \citenamefont {Duan},\ and\ \citenamefont {Zhou}}]{Bao2025}%
  \BibitemOpen
  \bibfield  {author} {\bibinfo {author} {\bibfnamefont {C.}~\bibnamefont
  {Bao}}, \bibinfo {author} {\bibfnamefont {H.}~\bibnamefont {Zhong}}, \bibinfo
  {author} {\bibfnamefont {B.}~\bibnamefont {Fan}}, \bibinfo {author}
  {\bibfnamefont {X.}~\bibnamefont {Cai}}, \bibinfo {author} {\bibfnamefont
  {F.}~\bibnamefont {Wang}}, \bibinfo {author} {\bibfnamefont {S.}~\bibnamefont
  {Zhou}}, \bibinfo {author} {\bibfnamefont {T.}~\bibnamefont {Lin}}, \bibinfo
  {author} {\bibfnamefont {H.}~\bibnamefont {Zhang}}, \bibinfo {author}
  {\bibfnamefont {P.}~\bibnamefont {Yu}}, \bibinfo {author} {\bibfnamefont
  {P.}~\bibnamefont {Tang}}, \bibinfo {author} {\bibfnamefont {W.}~\bibnamefont
  {Duan}},\ and\ \bibinfo {author} {\bibfnamefont {S.}~\bibnamefont {Zhou}},\
  }\href@noop {} {\bibfield  {journal} {\bibinfo  {journal} {Phys. Rev. B.}\
  }\textbf {\bibinfo {volume} {111}} (\bibinfo {year} {2025})}\BibitemShut
  {NoStop}%
\bibitem [{\citenamefont {Sobota}\ \emph {et~al.}(2021)\citenamefont {Sobota},
  \citenamefont {He},\ and\ \citenamefont {Shen}}]{Sobota2021}%
  \BibitemOpen
  \bibfield  {author} {\bibinfo {author} {\bibfnamefont {J.~A.}\ \bibnamefont
  {Sobota}}, \bibinfo {author} {\bibfnamefont {Y.}~\bibnamefont {He}},\ and\
  \bibinfo {author} {\bibfnamefont {Z.-X.}\ \bibnamefont {Shen}},\ }\bibfield
  {journal} {\bibinfo  {journal} {Reviews of Modern Physics}\ }\textbf
  {\bibinfo {volume} {93}},\ \href
  {https://doi.org/10.1103/revmodphys.93.025006} {10.1103/revmodphys.93.025006}
  (\bibinfo {year} {2021})\BibitemShut {NoStop}%
\bibitem [{\citenamefont {Boschini}\ \emph {et~al.}(2024)\citenamefont
  {Boschini}, \citenamefont {Zonno},\ and\ \citenamefont
  {Damascelli}}]{Boschini2024}%
  \BibitemOpen
  \bibfield  {author} {\bibinfo {author} {\bibfnamefont {F.}~\bibnamefont
  {Boschini}}, \bibinfo {author} {\bibfnamefont {M.}~\bibnamefont {Zonno}},\
  and\ \bibinfo {author} {\bibfnamefont {A.}~\bibnamefont {Damascelli}},\
  }\bibfield  {journal} {\bibinfo  {journal} {Reviews of Modern Physics}\
  }\textbf {\bibinfo {volume} {96}},\ \href
  {https://doi.org/10.1103/revmodphys.96.015003} {10.1103/revmodphys.96.015003}
  (\bibinfo {year} {2024})\BibitemShut {NoStop}%
\bibitem [{\citenamefont {Wang}\ \emph
  {et~al.}(2013{\natexlab{a}})\citenamefont {Wang}, \citenamefont {Steinberg},
  \citenamefont {Jarillo-Herrero},\ and\ \citenamefont {Gedik}}]{Wang2013}%
  \BibitemOpen
  \bibfield  {author} {\bibinfo {author} {\bibfnamefont {Y.~H.}\ \bibnamefont
  {Wang}}, \bibinfo {author} {\bibfnamefont {H.}~\bibnamefont {Steinberg}},
  \bibinfo {author} {\bibfnamefont {P.}~\bibnamefont {Jarillo-Herrero}},\ and\
  \bibinfo {author} {\bibfnamefont {N.}~\bibnamefont {Gedik}},\ }\href
  {https://doi.org/10.1126/science.1239834} {\bibfield  {journal} {\bibinfo
  {journal} {Science}\ }\textbf {\bibinfo {volume} {342}},\ \bibinfo {pages}
  {453–457} (\bibinfo {year} {2013}{\natexlab{a}})}\BibitemShut {NoStop}%
\bibitem [{\citenamefont {Mahmood}\ \emph {et~al.}(2016)\citenamefont
  {Mahmood}, \citenamefont {Chan}, \citenamefont {Alpichshev}, \citenamefont
  {Gardner}, \citenamefont {Lee}, \citenamefont {Lee},\ and\ \citenamefont
  {Gedik}}]{Mahmood2016}%
  \BibitemOpen
  \bibfield  {author} {\bibinfo {author} {\bibfnamefont {F.}~\bibnamefont
  {Mahmood}}, \bibinfo {author} {\bibfnamefont {C.-K.}\ \bibnamefont {Chan}},
  \bibinfo {author} {\bibfnamefont {Z.}~\bibnamefont {Alpichshev}}, \bibinfo
  {author} {\bibfnamefont {D.}~\bibnamefont {Gardner}}, \bibinfo {author}
  {\bibfnamefont {Y.}~\bibnamefont {Lee}}, \bibinfo {author} {\bibfnamefont
  {P.~A.}\ \bibnamefont {Lee}},\ and\ \bibinfo {author} {\bibfnamefont
  {N.}~\bibnamefont {Gedik}},\ }\href {https://doi.org/10.1038/nphys3609}
  {\bibfield  {journal} {\bibinfo  {journal} {Nature Physics}\ }\textbf
  {\bibinfo {volume} {12}},\ \bibinfo {pages} {306–310} (\bibinfo {year}
  {2016})}\BibitemShut {NoStop}%
\bibitem [{\citenamefont {De~Giovannini}\ \emph {et~al.}(2016)\citenamefont
  {De~Giovannini}, \citenamefont {H\"{u}bener},\ and\ \citenamefont
  {Rubio}}]{DeGiovannini2016}%
  \BibitemOpen
  \bibfield  {author} {\bibinfo {author} {\bibfnamefont {U.}~\bibnamefont
  {De~Giovannini}}, \bibinfo {author} {\bibfnamefont {H.}~\bibnamefont
  {H\"{u}bener}},\ and\ \bibinfo {author} {\bibfnamefont {A.}~\bibnamefont
  {Rubio}},\ }\href {https://doi.org/10.1021/acs.nanolett.6b04419} {\bibfield
  {journal} {\bibinfo  {journal} {Nano Letters}\ }\textbf {\bibinfo {volume}
  {16}},\ \bibinfo {pages} {7993–7998} (\bibinfo {year} {2016})}\BibitemShut
  {NoStop}%
\bibitem [{\citenamefont {Wang}\ \emph
  {et~al.}(2013{\natexlab{b}})\citenamefont {Wang}, \citenamefont {Steinberg},
  \citenamefont {Jarillo-Herrero},\ and\ \citenamefont {Gedik}}]{Gedik2013}%
  \BibitemOpen
  \bibfield  {author} {\bibinfo {author} {\bibfnamefont {Y.~H.}\ \bibnamefont
  {Wang}}, \bibinfo {author} {\bibfnamefont {H.}~\bibnamefont {Steinberg}},
  \bibinfo {author} {\bibfnamefont {P.}~\bibnamefont {Jarillo-Herrero}},\ and\
  \bibinfo {author} {\bibfnamefont {N.}~\bibnamefont {Gedik}},\ }\href
  {https://doi.org/10.1126/science.1239834} {\bibfield  {journal} {\bibinfo
  {journal} {Science}\ }\textbf {\bibinfo {volume} {342}},\ \bibinfo {pages}
  {453} (\bibinfo {year} {2013}{\natexlab{b}})},\ \Eprint
  {https://arxiv.org/abs/https://www.science.org/doi/pdf/10.1126/science.1239834}
  {https://www.science.org/doi/pdf/10.1126/science.1239834} \BibitemShut
  {NoStop}%
\bibitem [{\citenamefont {Aeschlimann}\ \emph {et~al.}(2021)\citenamefont
  {Aeschlimann}, \citenamefont {Sato}, \citenamefont {Krause}, \citenamefont
  {Ch\'avez-Cervantes}, \citenamefont {De~Giovannini}, \citenamefont
  {H\"ubener}, \citenamefont {Forti}, \citenamefont {Coletti}, \citenamefont
  {Hanff}, \citenamefont {Rossnagel}, \citenamefont {Rubio},\ and\
  \citenamefont {Gierz}}]{Gierz2021}%
  \BibitemOpen
  \bibfield  {author} {\bibinfo {author} {\bibfnamefont {S.}~\bibnamefont
  {Aeschlimann}}, \bibinfo {author} {\bibfnamefont {S.~A.}\ \bibnamefont
  {Sato}}, \bibinfo {author} {\bibfnamefont {R.}~\bibnamefont {Krause}},
  \bibinfo {author} {\bibfnamefont {M.}~\bibnamefont {Ch\'avez-Cervantes}},
  \bibinfo {author} {\bibfnamefont {U.}~\bibnamefont {De~Giovannini}}, \bibinfo
  {author} {\bibfnamefont {H.}~\bibnamefont {H\"ubener}}, \bibinfo {author}
  {\bibfnamefont {S.}~\bibnamefont {Forti}}, \bibinfo {author} {\bibfnamefont
  {C.}~\bibnamefont {Coletti}}, \bibinfo {author} {\bibfnamefont
  {K.}~\bibnamefont {Hanff}}, \bibinfo {author} {\bibfnamefont
  {K.}~\bibnamefont {Rossnagel}}, \bibinfo {author} {\bibfnamefont
  {A.}~\bibnamefont {Rubio}},\ and\ \bibinfo {author} {\bibfnamefont
  {I.}~\bibnamefont {Gierz}},\ }\href
  {https://doi.org/10.1021/acs.nanolett.1c00801} {\bibfield  {journal}
  {\bibinfo  {journal} {Nano Letters}\ }\textbf {\bibinfo {volume} {21}},\
  \bibinfo {pages} {5028–5035} (\bibinfo {year} {2021})}\BibitemShut
  {NoStop}%
\bibitem [{\citenamefont {Merboldt}\ \emph {et~al.}(2025)\citenamefont
  {Merboldt}, \citenamefont {Sch\"{u}ler}, \citenamefont {Schmitt},
  \citenamefont {Bange}, \citenamefont {Bennecke}, \citenamefont {Gadge},
  \citenamefont {Pierz}, \citenamefont {Schumacher}, \citenamefont {Momeni},
  \citenamefont {Steil}, \citenamefont {Manmana}, \citenamefont {Sentef},
  \citenamefont {Reutzel},\ and\ \citenamefont {Mathias}}]{Merboldt2025}%
  \BibitemOpen
  \bibfield  {author} {\bibinfo {author} {\bibfnamefont {M.}~\bibnamefont
  {Merboldt}}, \bibinfo {author} {\bibfnamefont {M.}~\bibnamefont
  {Sch\"{u}ler}}, \bibinfo {author} {\bibfnamefont {D.}~\bibnamefont
  {Schmitt}}, \bibinfo {author} {\bibfnamefont {J.~P.}\ \bibnamefont {Bange}},
  \bibinfo {author} {\bibfnamefont {W.}~\bibnamefont {Bennecke}}, \bibinfo
  {author} {\bibfnamefont {K.}~\bibnamefont {Gadge}}, \bibinfo {author}
  {\bibfnamefont {K.}~\bibnamefont {Pierz}}, \bibinfo {author} {\bibfnamefont
  {H.~W.}\ \bibnamefont {Schumacher}}, \bibinfo {author} {\bibfnamefont
  {D.}~\bibnamefont {Momeni}}, \bibinfo {author} {\bibfnamefont
  {D.}~\bibnamefont {Steil}}, \bibinfo {author} {\bibfnamefont {S.~R.}\
  \bibnamefont {Manmana}}, \bibinfo {author} {\bibfnamefont {M.~A.}\
  \bibnamefont {Sentef}}, \bibinfo {author} {\bibfnamefont {M.}~\bibnamefont
  {Reutzel}},\ and\ \bibinfo {author} {\bibfnamefont {S.}~\bibnamefont
  {Mathias}},\ }\bibfield  {journal} {\bibinfo  {journal} {Nature Physics}\
  }\href {https://doi.org/10.1038/s41567-025-02889-7}
  {10.1038/s41567-025-02889-7} (\bibinfo {year} {2025})\BibitemShut {NoStop}%
\bibitem [{\citenamefont {Choi}\ \emph {et~al.}(2025)\citenamefont {Choi},
  \citenamefont {Mogi}, \citenamefont {De~Giovannini}, \citenamefont {Azoury},
  \citenamefont {Lv}, \citenamefont {Su}, \citenamefont {H\"{u}bener},
  \citenamefont {Rubio},\ and\ \citenamefont {Gedik}}]{Choi2025}%
  \BibitemOpen
  \bibfield  {author} {\bibinfo {author} {\bibfnamefont {D.}~\bibnamefont
  {Choi}}, \bibinfo {author} {\bibfnamefont {M.}~\bibnamefont {Mogi}}, \bibinfo
  {author} {\bibfnamefont {U.}~\bibnamefont {De~Giovannini}}, \bibinfo {author}
  {\bibfnamefont {D.}~\bibnamefont {Azoury}}, \bibinfo {author} {\bibfnamefont
  {B.}~\bibnamefont {Lv}}, \bibinfo {author} {\bibfnamefont {Y.}~\bibnamefont
  {Su}}, \bibinfo {author} {\bibfnamefont {H.}~\bibnamefont {H\"{u}bener}},
  \bibinfo {author} {\bibfnamefont {A.}~\bibnamefont {Rubio}},\ and\ \bibinfo
  {author} {\bibfnamefont {N.}~\bibnamefont {Gedik}},\ }\bibfield  {journal}
  {\bibinfo  {journal} {Nature Physics}\ }\href
  {https://doi.org/10.1038/s41567-025-02888-8} {10.1038/s41567-025-02888-8}
  (\bibinfo {year} {2025})\BibitemShut {NoStop}%
\bibitem [{\citenamefont {Bielinski}\ \emph {et~al.}(2025)\citenamefont
  {Bielinski}, \citenamefont {Chari}, \citenamefont {May-Mann}, \citenamefont
  {Kim}, \citenamefont {Zwettler}, \citenamefont {Deng}, \citenamefont
  {Aishwarya}, \citenamefont {Roychowdhury}, \citenamefont {Shekhar},
  \citenamefont {Hashimoto}, \citenamefont {Lu}, \citenamefont {Yan},
  \citenamefont {Felser}, \citenamefont {Madhavan}, \citenamefont {Shen},
  \citenamefont {Hughes},\ and\ \citenamefont {Mahmood}}]{Bielinski2025}%
  \BibitemOpen
  \bibfield  {author} {\bibinfo {author} {\bibfnamefont {N.}~\bibnamefont
  {Bielinski}}, \bibinfo {author} {\bibfnamefont {R.}~\bibnamefont {Chari}},
  \bibinfo {author} {\bibfnamefont {J.}~\bibnamefont {May-Mann}}, \bibinfo
  {author} {\bibfnamefont {S.}~\bibnamefont {Kim}}, \bibinfo {author}
  {\bibfnamefont {J.}~\bibnamefont {Zwettler}}, \bibinfo {author}
  {\bibfnamefont {Y.}~\bibnamefont {Deng}}, \bibinfo {author} {\bibfnamefont
  {A.}~\bibnamefont {Aishwarya}}, \bibinfo {author} {\bibfnamefont
  {S.}~\bibnamefont {Roychowdhury}}, \bibinfo {author} {\bibfnamefont
  {C.}~\bibnamefont {Shekhar}}, \bibinfo {author} {\bibfnamefont
  {M.}~\bibnamefont {Hashimoto}}, \bibinfo {author} {\bibfnamefont
  {D.}~\bibnamefont {Lu}}, \bibinfo {author} {\bibfnamefont {J.}~\bibnamefont
  {Yan}}, \bibinfo {author} {\bibfnamefont {C.}~\bibnamefont {Felser}},
  \bibinfo {author} {\bibfnamefont {V.}~\bibnamefont {Madhavan}}, \bibinfo
  {author} {\bibfnamefont {Z.-X.}\ \bibnamefont {Shen}}, \bibinfo {author}
  {\bibfnamefont {T.~L.}\ \bibnamefont {Hughes}},\ and\ \bibinfo {author}
  {\bibfnamefont {F.}~\bibnamefont {Mahmood}},\ }\bibfield  {journal} {\bibinfo
   {journal} {Nature Physics}\ }\href
  {https://doi.org/10.1038/s41567-024-02769-6} {10.1038/s41567-024-02769-6}
  (\bibinfo {year} {2025})\BibitemShut {NoStop}%
\bibitem [{\citenamefont {Fragkos}\ \emph {et~al.}(2025)\citenamefont
  {Fragkos}, \citenamefont {Fabre}, \citenamefont {Tkach}, \citenamefont
  {Petit}, \citenamefont {Descamps}, \citenamefont {Sch\"{o}nhense},
  \citenamefont {Mairesse}, \citenamefont {Sch\"{u}ler},\ and\ \citenamefont
  {Beaulieu}}]{Fragkos2025}%
  \BibitemOpen
  \bibfield  {author} {\bibinfo {author} {\bibfnamefont {S.}~\bibnamefont
  {Fragkos}}, \bibinfo {author} {\bibfnamefont {B.}~\bibnamefont {Fabre}},
  \bibinfo {author} {\bibfnamefont {O.}~\bibnamefont {Tkach}}, \bibinfo
  {author} {\bibfnamefont {S.}~\bibnamefont {Petit}}, \bibinfo {author}
  {\bibfnamefont {D.}~\bibnamefont {Descamps}}, \bibinfo {author}
  {\bibfnamefont {G.}~\bibnamefont {Sch\"{o}nhense}}, \bibinfo {author}
  {\bibfnamefont {Y.}~\bibnamefont {Mairesse}}, \bibinfo {author}
  {\bibfnamefont {M.}~\bibnamefont {Sch\"{u}ler}},\ and\ \bibinfo {author}
  {\bibfnamefont {S.}~\bibnamefont {Beaulieu}},\ }\bibfield  {journal}
  {\bibinfo  {journal} {Nature Communications}\ }\textbf {\bibinfo {volume}
  {16}},\ \href {https://doi.org/10.1038/s41467-025-61076-7}
  {10.1038/s41467-025-61076-7} (\bibinfo {year} {2025})\BibitemShut {NoStop}%
\bibitem [{\citenamefont {Glover}\ \emph {et~al.}(1996)\citenamefont {Glover},
  \citenamefont {Schoenlein}, \citenamefont {Chin},\ and\ \citenamefont
  {Shank}}]{Glover1996}%
  \BibitemOpen
  \bibfield  {author} {\bibinfo {author} {\bibfnamefont {T.~E.}\ \bibnamefont
  {Glover}}, \bibinfo {author} {\bibfnamefont {R.~W.}\ \bibnamefont
  {Schoenlein}}, \bibinfo {author} {\bibfnamefont {A.~H.}\ \bibnamefont
  {Chin}},\ and\ \bibinfo {author} {\bibfnamefont {C.~V.}\ \bibnamefont
  {Shank}},\ }\href {https://doi.org/10.1103/physrevlett.76.2468} {\bibfield
  {journal} {\bibinfo  {journal} {Physical Review Letters}\ }\textbf {\bibinfo
  {volume} {76}},\ \bibinfo {pages} {2468–2471} (\bibinfo {year}
  {1996})}\BibitemShut {NoStop}%
\bibitem [{\citenamefont {Schins}\ \emph {et~al.}(1996)\citenamefont {Schins},
  \citenamefont {Breger}, \citenamefont {Agostini}, \citenamefont
  {Constantinescu}, \citenamefont {Muller}, \citenamefont {Bouhal},
  \citenamefont {Grillon}, \citenamefont {Antonetti},\ and\ \citenamefont
  {Mysyrowicz}}]{Schins1996}%
  \BibitemOpen
  \bibfield  {author} {\bibinfo {author} {\bibfnamefont {J.~M.}\ \bibnamefont
  {Schins}}, \bibinfo {author} {\bibfnamefont {P.}~\bibnamefont {Breger}},
  \bibinfo {author} {\bibfnamefont {P.}~\bibnamefont {Agostini}}, \bibinfo
  {author} {\bibfnamefont {R.~C.}\ \bibnamefont {Constantinescu}}, \bibinfo
  {author} {\bibfnamefont {H.~G.}\ \bibnamefont {Muller}}, \bibinfo {author}
  {\bibfnamefont {A.}~\bibnamefont {Bouhal}}, \bibinfo {author} {\bibfnamefont
  {G.}~\bibnamefont {Grillon}}, \bibinfo {author} {\bibfnamefont
  {A.}~\bibnamefont {Antonetti}},\ and\ \bibinfo {author} {\bibfnamefont
  {A.}~\bibnamefont {Mysyrowicz}},\ }\href
  {https://doi.org/10.1364/josab.13.000197} {\bibfield  {journal} {\bibinfo
  {journal} {Journal of the Optical Society of America B}\ }\textbf {\bibinfo
  {volume} {13}},\ \bibinfo {pages} {197} (\bibinfo {year} {1996})}\BibitemShut
  {NoStop}%
\bibitem [{\citenamefont {Saathoff}\ \emph {et~al.}(2008)\citenamefont
  {Saathoff}, \citenamefont {Miaja-Avila}, \citenamefont {Aeschlimann},
  \citenamefont {Murnane},\ and\ \citenamefont {Kapteyn}}]{Saathoff2008}%
  \BibitemOpen
  \bibfield  {author} {\bibinfo {author} {\bibfnamefont {G.}~\bibnamefont
  {Saathoff}}, \bibinfo {author} {\bibfnamefont {L.}~\bibnamefont
  {Miaja-Avila}}, \bibinfo {author} {\bibfnamefont {M.}~\bibnamefont
  {Aeschlimann}}, \bibinfo {author} {\bibfnamefont {M.~M.}\ \bibnamefont
  {Murnane}},\ and\ \bibinfo {author} {\bibfnamefont {H.~C.}\ \bibnamefont
  {Kapteyn}},\ }\bibfield  {journal} {\bibinfo  {journal} {Physical Review A}\
  }\textbf {\bibinfo {volume} {77}},\ \href
  {https://doi.org/10.1103/physreva.77.022903} {10.1103/physreva.77.022903}
  (\bibinfo {year} {2008})\BibitemShut {NoStop}%
\bibitem [{\citenamefont {Keunecke}\ \emph {et~al.}(2020)\citenamefont
  {Keunecke}, \citenamefont {Reutzel}, \citenamefont {Schmitt}, \citenamefont
  {Osterkorn}, \citenamefont {Mishra}, \citenamefont {M\"{o}ller},
  \citenamefont {Bennecke}, \citenamefont {Jansen}, \citenamefont {Steil},
  \citenamefont {Manmana}, \citenamefont {Steil}, \citenamefont {Kehrein},\
  and\ \citenamefont {Mathias}}]{Keunecke2020}%
  \BibitemOpen
  \bibfield  {author} {\bibinfo {author} {\bibfnamefont {M.}~\bibnamefont
  {Keunecke}}, \bibinfo {author} {\bibfnamefont {M.}~\bibnamefont {Reutzel}},
  \bibinfo {author} {\bibfnamefont {D.}~\bibnamefont {Schmitt}}, \bibinfo
  {author} {\bibfnamefont {A.}~\bibnamefont {Osterkorn}}, \bibinfo {author}
  {\bibfnamefont {T.~A.}\ \bibnamefont {Mishra}}, \bibinfo {author}
  {\bibfnamefont {C.}~\bibnamefont {M\"{o}ller}}, \bibinfo {author}
  {\bibfnamefont {W.}~\bibnamefont {Bennecke}}, \bibinfo {author}
  {\bibfnamefont {G.~S.~M.}\ \bibnamefont {Jansen}}, \bibinfo {author}
  {\bibfnamefont {D.}~\bibnamefont {Steil}}, \bibinfo {author} {\bibfnamefont
  {S.~R.}\ \bibnamefont {Manmana}}, \bibinfo {author} {\bibfnamefont
  {S.}~\bibnamefont {Steil}}, \bibinfo {author} {\bibfnamefont
  {S.}~\bibnamefont {Kehrein}},\ and\ \bibinfo {author} {\bibfnamefont
  {S.}~\bibnamefont {Mathias}},\ }\bibfield  {journal} {\bibinfo  {journal}
  {Physical Review B}\ }\textbf {\bibinfo {volume} {102}},\ \href
  {https://doi.org/10.1103/physrevb.102.161403} {10.1103/physrevb.102.161403}
  (\bibinfo {year} {2020})\BibitemShut {NoStop}%
\bibitem [{\citenamefont {Park}(2014)}]{Park2014}%
  \BibitemOpen
  \bibfield  {author} {\bibinfo {author} {\bibfnamefont {S.~T.}\ \bibnamefont
  {Park}},\ }\bibfield  {journal} {\bibinfo  {journal} {Physical Review A}\
  }\textbf {\bibinfo {volume} {90}},\ \href
  {https://doi.org/10.1103/physreva.90.013420} {10.1103/physreva.90.013420}
  (\bibinfo {year} {2014})\BibitemShut {NoStop}%
\bibitem [{\citenamefont {Yen}\ \emph {et~al.}(2025)\citenamefont {Yen},
  \citenamefont {Reutzel}, \citenamefont {Li}, \citenamefont {Wang},
  \citenamefont {Petek},\ and\ \citenamefont {Schüler}}]{Yen2025}%
  \BibitemOpen
  \bibfield  {author} {\bibinfo {author} {\bibfnamefont {Y.}~\bibnamefont
  {Yen}}, \bibinfo {author} {\bibfnamefont {M.}~\bibnamefont {Reutzel}},
  \bibinfo {author} {\bibfnamefont {A.}~\bibnamefont {Li}}, \bibinfo {author}
  {\bibfnamefont {Z.}~\bibnamefont {Wang}}, \bibinfo {author} {\bibfnamefont
  {H.}~\bibnamefont {Petek}},\ and\ \bibinfo {author} {\bibfnamefont
  {M.}~\bibnamefont {Schüler}},\ }\href {https://arxiv.org/abs/2503.04431}
  {\bibinfo {title} {Observation of non-adiabatic landau-zener tunneling among
  floquet states}} (\bibinfo {year} {2025}),\ \Eprint
  {https://arxiv.org/abs/2503.04431} {arXiv:2503.04431 [cond-mat.mes-hall]}
  \BibitemShut {NoStop}%
\bibitem [{\citenamefont {Freericks}\ \emph {et~al.}(2015)\citenamefont
  {Freericks}, \citenamefont {Krishnamurthy}, \citenamefont {Sentef},\ and\
  \citenamefont {Devereaux}}]{Freericks2015}%
  \BibitemOpen
  \bibfield  {author} {\bibinfo {author} {\bibfnamefont {J.~K.}\ \bibnamefont
  {Freericks}}, \bibinfo {author} {\bibfnamefont {H.~R.}\ \bibnamefont
  {Krishnamurthy}}, \bibinfo {author} {\bibfnamefont {M.~A.}\ \bibnamefont
  {Sentef}},\ and\ \bibinfo {author} {\bibfnamefont {T.~P.}\ \bibnamefont
  {Devereaux}},\ }\href {https://doi.org/10.1088/0031-8949/2015/t165/014012}
  {\bibfield  {journal} {\bibinfo  {journal} {Physica Scripta}\ }\textbf
  {\bibinfo {volume} {T165}},\ \bibinfo {pages} {014012} (\bibinfo {year}
  {2015})}\BibitemShut {NoStop}%
\bibitem [{\citenamefont {Sentef}\ \emph {et~al.}(2015)\citenamefont {Sentef},
  \citenamefont {Claassen}, \citenamefont {Kemper}, \citenamefont {Moritz},
  \citenamefont {Oka}, \citenamefont {Freericks},\ and\ \citenamefont
  {Devereaux}}]{Sentef2015}%
  \BibitemOpen
  \bibfield  {author} {\bibinfo {author} {\bibfnamefont {M.}~\bibnamefont
  {Sentef}}, \bibinfo {author} {\bibfnamefont {M.}~\bibnamefont {Claassen}},
  \bibinfo {author} {\bibfnamefont {A.}~\bibnamefont {Kemper}}, \bibinfo
  {author} {\bibfnamefont {B.}~\bibnamefont {Moritz}}, \bibinfo {author}
  {\bibfnamefont {T.}~\bibnamefont {Oka}}, \bibinfo {author} {\bibfnamefont
  {J.}~\bibnamefont {Freericks}},\ and\ \bibinfo {author} {\bibfnamefont
  {T.}~\bibnamefont {Devereaux}},\ }\bibfield  {journal} {\bibinfo  {journal}
  {Nature Communications}\ }\textbf {\bibinfo {volume} {6}},\ \href
  {https://doi.org/10.1038/ncomms8047} {10.1038/ncomms8047} (\bibinfo {year}
  {2015})\BibitemShut {NoStop}%
\bibitem [{\citenamefont {Ito}\ \emph {et~al.}(2023)\citenamefont {Ito},
  \citenamefont {Sch\"{u}ler}, \citenamefont {Meierhofer}, \citenamefont
  {Schlauderer}, \citenamefont {Freudenstein}, \citenamefont {Reimann},
  \citenamefont {Afanasiev}, \citenamefont {Kokh}, \citenamefont
  {Tereshchenko}, \citenamefont {G\"{u}dde}, \citenamefont {Sentef},
  \citenamefont {H\"{o}fer},\ and\ \citenamefont {Huber}}]{Ito2023}%
  \BibitemOpen
  \bibfield  {author} {\bibinfo {author} {\bibfnamefont {S.}~\bibnamefont
  {Ito}}, \bibinfo {author} {\bibfnamefont {M.}~\bibnamefont {Sch\"{u}ler}},
  \bibinfo {author} {\bibfnamefont {M.}~\bibnamefont {Meierhofer}}, \bibinfo
  {author} {\bibfnamefont {S.}~\bibnamefont {Schlauderer}}, \bibinfo {author}
  {\bibfnamefont {J.}~\bibnamefont {Freudenstein}}, \bibinfo {author}
  {\bibfnamefont {J.}~\bibnamefont {Reimann}}, \bibinfo {author} {\bibfnamefont
  {D.}~\bibnamefont {Afanasiev}}, \bibinfo {author} {\bibfnamefont {K.~A.}\
  \bibnamefont {Kokh}}, \bibinfo {author} {\bibfnamefont {O.~E.}\ \bibnamefont
  {Tereshchenko}}, \bibinfo {author} {\bibfnamefont {J.}~\bibnamefont
  {G\"{u}dde}}, \bibinfo {author} {\bibfnamefont {M.~A.}\ \bibnamefont
  {Sentef}}, \bibinfo {author} {\bibfnamefont {U.}~\bibnamefont {H\"{o}fer}},\
  and\ \bibinfo {author} {\bibfnamefont {R.}~\bibnamefont {Huber}},\ }\href
  {https://doi.org/10.1038/s41586-023-05850-x} {\bibfield  {journal} {\bibinfo
  {journal} {Nature}\ }\textbf {\bibinfo {volume} {616}},\ \bibinfo {pages}
  {696–701} (\bibinfo {year} {2023})}\BibitemShut {NoStop}%
\bibitem [{\citenamefont {Sch\"{u}ler}\ and\ \citenamefont
  {Sentef}(2021)}]{Schler2021}%
  \BibitemOpen
  \bibfield  {author} {\bibinfo {author} {\bibfnamefont {M.}~\bibnamefont
  {Sch\"{u}ler}}\ and\ \bibinfo {author} {\bibfnamefont {M.~A.}\ \bibnamefont
  {Sentef}},\ }\href {https://doi.org/10.1016/j.elspec.2021.147121} {\bibfield
  {journal} {\bibinfo  {journal} {Journal of Electron Spectroscopy and Related
  Phenomena}\ }\textbf {\bibinfo {volume} {253}},\ \bibinfo {pages} {147121}
  (\bibinfo {year} {2021})}\BibitemShut {NoStop}%
\bibitem [{\citenamefont {Bostwick}\ \emph {et~al.}(2006)\citenamefont
  {Bostwick}, \citenamefont {Ohta}, \citenamefont {Seyller}, \citenamefont
  {Horn},\ and\ \citenamefont {Rotenberg}}]{Bostwick2006}%
  \BibitemOpen
  \bibfield  {author} {\bibinfo {author} {\bibfnamefont {A.}~\bibnamefont
  {Bostwick}}, \bibinfo {author} {\bibfnamefont {T.}~\bibnamefont {Ohta}},
  \bibinfo {author} {\bibfnamefont {T.}~\bibnamefont {Seyller}}, \bibinfo
  {author} {\bibfnamefont {K.}~\bibnamefont {Horn}},\ and\ \bibinfo {author}
  {\bibfnamefont {E.}~\bibnamefont {Rotenberg}},\ }\href
  {https://doi.org/10.1038/nphys477} {\bibfield  {journal} {\bibinfo  {journal}
  {Nature Physics}\ }\textbf {\bibinfo {volume} {3}},\ \bibinfo {pages}
  {36–40} (\bibinfo {year} {2006})}\BibitemShut {NoStop}%
\bibitem [{\citenamefont {Castro~Neto}\ \emph {et~al.}(2009)\citenamefont
  {Castro~Neto}, \citenamefont {Guinea}, \citenamefont {Peres}, \citenamefont
  {Novoselov},\ and\ \citenamefont {Geim}}]{CastroNeto2009}%
  \BibitemOpen
  \bibfield  {author} {\bibinfo {author} {\bibfnamefont {A.~H.}\ \bibnamefont
  {Castro~Neto}}, \bibinfo {author} {\bibfnamefont {F.}~\bibnamefont {Guinea}},
  \bibinfo {author} {\bibfnamefont {N.~M.~R.}\ \bibnamefont {Peres}}, \bibinfo
  {author} {\bibfnamefont {K.~S.}\ \bibnamefont {Novoselov}},\ and\ \bibinfo
  {author} {\bibfnamefont {A.~K.}\ \bibnamefont {Geim}},\ }\href
  {https://doi.org/10.1103/revmodphys.81.109} {\bibfield  {journal} {\bibinfo
  {journal} {Reviews of Modern Physics}\ }\textbf {\bibinfo {volume} {81}},\
  \bibinfo {pages} {109–162} (\bibinfo {year} {2009})}\BibitemShut {NoStop}%
\bibitem [{\citenamefont {Shirley}(1965)}]{Shirley1965}%
  \BibitemOpen
  \bibfield  {author} {\bibinfo {author} {\bibfnamefont {J.~H.}\ \bibnamefont
  {Shirley}},\ }\href {https://doi.org/10.1103/physrev.138.b979} {\bibfield
  {journal} {\bibinfo  {journal} {Physical Review}\ }\textbf {\bibinfo {volume}
  {138}},\ \bibinfo {pages} {B979–B987} (\bibinfo {year} {1965})}\BibitemShut
  {NoStop}%
\bibitem [{\citenamefont {Reiss}(1980)}]{Reiss1980}%
  \BibitemOpen
  \bibfield  {author} {\bibinfo {author} {\bibfnamefont {H.~R.}\ \bibnamefont
  {Reiss}},\ }\href {https://doi.org/10.1103/PhysRevA.22.1786} {\bibfield
  {journal} {\bibinfo  {journal} {Phys. Rev. A}\ }\textbf {\bibinfo {volume}
  {22}},\ \bibinfo {pages} {1786} (\bibinfo {year} {1980})}\BibitemShut
  {NoStop}%
\bibitem [{\citenamefont {Miaja-Avila}\ \emph {et~al.}(2006)\citenamefont
  {Miaja-Avila}, \citenamefont {Lei}, \citenamefont {Aeschlimann},
  \citenamefont {Gland}, \citenamefont {Murnane}, \citenamefont {Kapteyn},\
  and\ \citenamefont {Saathoff}}]{MiajaAvila2006}%
  \BibitemOpen
  \bibfield  {author} {\bibinfo {author} {\bibfnamefont {L.}~\bibnamefont
  {Miaja-Avila}}, \bibinfo {author} {\bibfnamefont {C.-F.}\ \bibnamefont
  {Lei}}, \bibinfo {author} {\bibfnamefont {M.}~\bibnamefont {Aeschlimann}},
  \bibinfo {author} {\bibfnamefont {J.}~\bibnamefont {Gland}}, \bibinfo
  {author} {\bibfnamefont {M.~M.}\ \bibnamefont {Murnane}}, \bibinfo {author}
  {\bibfnamefont {H.~C.}\ \bibnamefont {Kapteyn}},\ and\ \bibinfo {author}
  {\bibfnamefont {G.}~\bibnamefont {Saathoff}},\ }\href
  {https://doi.org/10.1103/PhysRevLett.97.113604} {\bibfield  {journal}
  {\bibinfo  {journal} {Phys. Rev. Lett.}\ }\textbf {\bibinfo {volume} {97}},\
  \bibinfo {pages} {113604} (\bibinfo {year} {2006})}\BibitemShut {NoStop}%
\bibitem [{\citenamefont {Hwang}\ \emph {et~al.}(2011)\citenamefont {Hwang},
  \citenamefont {Park}, \citenamefont {Siegel}, \citenamefont {Fedorov},
  \citenamefont {Louie},\ and\ \citenamefont {Lanzara}}]{Hwang2011}%
  \BibitemOpen
  \bibfield  {author} {\bibinfo {author} {\bibfnamefont {C.}~\bibnamefont
  {Hwang}}, \bibinfo {author} {\bibfnamefont {C.-H.}\ \bibnamefont {Park}},
  \bibinfo {author} {\bibfnamefont {D.~A.}\ \bibnamefont {Siegel}}, \bibinfo
  {author} {\bibfnamefont {A.~V.}\ \bibnamefont {Fedorov}}, \bibinfo {author}
  {\bibfnamefont {S.~G.}\ \bibnamefont {Louie}},\ and\ \bibinfo {author}
  {\bibfnamefont {A.}~\bibnamefont {Lanzara}},\ }\bibfield  {journal} {\bibinfo
   {journal} {Physical Review B}\ }\textbf {\bibinfo {volume} {84}},\ \href
  {https://doi.org/10.1103/physrevb.84.125422} {10.1103/physrevb.84.125422}
  (\bibinfo {year} {2011})\BibitemShut {NoStop}%
\bibitem [{\citenamefont {Moser}(2017)}]{Moser2017}%
  \BibitemOpen
  \bibfield  {author} {\bibinfo {author} {\bibfnamefont {S.}~\bibnamefont
  {Moser}},\ }\href {https://doi.org/10.1016/j.elspec.2016.11.007} {\bibfield
  {journal} {\bibinfo  {journal} {J. Electron Spectrosc. Relat. Phenom.}\
  }\textbf {\bibinfo {volume} {214}},\ \bibinfo {pages} {29} (\bibinfo {year}
  {2017})}\BibitemShut {NoStop}%
\bibitem [{\citenamefont {Krasovskii}(2021)}]{Krasovskii2021}%
  \BibitemOpen
  \bibfield  {author} {\bibinfo {author} {\bibfnamefont {E.}~\bibnamefont
  {Krasovskii}},\ }\href {https://doi.org/10.3390/nano11051212} {\bibfield
  {journal} {\bibinfo  {journal} {Nanomaterials}\ }\textbf {\bibinfo {volume}
  {11}},\ \bibinfo {pages} {1212} (\bibinfo {year} {2021})}\BibitemShut
  {NoStop}%
\bibitem [{\citenamefont {Neppl}\ \emph {et~al.}(2015)\citenamefont {Neppl},
  \citenamefont {Ernstorfer}, \citenamefont {Cavalieri}, \citenamefont
  {Lemell}, \citenamefont {Wachter}, \citenamefont {Magerl}, \citenamefont
  {Bothschafter}, \citenamefont {Jobst}, \citenamefont {Hofstetter},
  \citenamefont {Kleineberg}, \citenamefont {Barth}, \citenamefont {Menzel},
  \citenamefont {Burgd\"{o}rfer}, \citenamefont {Feulner}, \citenamefont
  {Krausz},\ and\ \citenamefont {Kienberger}}]{Neppl2015}%
  \BibitemOpen
  \bibfield  {author} {\bibinfo {author} {\bibfnamefont {S.}~\bibnamefont
  {Neppl}}, \bibinfo {author} {\bibfnamefont {R.}~\bibnamefont {Ernstorfer}},
  \bibinfo {author} {\bibfnamefont {A.~L.}\ \bibnamefont {Cavalieri}}, \bibinfo
  {author} {\bibfnamefont {C.}~\bibnamefont {Lemell}}, \bibinfo {author}
  {\bibfnamefont {G.}~\bibnamefont {Wachter}}, \bibinfo {author} {\bibfnamefont
  {E.}~\bibnamefont {Magerl}}, \bibinfo {author} {\bibfnamefont {E.~M.}\
  \bibnamefont {Bothschafter}}, \bibinfo {author} {\bibfnamefont
  {M.}~\bibnamefont {Jobst}}, \bibinfo {author} {\bibfnamefont
  {M.}~\bibnamefont {Hofstetter}}, \bibinfo {author} {\bibfnamefont
  {U.}~\bibnamefont {Kleineberg}}, \bibinfo {author} {\bibfnamefont {J.~V.}\
  \bibnamefont {Barth}}, \bibinfo {author} {\bibfnamefont {D.}~\bibnamefont
  {Menzel}}, \bibinfo {author} {\bibfnamefont {J.}~\bibnamefont
  {Burgd\"{o}rfer}}, \bibinfo {author} {\bibfnamefont {P.}~\bibnamefont
  {Feulner}}, \bibinfo {author} {\bibfnamefont {F.}~\bibnamefont {Krausz}},\
  and\ \bibinfo {author} {\bibfnamefont {R.}~\bibnamefont {Kienberger}},\
  }\href {https://doi.org/10.1038/nature14094} {\bibfield  {journal} {\bibinfo
  {journal} {Nature}\ }\textbf {\bibinfo {volume} {517}},\ \bibinfo {pages}
  {342–346} (\bibinfo {year} {2015})}\BibitemShut {NoStop}%
\bibitem [{\citenamefont {Chen}\ \emph {et~al.}(2017)\citenamefont {Chen},
  \citenamefont {Tao}, \citenamefont {Carr}, \citenamefont {Matyba},
  \citenamefont {Szilvási}, \citenamefont {Emmerich}, \citenamefont {Piecuch},
  \citenamefont {Keller}, \citenamefont {Zusin}, \citenamefont {Eich},
  \citenamefont {Rollinger}, \citenamefont {You}, \citenamefont {Mathias},
  \citenamefont {Thumm}, \citenamefont {Mavrikakis}, \citenamefont
  {Aeschlimann}, \citenamefont {Oppeneer}, \citenamefont {Kapteyn},\ and\
  \citenamefont {Murnane}}]{Chen2017}%
  \BibitemOpen
  \bibfield  {author} {\bibinfo {author} {\bibfnamefont {C.}~\bibnamefont
  {Chen}}, \bibinfo {author} {\bibfnamefont {Z.}~\bibnamefont {Tao}}, \bibinfo
  {author} {\bibfnamefont {A.}~\bibnamefont {Carr}}, \bibinfo {author}
  {\bibfnamefont {P.}~\bibnamefont {Matyba}}, \bibinfo {author} {\bibfnamefont
  {T.}~\bibnamefont {Szilvási}}, \bibinfo {author} {\bibfnamefont
  {S.}~\bibnamefont {Emmerich}}, \bibinfo {author} {\bibfnamefont
  {M.}~\bibnamefont {Piecuch}}, \bibinfo {author} {\bibfnamefont
  {M.}~\bibnamefont {Keller}}, \bibinfo {author} {\bibfnamefont
  {D.}~\bibnamefont {Zusin}}, \bibinfo {author} {\bibfnamefont
  {S.}~\bibnamefont {Eich}}, \bibinfo {author} {\bibfnamefont {M.}~\bibnamefont
  {Rollinger}}, \bibinfo {author} {\bibfnamefont {W.}~\bibnamefont {You}},
  \bibinfo {author} {\bibfnamefont {S.}~\bibnamefont {Mathias}}, \bibinfo
  {author} {\bibfnamefont {U.}~\bibnamefont {Thumm}}, \bibinfo {author}
  {\bibfnamefont {M.}~\bibnamefont {Mavrikakis}}, \bibinfo {author}
  {\bibfnamefont {M.}~\bibnamefont {Aeschlimann}}, \bibinfo {author}
  {\bibfnamefont {P.~M.}\ \bibnamefont {Oppeneer}}, \bibinfo {author}
  {\bibfnamefont {H.}~\bibnamefont {Kapteyn}},\ and\ \bibinfo {author}
  {\bibfnamefont {M.}~\bibnamefont {Murnane}},\ }\bibfield  {journal} {\bibinfo
   {journal} {Proceedings of the National Academy of Sciences}\ }\textbf
  {\bibinfo {volume} {114}},\ \href {https://doi.org/10.1073/pnas.1706466114}
  {10.1073/pnas.1706466114} (\bibinfo {year} {2017})\BibitemShut {NoStop}%
\bibitem [{\citenamefont {Tao}\ \emph {et~al.}(2016)\citenamefont {Tao},
  \citenamefont {Chen}, \citenamefont {Szilvási}, \citenamefont {Keller},
  \citenamefont {Mavrikakis}, \citenamefont {Kapteyn},\ and\ \citenamefont
  {Murnane}}]{Tao2016}%
  \BibitemOpen
  \bibfield  {author} {\bibinfo {author} {\bibfnamefont {Z.}~\bibnamefont
  {Tao}}, \bibinfo {author} {\bibfnamefont {C.}~\bibnamefont {Chen}}, \bibinfo
  {author} {\bibfnamefont {T.}~\bibnamefont {Szilvási}}, \bibinfo {author}
  {\bibfnamefont {M.}~\bibnamefont {Keller}}, \bibinfo {author} {\bibfnamefont
  {M.}~\bibnamefont {Mavrikakis}}, \bibinfo {author} {\bibfnamefont
  {H.}~\bibnamefont {Kapteyn}},\ and\ \bibinfo {author} {\bibfnamefont
  {M.}~\bibnamefont {Murnane}},\ }\href
  {https://doi.org/10.1126/science.aaf6793} {\bibfield  {journal} {\bibinfo
  {journal} {Science}\ }\textbf {\bibinfo {volume} {353}},\ \bibinfo {pages}
  {62–67} (\bibinfo {year} {2016})}\BibitemShut {NoStop}%
\bibitem [{\citenamefont {Floquet}(1883)}]{Floquet1883}%
  \BibitemOpen
  \bibfield  {author} {\bibinfo {author} {\bibfnamefont {G.}~\bibnamefont
  {Floquet}},\ }\href {https://doi.org/10.24033/asens.220} {\bibfield
  {journal} {\bibinfo  {journal} {Annales scientifiques de l’École normale
  supérieure}\ }\textbf {\bibinfo {volume} {12}},\ \bibinfo {pages} {47–88}
  (\bibinfo {year} {1883})}\BibitemShut {NoStop}%
\bibitem [{\citenamefont {Gumhalter}(2025)}]{GUMHALTER2025100768}%
  \BibitemOpen
  \bibfield  {author} {\bibinfo {author} {\bibfnamefont {B.}~\bibnamefont
  {Gumhalter}},\ }\href
  {https://doi.org/https://doi.org/10.1016/j.progsurf.2025.100768} {\bibfield
  {journal} {\bibinfo  {journal} {Progress in Surface Science}\ }\textbf
  {\bibinfo {volume} {100}},\ \bibinfo {pages} {100768} (\bibinfo {year}
  {2025})}\BibitemShut {NoStop}%
\bibitem [{\citenamefont {Pazourek}\ \emph {et~al.}(2013)\citenamefont
  {Pazourek}, \citenamefont {Nagele},\ and\ \citenamefont
  {Burgdörfer}}]{Pazourek2013}%
  \BibitemOpen
  \bibfield  {author} {\bibinfo {author} {\bibfnamefont {R.}~\bibnamefont
  {Pazourek}}, \bibinfo {author} {\bibfnamefont {S.}~\bibnamefont {Nagele}},\
  and\ \bibinfo {author} {\bibfnamefont {J.}~\bibnamefont {Burgdörfer}},\
  }\href {https://doi.org/10.1039/C3FD00004D} {\bibfield  {journal} {\bibinfo
  {journal} {Faraday Discuss.}\ }\textbf {\bibinfo {volume} {163}},\ \bibinfo
  {pages} {353} (\bibinfo {year} {2013})}\BibitemShut {NoStop}%
\bibitem [{\citenamefont {Bergou}\ and\ \citenamefont
  {Varro}(1980)}]{Bergou1980}%
  \BibitemOpen
  \bibfield  {author} {\bibinfo {author} {\bibfnamefont {J.}~\bibnamefont
  {Bergou}}\ and\ \bibinfo {author} {\bibfnamefont {S.}~\bibnamefont {Varro}},\
  }\href {https://doi.org/10.1088/0305-4470/13/11/025} {\bibfield  {journal}
  {\bibinfo  {journal} {Journal of Physics A: Mathematical and General}\
  }\textbf {\bibinfo {volume} {13}},\ \bibinfo {pages} {3553–3559} (\bibinfo
  {year} {1980})}\BibitemShut {NoStop}%
\bibitem [{\citenamefont {Madsen}(2004)}]{Madsen2005bl}%
  \BibitemOpen
  \bibfield  {author} {\bibinfo {author} {\bibfnamefont {L.~B.}\ \bibnamefont
  {Madsen}},\ }\href {https://doi.org/10.1119/1.1796791} {\bibfield  {journal}
  {\bibinfo  {journal} {American Journal of Physics}\ }\textbf {\bibinfo
  {volume} {73}},\ \bibinfo {pages} {57–62} (\bibinfo {year}
  {2004})}\BibitemShut {NoStop}%
\bibitem [{\citenamefont {Sch\"{u}ler}\ \emph {et~al.}(2020)\citenamefont
  {Sch\"{u}ler}, \citenamefont {De~Giovannini}, \citenamefont {H\"{u}bener},
  \citenamefont {Rubio}, \citenamefont {Sentef}, \citenamefont {Devereaux},\
  and\ \citenamefont {Werner}}]{Schler2020}%
  \BibitemOpen
  \bibfield  {author} {\bibinfo {author} {\bibfnamefont {M.}~\bibnamefont
  {Sch\"{u}ler}}, \bibinfo {author} {\bibfnamefont {U.}~\bibnamefont
  {De~Giovannini}}, \bibinfo {author} {\bibfnamefont {H.}~\bibnamefont
  {H\"{u}bener}}, \bibinfo {author} {\bibfnamefont {A.}~\bibnamefont {Rubio}},
  \bibinfo {author} {\bibfnamefont {M.~A.}\ \bibnamefont {Sentef}}, \bibinfo
  {author} {\bibfnamefont {T.~P.}\ \bibnamefont {Devereaux}},\ and\ \bibinfo
  {author} {\bibfnamefont {P.}~\bibnamefont {Werner}},\ }\bibfield  {journal}
  {\bibinfo  {journal} {Physical Review X}\ }\textbf {\bibinfo {volume} {10}},\
  \href {https://doi.org/10.1103/physrevx.10.041013}
  {10.1103/physrevx.10.041013} (\bibinfo {year} {2020})\BibitemShut {NoStop}%
\bibitem [{\citenamefont {Freericks}\ \emph {et~al.}(2009)\citenamefont
  {Freericks}, \citenamefont {Krishnamurthy},\ and\ \citenamefont
  {Pruschke}}]{PruschkePRL2009}%
  \BibitemOpen
  \bibfield  {author} {\bibinfo {author} {\bibfnamefont {J.~K.}\ \bibnamefont
  {Freericks}}, \bibinfo {author} {\bibfnamefont {H.~R.}\ \bibnamefont
  {Krishnamurthy}},\ and\ \bibinfo {author} {\bibfnamefont {T.}~\bibnamefont
  {Pruschke}},\ }\href {https://doi.org/10.1103/PhysRevLett.102.136401}
  {\bibfield  {journal} {\bibinfo  {journal} {Phys. Rev. Lett.}\ }\textbf
  {\bibinfo {volume} {102}},\ \bibinfo {pages} {136401} (\bibinfo {year}
  {2009})}\BibitemShut {NoStop}%
\bibitem [{\citenamefont {Shirley}\ \emph {et~al.}(1995)\citenamefont
  {Shirley}, \citenamefont {Terminello}, \citenamefont {Santoni},\ and\
  \citenamefont {Himpsel}}]{Shirley1995}%
  \BibitemOpen
  \bibfield  {author} {\bibinfo {author} {\bibfnamefont {E.~L.}\ \bibnamefont
  {Shirley}}, \bibinfo {author} {\bibfnamefont {L.~J.}\ \bibnamefont
  {Terminello}}, \bibinfo {author} {\bibfnamefont {A.}~\bibnamefont
  {Santoni}},\ and\ \bibinfo {author} {\bibfnamefont {F.~J.}\ \bibnamefont
  {Himpsel}},\ }\href {https://doi.org/10.1103/physrevb.51.13614} {\bibfield
  {journal} {\bibinfo  {journal} {Physical Review B}\ }\textbf {\bibinfo
  {volume} {51}},\ \bibinfo {pages} {13614–13622} (\bibinfo {year}
  {1995})}\BibitemShut {NoStop}%
\bibitem [{\citenamefont {Gierz}\ \emph {et~al.}(2011)\citenamefont {Gierz},
  \citenamefont {Henk}, \citenamefont {H\"{o}chst}, \citenamefont {Ast},\ and\
  \citenamefont {Kern}}]{Gierz2011}%
  \BibitemOpen
  \bibfield  {author} {\bibinfo {author} {\bibfnamefont {I.}~\bibnamefont
  {Gierz}}, \bibinfo {author} {\bibfnamefont {J.}~\bibnamefont {Henk}},
  \bibinfo {author} {\bibfnamefont {H.}~\bibnamefont {H\"{o}chst}}, \bibinfo
  {author} {\bibfnamefont {C.~R.}\ \bibnamefont {Ast}},\ and\ \bibinfo {author}
  {\bibfnamefont {K.}~\bibnamefont {Kern}},\ }\bibfield  {journal} {\bibinfo
  {journal} {Physical Review B}\ }\textbf {\bibinfo {volume} {83}},\ \href
  {https://doi.org/10.1103/physrevb.83.121408} {10.1103/physrevb.83.121408}
  (\bibinfo {year} {2011})\BibitemShut {NoStop}%
\bibitem [{\citenamefont {Kuemmeth}\ and\ \citenamefont
  {Rashba}(2009)}]{Kuemmeth2009}%
  \BibitemOpen
  \bibfield  {author} {\bibinfo {author} {\bibfnamefont {F.}~\bibnamefont
  {Kuemmeth}}\ and\ \bibinfo {author} {\bibfnamefont {E.~I.}\ \bibnamefont
  {Rashba}},\ }\bibfield  {journal} {\bibinfo  {journal} {Physical Review B}\
  }\textbf {\bibinfo {volume} {80}},\ \href
  {https://doi.org/10.1103/physrevb.80.241409} {10.1103/physrevb.80.241409}
  (\bibinfo {year} {2009})\BibitemShut {NoStop}%
\bibitem [{\citenamefont {Mucha-Kruczyński}\ \emph {et~al.}(2008)\citenamefont
  {Mucha-Kruczyński}, \citenamefont {Tsyplyatyev}, \citenamefont {Grishin},
  \citenamefont {McCann}, \citenamefont {Fal’ko}, \citenamefont {Bostwick},\
  and\ \citenamefont {Rotenberg}}]{MuchaKruczyski2008}%
  \BibitemOpen
  \bibfield  {author} {\bibinfo {author} {\bibfnamefont {M.}~\bibnamefont
  {Mucha-Kruczyński}}, \bibinfo {author} {\bibfnamefont {O.}~\bibnamefont
  {Tsyplyatyev}}, \bibinfo {author} {\bibfnamefont {A.}~\bibnamefont
  {Grishin}}, \bibinfo {author} {\bibfnamefont {E.}~\bibnamefont {McCann}},
  \bibinfo {author} {\bibfnamefont {V.~I.}\ \bibnamefont {Fal’ko}}, \bibinfo
  {author} {\bibfnamefont {A.}~\bibnamefont {Bostwick}},\ and\ \bibinfo
  {author} {\bibfnamefont {E.}~\bibnamefont {Rotenberg}},\ }\bibfield
  {journal} {\bibinfo  {journal} {Physical Review B}\ }\textbf {\bibinfo
  {volume} {77}},\ \href {https://doi.org/10.1103/physrevb.77.195403}
  {10.1103/physrevb.77.195403} (\bibinfo {year} {2008})\BibitemShut {NoStop}%
\bibitem [{\citenamefont {Syzranov}\ \emph {et~al.}(2008)\citenamefont
  {Syzranov}, \citenamefont {Fistul},\ and\ \citenamefont
  {Efetov}}]{Syzranov2008}%
  \BibitemOpen
  \bibfield  {author} {\bibinfo {author} {\bibfnamefont {S.~V.}\ \bibnamefont
  {Syzranov}}, \bibinfo {author} {\bibfnamefont {M.~V.}\ \bibnamefont
  {Fistul}},\ and\ \bibinfo {author} {\bibfnamefont {K.~B.}\ \bibnamefont
  {Efetov}},\ }\bibfield  {journal} {\bibinfo  {journal} {Physical Review B}\
  }\textbf {\bibinfo {volume} {78}},\ \href
  {https://doi.org/10.1103/physrevb.78.045407} {10.1103/physrevb.78.045407}
  (\bibinfo {year} {2008})\BibitemShut {NoStop}%
\bibitem [{\citenamefont {Oka}\ and\ \citenamefont {Aoki}(2009)}]{Oka2009}%
  \BibitemOpen
  \bibfield  {author} {\bibinfo {author} {\bibfnamefont {T.}~\bibnamefont
  {Oka}}\ and\ \bibinfo {author} {\bibfnamefont {H.}~\bibnamefont {Aoki}},\
  }\bibfield  {journal} {\bibinfo  {journal} {Physical Review B}\ }\textbf
  {\bibinfo {volume} {79}},\ \href {https://doi.org/10.1103/physrevb.79.081406}
  {10.1103/physrevb.79.081406} (\bibinfo {year} {2009})\BibitemShut {NoStop}%
\bibitem [{\citenamefont {Schollw\"{o}ck}(2011)}]{Schollwck2011}%
  \BibitemOpen
  \bibfield  {author} {\bibinfo {author} {\bibfnamefont {U.}~\bibnamefont
  {Schollw\"{o}ck}},\ }\href {https://doi.org/10.1016/j.aop.2010.09.012}
  {\bibfield  {journal} {\bibinfo  {journal} {Annals of Physics}\ }\textbf
  {\bibinfo {volume} {326}},\ \bibinfo {pages} {96–192} (\bibinfo {year}
  {2011})}\BibitemShut {NoStop}%
\bibitem [{\citenamefont {Paeckel}\ \emph {et~al.}(2019)\citenamefont
  {Paeckel}, \citenamefont {K\"{o}hler}, \citenamefont {Swoboda}, \citenamefont
  {Manmana}, \citenamefont {Schollw\"{o}ck},\ and\ \citenamefont
  {Hubig}}]{Paeckel2019}%
  \BibitemOpen
  \bibfield  {author} {\bibinfo {author} {\bibfnamefont {S.}~\bibnamefont
  {Paeckel}}, \bibinfo {author} {\bibfnamefont {T.}~\bibnamefont {K\"{o}hler}},
  \bibinfo {author} {\bibfnamefont {A.}~\bibnamefont {Swoboda}}, \bibinfo
  {author} {\bibfnamefont {S.~R.}\ \bibnamefont {Manmana}}, \bibinfo {author}
  {\bibfnamefont {U.}~\bibnamefont {Schollw\"{o}ck}},\ and\ \bibinfo {author}
  {\bibfnamefont {C.}~\bibnamefont {Hubig}},\ }\href
  {https://doi.org/10.1016/j.aop.2019.167998} {\bibfield  {journal} {\bibinfo
  {journal} {Annals of Physics}\ }\textbf {\bibinfo {volume} {411}},\ \bibinfo
  {pages} {167998} (\bibinfo {year} {2019})}\BibitemShut {NoStop}%
\bibitem [{\citenamefont {Gadge}\ and\ \citenamefont
  {Manmana}(2025)}]{Gadge2025}%
  \BibitemOpen
  \bibfield  {author} {\bibinfo {author} {\bibfnamefont {K.}~\bibnamefont
  {Gadge}}\ and\ \bibinfo {author} {\bibfnamefont {S.~R.}\ \bibnamefont
  {Manmana}},\ }\bibfield  {journal} {\bibinfo  {journal} {Physical Review B}\
  }\textbf {\bibinfo {volume} {112}},\ \href
  {https://doi.org/10.1103/7fzq-zkbv} {10.1103/7fzq-zkbv} (\bibinfo {year}
  {2025})\BibitemShut {NoStop}%
\bibitem [{\citenamefont {Gadge}\ and\ \citenamefont
  {Manmana}(2026)}]{Gadge_data_2026}%
  \BibitemOpen
  \bibfield  {author} {\bibinfo {author} {\bibfnamefont {K.}~\bibnamefont
  {Gadge}}\ and\ \bibinfo {author} {\bibfnamefont {S.~R.}\ \bibnamefont
  {Manmana}},\ }\href {https://doi.org/10.5281/zenodo.18315671} {\bibinfo
  {title} {A comparative study of perturbative and nonequilibrium green's
  function approaches for floquet sidebands in periodically driven quantum
  systems[data set]}} (\bibinfo {year} {2026})\BibitemShut {NoStop}%
\bibitem [{\citenamefont {Lindeberg}(1994)}]{Lindeberg1994}%
  \BibitemOpen
  \bibfield  {author} {\bibinfo {author} {\bibfnamefont {T.}~\bibnamefont
  {Lindeberg}},\ }\href {https://doi.org/10.1007/978-1-4757-6465-9} {\emph
  {\bibinfo {title} {Scale-Space Theory in Computer Vision}}}\ (\bibinfo
  {publisher} {Springer US},\ \bibinfo {year} {1994})\BibitemShut {NoStop}%
\bibitem [{\citenamefont {Gonzalez}\ and\ \citenamefont
  {Woods}(2017)}]{Gonzalez2017}%
  \BibitemOpen
  \bibfield  {author} {\bibinfo {author} {\bibfnamefont {R.~C.}\ \bibnamefont
  {Gonzalez}}\ and\ \bibinfo {author} {\bibfnamefont {R.~E.}\ \bibnamefont
  {Woods}},\ }\href@noop {} {\emph {\bibinfo {title} {Digital Image Processing,
  Global Edition}}},\ \bibinfo {edition} {4th}\ ed.\ (\bibinfo  {publisher}
  {Pearson Education},\ \bibinfo {address} {London, England},\ \bibinfo {year}
  {2017})\BibitemShut {NoStop}%
\end{thebibliography}%
\end{document}